\documentclass{article}

\usepackage{arxiv}

\usepackage[utf8]{inputenc} 
\usepackage[T1]{fontenc}    
\usepackage{hyperref}       
\usepackage{url}            
\usepackage{booktabs}       
\usepackage{amsfonts}       
\usepackage{nicefrac}       
\usepackage{microtype}      
\usepackage{lipsum}		
\usepackage{graphicx}
\usepackage[numbers]{natbib}
\usepackage{doi}

\graphicspath{ {./figures/} }

\usepackage{amsmath}
\usepackage{calc}
\usepackage{float}
\usepackage{subfig}
\usepackage[usenames,dvipsnames,svgnames,table]{xcolor}
\usepackage{amsthm}
\usepackage[utf8]{inputenc}
\usepackage{listings}
\usepackage{xcolor}
\usepackage{appendix}

\title{Uncertainty Quantification for Transport in Porous media using Parameterized Physics Informed neural Networks}

\author{
  Cedric G. Fraces \\
  Department of Energy Resources Engineering\\
  Stanford University\\
  Stanford, CA 94305 \\
  \texttt{cfraces@stanford.edu} \\
   \And
 Hamdi Tchelepi \\
  Department of Energy Resources Engineering\\
  Stanford University\\
  Stanford, CA 94305 \\
  \texttt{tchelepi@stanford.edu} \\
}

\renewcommand{\headeright}{}

\begin{document}
\maketitle

\begin{abstract}
	We present a Parametrization of the Physics Informed Neural Network (P-PINN) approach to tackle the problem of uncertainty quantification in reservoir engineering problems. We demonstrate the approach with the immiscible two phase flow displacement (Buckley-Leverett problem) in heterogeneous porous medium. The reservoir properties (porosity, permeability) are treated as random variables. The distribution of these properties can affect dynamic properties such as the fluids saturation, front propagation speed or breakthrough time. We explore and use to our advantage the ability of networks to interpolate complex high dimensional functions. We observe that the additional dimensions resulting from a stochastic treatment of the partial differential equations tend to produce smoother solutions on quantities of interest (distributions parameters) which is shown to improve the performance of PINNS. We show that provided a proper parameterization of the uncertainty space, PINN can produce solutions that match closely both the ensemble realizations and the stochastic moments. We demonstrate applications for both homogeneous and heterogeneous fields of properties. We are able to solve problems that can be challenging for classical methods. This approach gives rise to trained models that are both more robust to variations in the input space and can compete in performance with traditional stochastic sampling methods. 
\end{abstract}


\section{Introduction: Stochastic Riemann problem}
\label{sec:stoch_riemann}
We present new results using the P-PINN approach. So far, the forward problem was used to resolve a single set of input parameters (boundary conditions and physical properties of the system). We know that these can be highly uncertain as our knowledge of the subsurface is incomplete. We show how PINN can be extended to solve ensemble of solutions. Sampling methods in general allow to treat various components of the problem in a probabilistic manner and get solutions for distributions of input. 
For hyperbolic problems in particular where shocks can develop, we find that stochastic solutions tend to smooth out the solutions and help with the convergence of PINNs. 

The problem presented here is that of incompressible, immiscible displacement of two phases in a porous medium. This is also referred to as the transport problem and has been delineated in various forms over the years. It has been applied to the displacement of oil by water for waterflood applications in reservoir for over half a century (Buckley-Leverett, \cite{BuckleyLeverett1942}) or more recently to the displacement of brine by CO2 in Carbon Sequestration applications, \cite{Orr225}. 
We assume that a wetting phase (w) is displacing a non-wetting phase (n). We remind that the wettability is the preference of a fluid to be in contact with a solid surrounded by another fluid (example, water is wetting on most surfaces vs air) and conservation of mass applies to both phases. For the wetting phase:

The 1D conservation for the two phase (water-$CO_2$) becomes:

\begin{equation}
    \phi \frac{\partial S_w}{\partial t} + \nabla q_w = 0
    \label{eq::Conservation_w}
\end{equation}

Where $\phi$ is the porosity, $S_w$ is the saturation (or concentration) of that wetting phase (while $S_n=1-S_w$ is the saturation of the non-wetting phase). The flow rate of the wetting phase $q_w$ can be written:

\begin{equation}
    q_w = -\frac{kk_{rw}(S_w)}{\mu_w}\nabla p
\end{equation}

We can re-write equation \ref{eq::Conservation_w} as a function of the total flux. This gives rise to the fractional flow:

\begin{equation}
    f_w = \frac{q_w}{q} = \frac{1}{1+k_{rn}\mu_w/k_{rw}\mu_n}
\end{equation}

The conservation equation can now be written:

\begin{equation}
    \label{eq:conservation_phi}
    \frac{\partial S_w}{\partial t} + \frac{q(x,t)}{\phi(x)}\nabla f_w = 0
\end{equation}

Where the fractional flow is a nonlinear equation defined as:

\begin{equation}
    f_w(S) = \frac{(S - S_{wc})^2}{(S - S_{wc})^2 + (1 - S - S_{nr})^2/M}
\end{equation}

Where $S_{wc}$ and $S_{nr}$ are the residual (irreducible) saturations for the wetting and non-wetting phases resulting from trapping mechanisms and $M$ is the end point mobility ratio defined as the ratio of end-point relative permeability and viscosity of both phases. We use Corey and Brooks relative permeability relationship.
The partial differential equation solved here is hyperbolic of order 1 and the fractional flow term is non-convex. It belongs to the class of Riemann conservation problem that is typically solved using finite volume methods (\cite{Lax_Hyperbolic}). Coincidentally, the Method of Characteristics (MOC) can be used to find an analytical solution to this problem if the boundary conditions are uniform.

\begin{eqnarray}
    S(x=0,t) &= S_{inj}\\ 
    S(x,t=0) &= S_{wc}
\end{eqnarray}

\begin{figure}[h!]%
    \centering
    \includegraphics[width=0.5\linewidth]{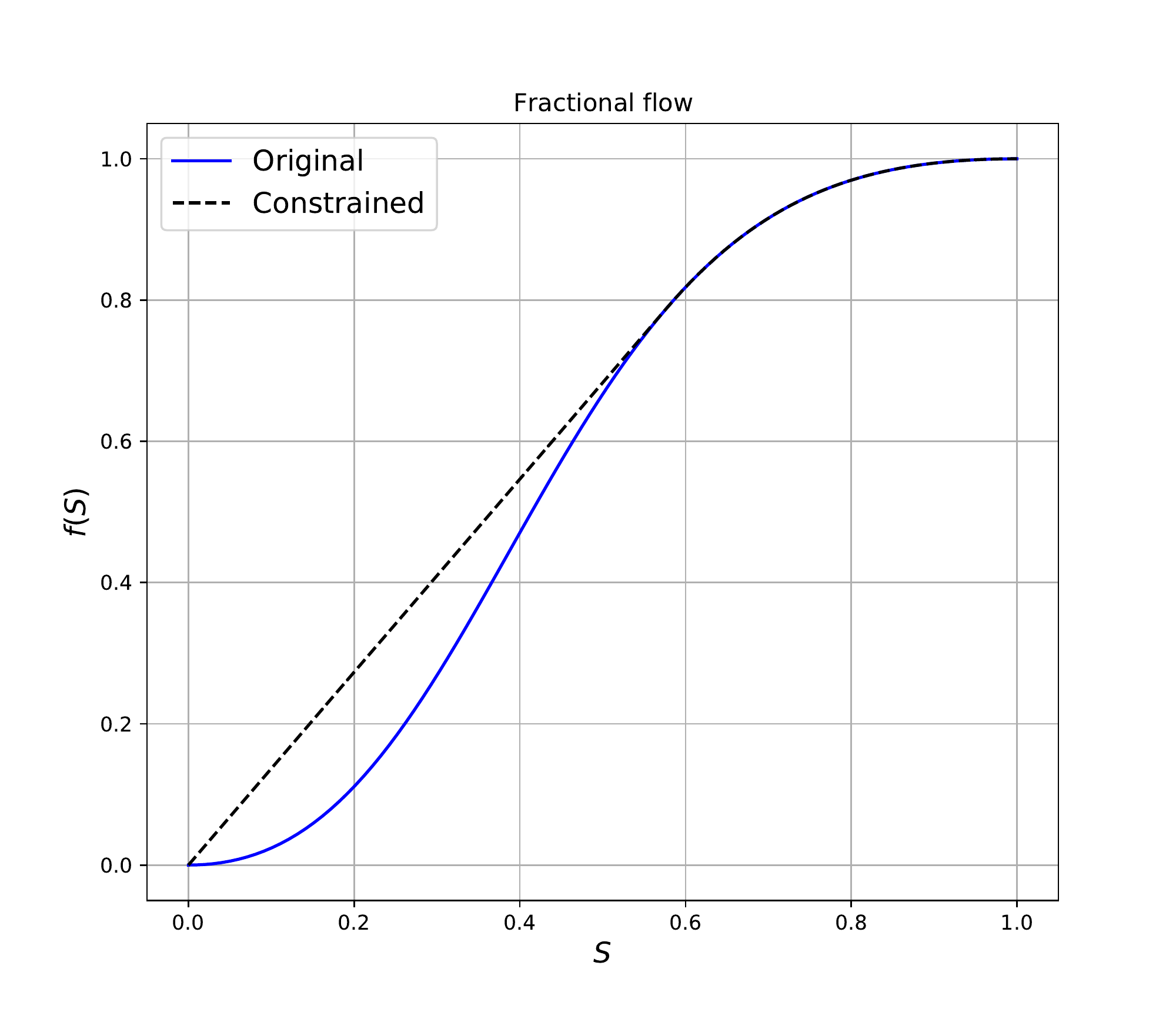}%
    \caption{Fractional flow curve (blue) along with Welge construction (dotted black) for case with $S_{wc}=S_{or}=0$ (ref. \cite{gasmi2021physics})}%
    \label{fig:frac_flow_welge}%
\end{figure}

For the MOC or any finite volume method to be conservative, one needs to modify the fractional flow term as shown in Figure \ref{fig:frac_flow_welge}.

\cite{Zhang_tchelepi1999} presents a stochastic analysis of the Buckley-Leverett problem. In this formulation, the porosity $\phi(x)$ is non-uniform and distributed randomly across space. If we assume that the Darcy flow rate $q(x,t)=q(x)$ is uniform with time, we can re-write Eq.~\ref{eq:conservation_phi}:

\begin{equation}
    \label{eq:conservation_vd}
    \frac{\partial S_w}{\partial t} + v_d f_w^{\prime}(S) \frac{\partial S_w}{\partial x} = 0
\end{equation}

Where $v_d=q(x)/\phi(x)$ is the total (Darcy) velocity vector. Since $\phi$ is a random function, this has an effect on $v_d$ and $S_w$ which are treated as a random variables. We will explore the effect of varying the distribution parameters for $v_d$ on $S_w$.

\subsection{Related work}
In the family of numerical methods proposed to solve random PDEs, Monte Carlo Simulation (MCS) and its derivatives are the most commonly used. Several approaches have been proposed to improve upon MCS for the problem of flow and transport in porous media. 

Some of them focus on establishing a more efficient sampling strategy in order to improve upon the convergence rate of classical MCS. \cite{Muller_Jenny2013} apply a Multi-level Monte Carlo approach to the flow and transport problem using a streamline based solver on coarse grids to guide the generation of sample on finer grids decreasing the need for large sample number. They present numerical results on Gaussian permeability fields that can be run an order of magnitude faster than MCS without loss of accuracy. 

\cite{phdPipat}, \cite{Zhang_tchelepi1999} propose a method based on the resolution of statistical moments where random input and output are expanded using Taylor series approximation around their mean. The stochastic PDEs can be transformed into deterministic PDEs that describe the behavior of the output moments. Example where closure for the mean and variance are derived are presented. \cite{Jarman2000}, \cite{Jarman2006}, \cite{Langlo1994} develop a solution for correlated fields where the correlation structure of the input random fields is used to derive PDEs where the randomness is equated to the addition of a diffusive term whose effects can be compared to a physical macro-dispersion. 

\cite{wang2013}, \cite{phdFay}, \cite{fuksUQ} propose a probability distribution method based on frozen streamlines that relies on the monotonicity of the mapping between time of flight and solute concentration cumulative distribution functions. It propagates uncertainty using an analytical method and reconstruct the statistics of the QOI for non-linear transport. 

Polynomial Chaos and Spectral methods (\cite{StochasticFE}) rely on the projection of random variables on orthogonal basis of polynomials using Galerkin-like decomposition. \cite{petterson2014} proposes such an approach for the Buckley-Leverett function and studies convergence based on the number of basis functions used to represent the solution. These techniques of projection have been the starting point of neural network based approximations. 

A new class of methods relying on deep learning frameworks have recently been proposed to solve partial differential equations with random operators. \cite{NABIAN201914} uses a residual neural network to solve random parabolic and elliptic equations. 
\cite{fraces2020physics} presents a method of variational inference to quantify the uncertainty in the Buckley Leverett problem. It uses Generative Adversarial Networks (GAN) to solve the the problem in the presence of noisy data. GANs have been known to provide good maximum likelihood estimators in the case on non parametric, non tractable density distributions. However, they are sensitive to hyper-parameter during training and suffer from mode collapse.

\section{Paramterization of Physics Informed Neural Networks}
We describe the formalism of the physics informed approach applied to the stochastic problem. We remind that for the deterministic formulation, we solve the following equation:
\begin{equation}
    \mathcal{R}(\mathbf{X}, t, \mathbf{\nu}, \mathbf{\nu}_t, \nabla\mathbf{\nu}, \nabla^2\mathbf{\nu},\dots) = 0
\end{equation}
Where $\mathcal{R}$ is a non-linear differential operator, $\mathbf{\nu}$ is the state variables such as the reservoir pressure and saturation $\mathbf{\nu}=[p, S]^T$ and $\mathbf{X}$ are the space coordinates ($(x,y,z)$ in 3D although we will mostly show application in one space dimension).
We assume that it is represented by a multi-layer perceptron of the form:
\begin{equation}
\label{eq:ffwd_form_base}
    \mathbf{\nu} \approx \mathbf{\nu}_{\theta} = \sigma\left[\mathbf{W}^{[n]}\times \sigma(\mathbf{W}^{[n-1]}(\dots\sigma(\mathbf{W}^{[0]}[\mathbf{X},t]^T + \mathbf{b}^{[0]}))\dots + \mathbf{b}^{[n-1]}) + \mathbf{b}^{[n]}\right]
\end{equation}
Where $\sigma$ is a nonlinear activation function (sigmoid, tanh, ReLu,...), $\mathbf{W}_i$ are weight matrices and $\mathbf{b_i}$ are bias vectors for layer $i$. 

$\nu_{\theta}(\mathbf{X},t)$ is a continuous and differentiable representation of the state variables. For any $\mathbf{X},t$ we can compute an output saturation and pressure along with their derivatives with respect to any coordinates. This allows us to construct the residual $\mathcal{R}$ of the PDE using automatic differentiation (\cite{tensorflow2015-whitepaper}).

\section{Homogeneous 1D case}

We present the 1D Buckley Leverett solution with total velocity $v_d$ constraint where the velocity is constant for a given realization but varies from one realization to the next. With a single training using the parameterized PINN (P-PINN) approach , we intend to produce a solution that will honor solutions for all realizations within the bounds of the distribution initially used. This formulation allows to explore the stability of PINNs to variation in the input space.

\begin{figure}[h!]
    \centering
    \includegraphics[width=0.8\linewidth]{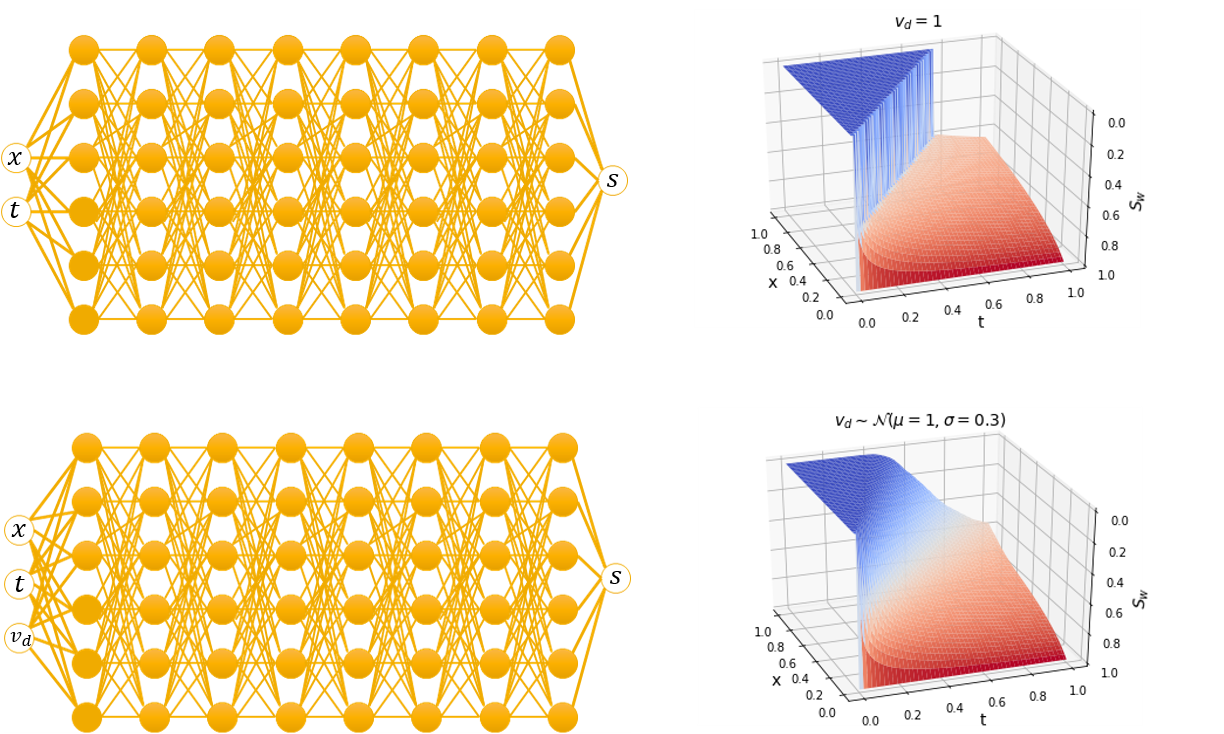}
    \caption{Representation of the neural network model (input/output) and solution for the deterministic (top) and probabilistic (bottom) Buckley-Leverett problem using PINNs. The saturation distributions surfaces $(x, t, S_w)$ for a deterministic (top) and stochastic (bottom) Total velocity are represented on the right.}
    \label{fig:mlp_probabilistic}
\end{figure}

Figure~\ref{fig:mlp_probabilistic} shows the difference between the deterministic and stochastic models and solutions.

We see that while the deterministic solution requires a lower dimensional input space, it features sharp gradients (discontinuity, shocks). The stochastic solution on the other hand is smoother. This characteristic of stochastic solutions is interesting for training as we have empirically observed (\cite{Fuks2020}, \cite{gasmi2021physics}) that solutions with steep gradients can be challenging when trying to solve PDEs using PINNs.

\subsection{Implementation}
Just like in Eq.\ref{eq:ffwd_form_base}, the model is a dense neural network of the form:
\begin{equation}
    \mathbf{\nu} \approx \mathbf{\nu}_{\theta} = \sigma\left[\mathbf{W}^{[n]}\times \sigma(\mathbf{W}^{[n-1]}(\dots\sigma(\mathbf{W}^{[0]}[\mathbf{X},t, v_d]^T + \mathbf{b}^{[0]}))\dots + \mathbf{b}^{[n-1]}) + \mathbf{b}^{[n]}\right]
\end{equation}

We note that this model takes three arguments as input ($[\mathbf{X},t, v_d]$) versus two in the previous form. Indeed, the space time and velocity dimensions can be considered independently. In practice, the addition of a third dimension has a smoothing effect on the form to be interpolated as shown in Figure \ref{fig:mlp_probabilistic}.

During training, the three variables are uniformly sampled across a wide enough parameter range (\ref{eq:distrib_xtvd}):
\begin{equation}
\label{eq:distrib_xtvd}
\begin{split}
    x\sim \mathcal{U}(low=0., up=1)\\
    t\sim \mathcal{U}(low=0., up=1)\\
    v_d\sim \mathcal{U}(low=0., up=10)
\end{split}
\end{equation}
 An entropy constrained fractional flow curve is used (figure~\ref{fig:frac_flow_welge}). The loss functions representing initial, boundary and interior conditions are implemented:
 
Initial condition:
\begin{equation}
\label{eq:IC}
        \mathcal{L}_{IC} = \sum_{i=1}^{N}\left\Vert S_w(x^{(i)}, 0, v_d^{(i)})\right\Vert^2 + \left\Vert \frac{\partial S_w(x^{(i)}, 0, v_d^{(i)})}{\partial t^{(i)}}\right\Vert^2
\end{equation}
The addition of the derivative at initial condition has little influence to the final result. It is nonetheless possible to over-constrain this problem since we minimize the least square function defined in eq.~\ref{eq:IC}

Boundary condition:
\begin{equation}
\label{eq:BC}
        \mathcal{L}_{BC} = \sum_{i=1}^{N}\left\Vert S_w(0, t^{(i)}, v_d^{(i)}) - 1\right\Vert^2
\end{equation}

Interior/collocation:
\begin{equation}
\label{eq:collocation}
    \mathcal{L}_{R} = \sum_{i=1}^{N}\left\Vert\frac{\partial S_w(x^{(i)},t^{(i)},v_d^{(i)})}{\partial t^{(i)}} + f_w^{\prime}(S_w(x^{(i)},t^{(i)},v_d^{(i)}))v_d^{(i)}\frac{\partial S_w(x^{(i)},t^{(i)},v_d^{(i)})}{\partial x^{(i)}}\right\Vert^2
\end{equation}

Note that we use $N=5000$ sample points for training and apply multipliers of $1$ for all three losses. We want to keep the hyper-parameter of the model to the simplest and do not intend to perform much optimization on that aspect of the process. We will present the rationale behind this choice later. Note that the training occurs once.

At inference, we load the model trained and simulate saturation realizations based on series of input $(x,t,v_d)$. We then perform MCS using the trained network as forward model. We can use non uniform sampling distribution for the velocity $v_d$ as shown in subsequent sections.

\subsection{Normal distribution}
We assume a Normal Darcy velocity $v_d$ distribution:

\begin{equation}
    \label{eq:distrib_v_d_narrow}
    v_d\sim \mathcal{N}(\mu=1,\,\sigma=0.3, low=0.5, up=2)
\end{equation}

We truncate the distribution between $0.5$ and $2$ for physical consistency (positive velocity) and to ensure that the shock front remains within spatial bounds for most of the simulation duration

\begin{figure}[h!]%
    \centering
    \includegraphics[width=0.5\linewidth]{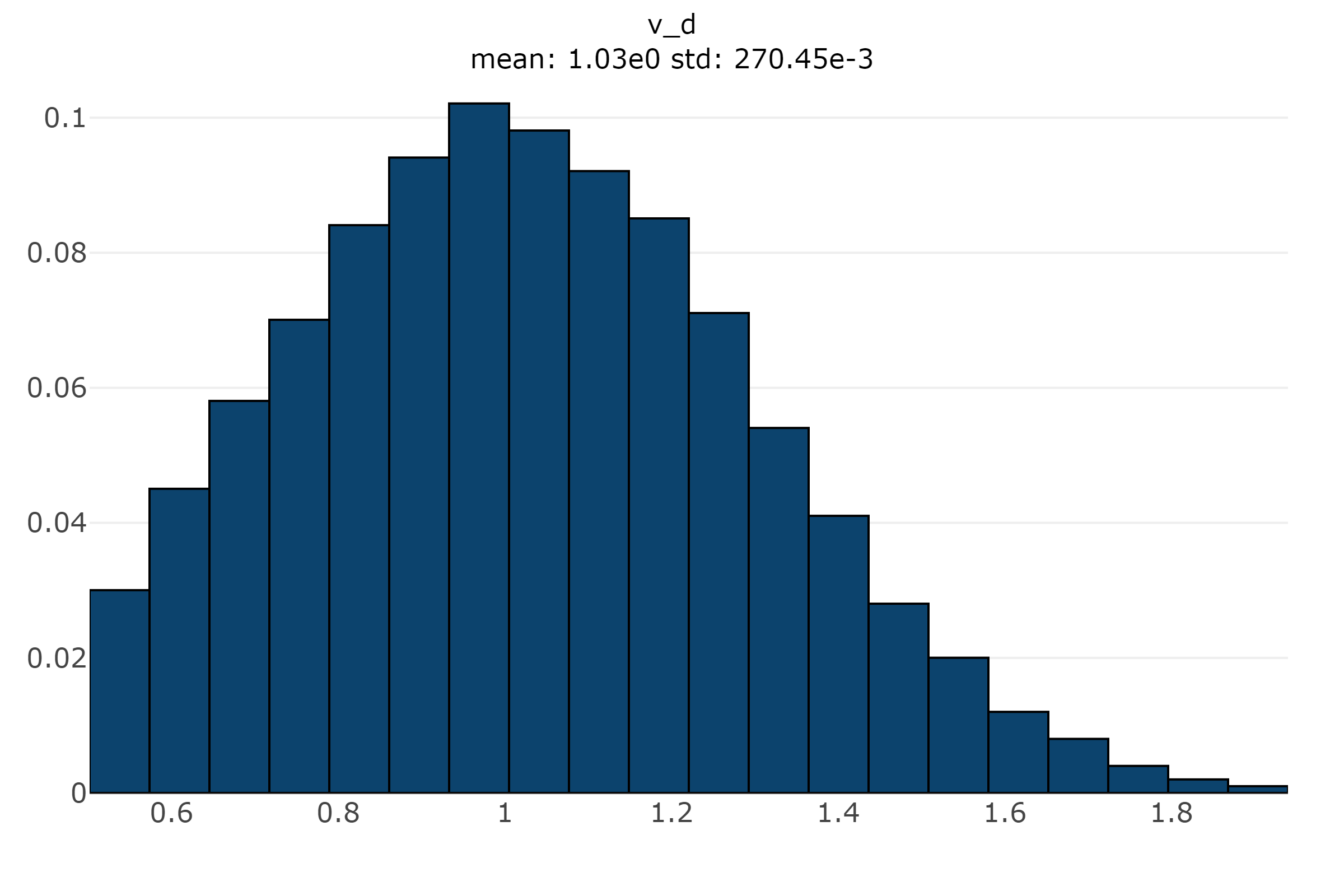}%
    \caption{Total (Darcy) velocity distribution for stochastic resolution of Buckley Leverett equation}%
    \label{fig:v_d_narrow}%
\end{figure}

\begin{figure}[h!]%
    \centering
    \includegraphics[width=1\linewidth]{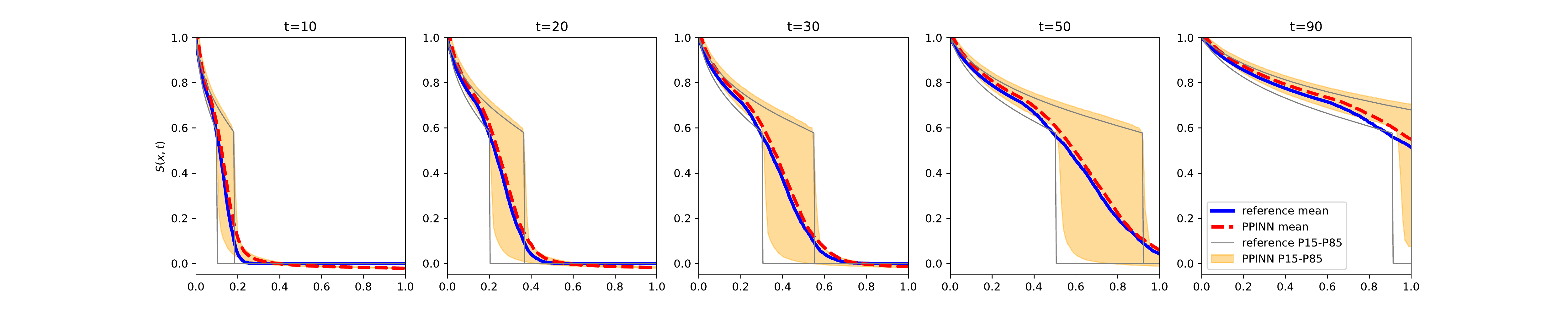}%
    \caption{Comparison of saturation distributions profiles at five different time steps for a Total velocity $v_d\sim \mathcal{N}(\mu=1,\,\sigma=0.3, low=0.5, up=2)$. The reference mean saturation (computed through Monte Carlo simulation of MOC) is in solid blue while the one computed with P-PINNs is in dashed red. The P15-P85 envelopes are represented for both reference and P-PINN}%
    \label{fig:sat_dist_vd_narrow}%
\end{figure}

Figure \ref{fig:sat_dist_vd_narrow} shows a comparison of the probabilistic realizations obtained with P-PINN and the Method of Characteristics (MOC) forward models. We arbitrarily choose 1000 samples for both cases. We represent the mean of all the realizations for both the P-PINN and a reference MOC approach. The uncertainty envelope around the profile represents the $15$th to $85$th percentiles

\begin{figure}[h!]%
    \centering
    \includegraphics[width=1\linewidth]{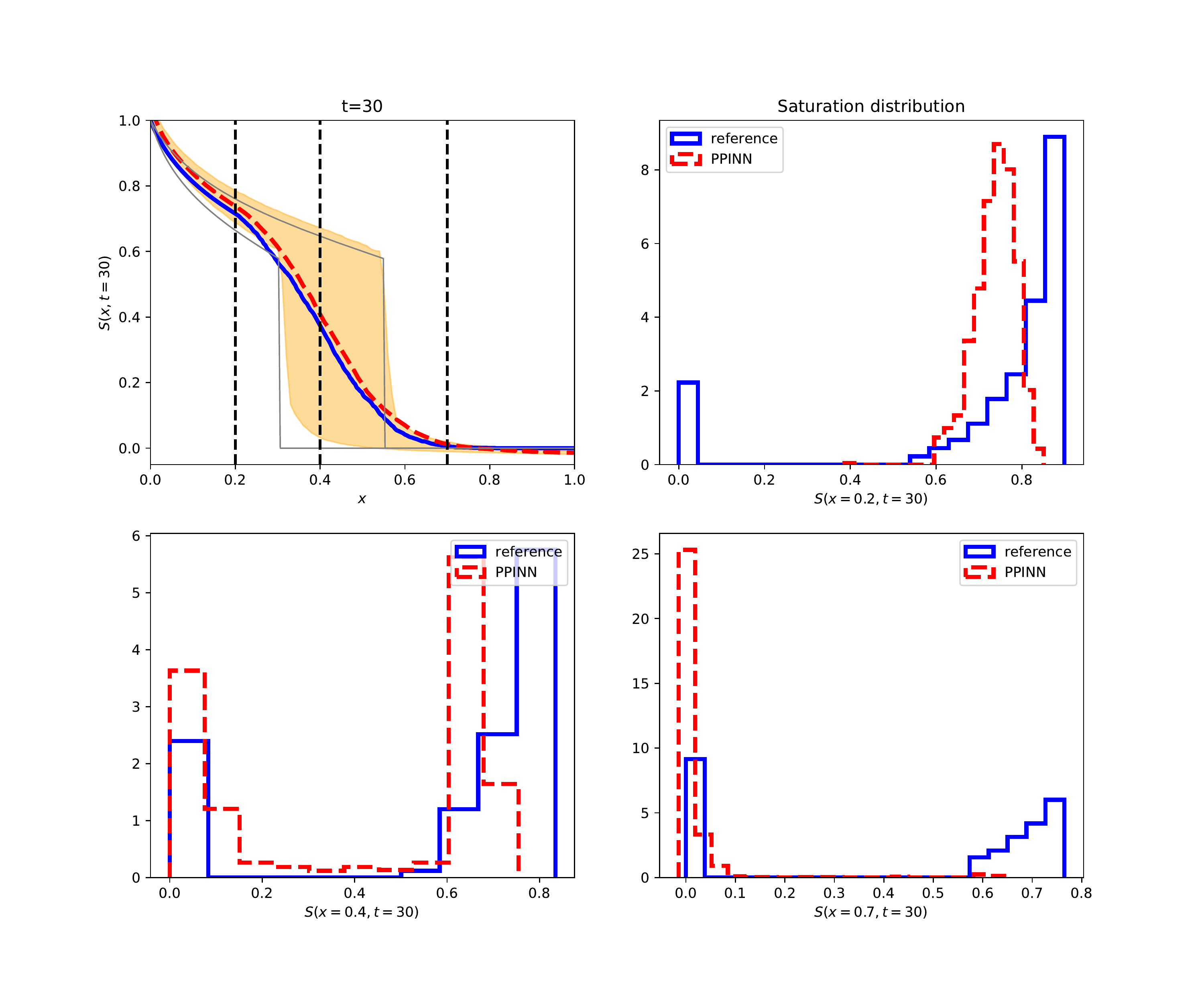}%
    \caption{Comparison of saturation distributions profiles at three different locations (represented by the black dotted line on the top left) for a Total velocity $v_d\sim \mathcal{N}(\mu=1,\,\sigma=0.3, low=0.5, up=2)$. The reference distributions (blue) as well as the one obtained with P-PINN (red) are represented at $x=(0.2, 0.4, 0.7)$}%
    \label{fig:sat_dist_t_30}%
\end{figure}

We represent the saturation distributions at different locations for a given time (Figure \ref{fig:sat_dist_t_30}). This highlights the error around the shock region. Although the mean value of the saturation for all realizations is well preserved across time, the quantiles display some error as shown in Figure \ref{fig:sat_dist_t_30}.

\begin{figure}[h!]%
    \centering
    \includegraphics[width=1\linewidth]{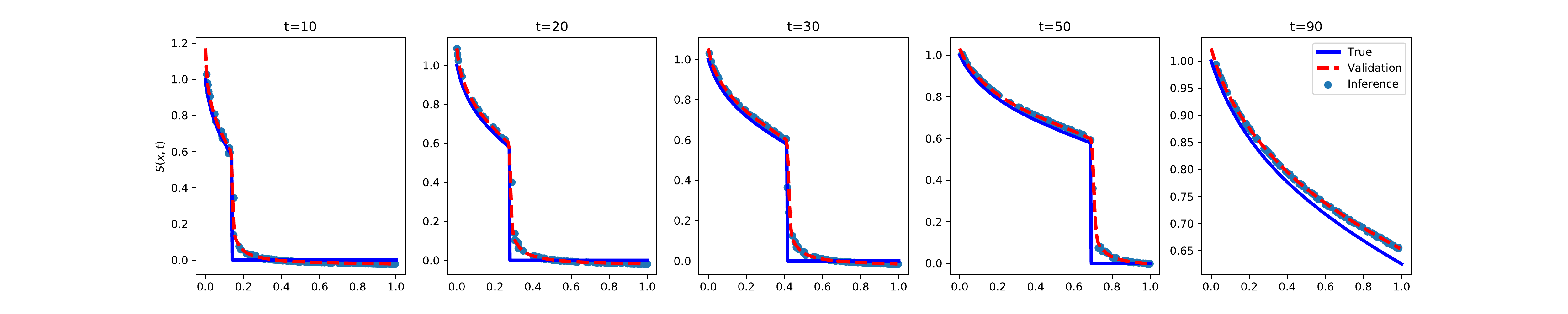}%
    \caption{Comparison of saturation profile at five different time steps for a Total velocity $v_d=1$. The "True" saturation (computed through MOC) is in solid blue while the validation computed with P-PINN is in dashed red. The result of inference is dotted}%
    \label{fig:saturation_vd_narrow}%
\end{figure}

The saturation results we get for one realization are presented in Figure \ref{fig:saturation_vd_narrow}. The individual realizations do not match the analytical solution as well as for the deterministic cases presented in \cite{gasmi2021physics} (slight smearing of the shock). This is partly due to the interpolation error that occurs as a result of computing a numerical solution over the entire distribution of total velocities. This smearing has consequences on the model performances just like in numerical simulation. The error does not grow with time though as we operate in a totally mesh free (both space and time) setup.

\begin{figure}[h!]%
    \centering
    \includegraphics[width=1\linewidth]{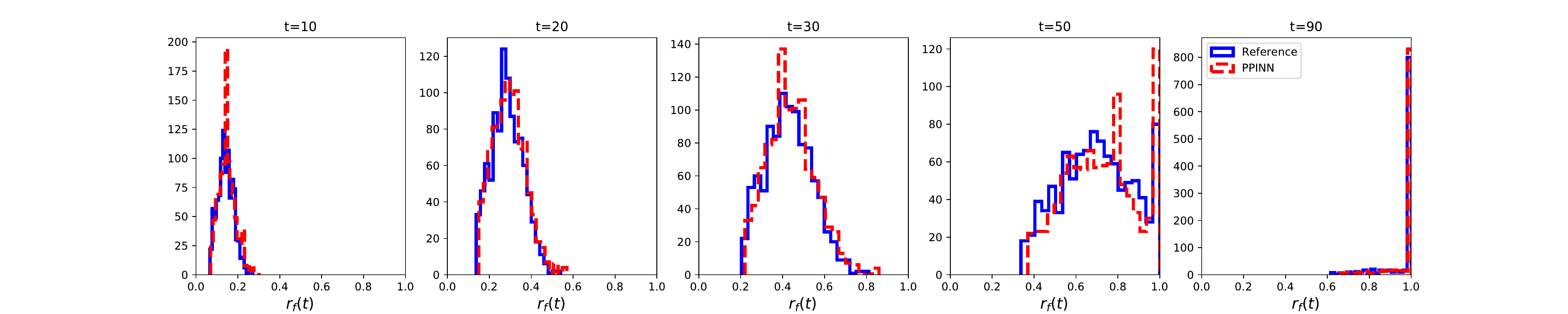}%
    \caption{Comparison of front radius distributions profiles at five different time steps for a Total velocity $v_d\sim \mathcal{N}(\mu=1,\,\sigma=0.3, low=0.5, up=2)$. The reference front radius (computed through Monte Carlo simulation with MOC) is in solid blue while the one computed with P-PINNs is in dashed red.}%
    \label{fig:front_radius_vd_narrow}%
\end{figure}

The Buckley-Leverett model is typically used to predict:
\begin{itemize}
    \item The extent of the multi-phase flow region and its progression with time
    \item the injected fluid's breakthrough time at given locations
\end{itemize}  
We represent the associated distributions for these two quantities of interest (QOI) and compare them with the reference analytical solution. The distributions of shock radii are shown in Figure \ref{fig:front_radius_vd_narrow}. They present close matches with 

\begin{figure}[h!]%
    \centering
    \includegraphics[width=1\linewidth]{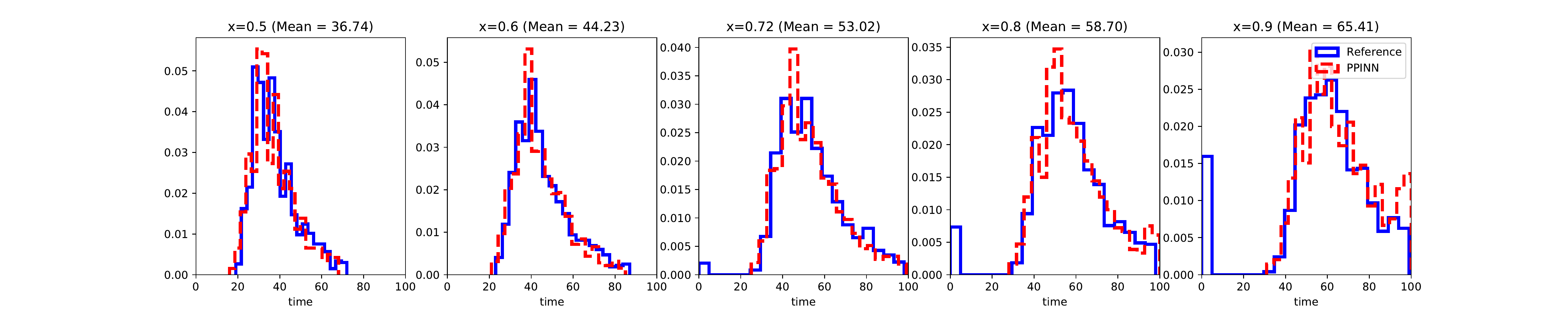}%
    \caption{Comparison of breakthrough time distributions profiles at five different locations in space for a Total velocity $v_d\sim \mathcal{N}(\mu=1,\,\sigma=0.3, low=0.5, up=2)$. The reference breakthrough time (computed through Monte Carlo simulation with MOC) is in solid blue while the one computed with P-PINNs is in dashed red.}%
    \label{fig:bt_vd_narrow}%
\end{figure}

The distribution of the breakthrough times follow closely the reference produced using MCS with MOC forward model as shown in Figure \ref{fig:bt_vd_narrow}

We observe that the QOIs such as breakthrough time and front radius are rather forgiving to the front smearing effect we highlighted for single realizations as well as for saturation distributions at specific space and time coordinates.

We compare the output distribution. Several options are available including the Kullback–Leibler (KL) divergence (\cite{kl1951}, \cite{kullback1997information}) defined as
\begin{equation}
\label{eq:kl_div}
    {\displaystyle D_{\text{KL}}(P_{ref}\parallel P)=\sum _{x\in {\mathcal {X}}}P_{ref}(x)\log \left({\frac {P_{ref}(x)}{P(x)}}\right).}
\end{equation}
It is a statistical distance between two distributions $P_{ref}$ and $P$. or the Jensen-Shannon divergence (\cite{Shannon_1948}) which is a is a symmetrized and smoothed version of the Kullback–Leibler divergence defined as:
\begin{equation}
\label{eq:kl_div}
    {\displaystyle JSD(P_{ref}\parallel P)=\frac{1}{2}D_{\text{KL}}(P_{ref}\parallel M)+\frac{1}{2}D_{\text{KL}}(P\parallel M)}
\end{equation}
Where $M = \frac{1}{2}(P_{ref}+P)$. Both measures of the distance required some normalization and provide a measure that can be hard to interpret.
For this analysis, we choose to use the Wasserstein distance (\cite{wasserstein1969}) defined as:
\begin{equation}
    {\displaystyle W_p(P_{ref}, P) = \int_{-\infty}^{\infty} \|\Phi(P)-\Phi(P_{ref})\|}
\end{equation}
Where $\Phi$ is the cdf function. \cite{ramdas2015wasserstein} describes a development of the distance. It can be seen as the minimum amount of distribution weight that must be moved times the distance that it must be moved to transform $P$ into $P_{ref}$. Table~\ref{tab:vd_narrow} shows the average Wasserstein distances between distributions computed using P-PINNS and the reference for  both front radius and breakthrough times.

\begin{table}[h!]
	\caption{Wasserstein distance average for distributions of QOIs for a Total velocity $v_d\sim \mathcal{N}(\mu=1,\,\sigma=0.3, low=0.5, up=2)$. U is a uniform distribution.}
	\centering
	\begin{tabular}{ccc}
		\toprule
		    & Front Radius     & Breakthrough time \\
		\midrule
		$W_p(P_{MOC}, P_{PINN})$ & 0.01  & 2.39 \\
        $W_p(P_{MOC}, U)$     & 0.29 & 49.91 \\
        \midrule
        Relative difference & 4.2\% & 4.8\% \\
		\bottomrule
	\end{tabular}
	\label{tab:vd_narrow}
\end{table}

\subsection{Normal distribution with extended support}

\begin{figure}[h!]%
    \centering
    \includegraphics[width=0.5\linewidth]{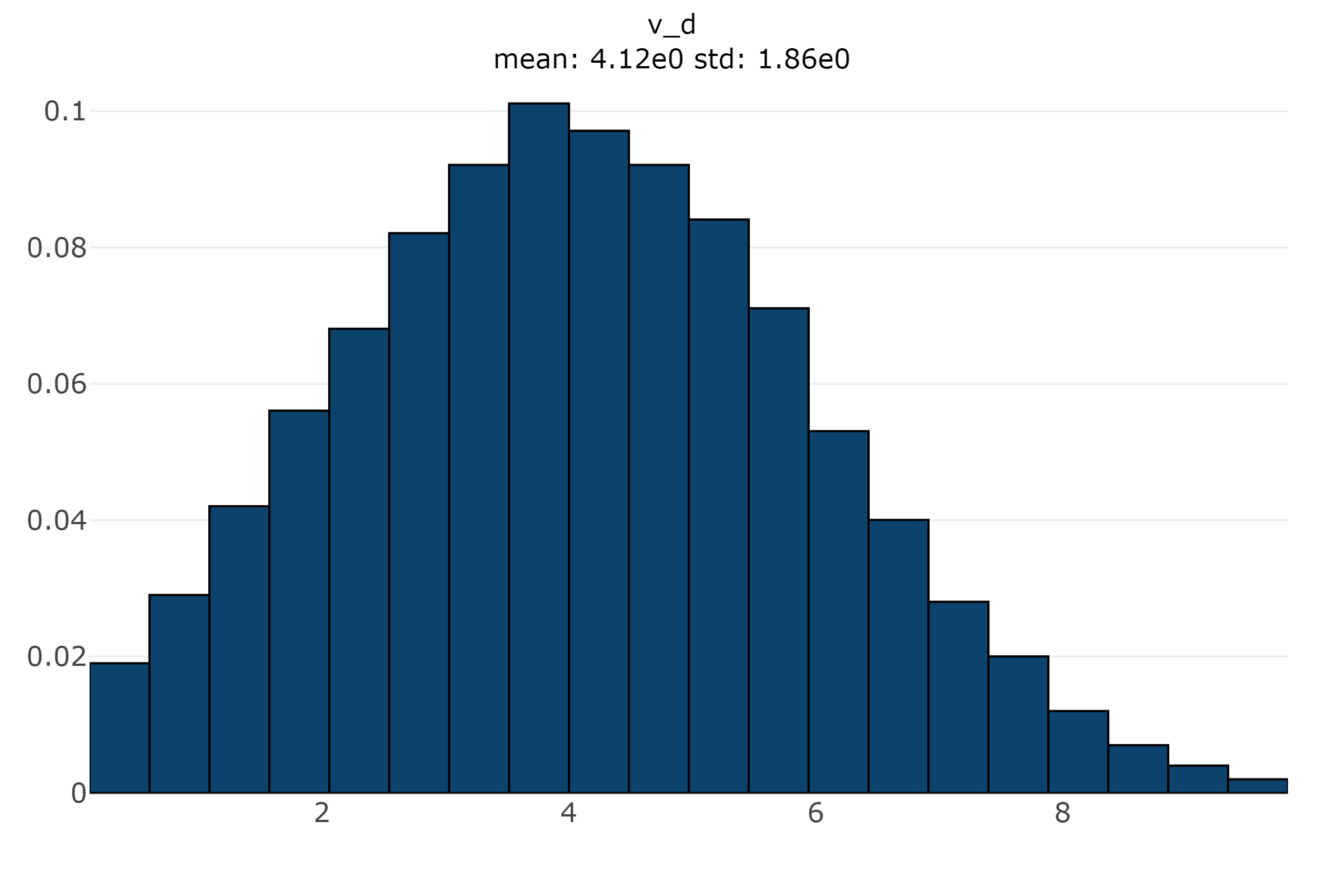}%
    \caption{Total (Darcy) velocity distribution with extended support for stochastic resolution of Buckley Leverett equation}%
    \label{fig:v_d_wide}%
\end{figure}

We test how sensitive the method is to wider distribution supports. In this setup the total velocity $v_d$ follows a Normal distribution defined by Equation~\ref{eq::distrib_vd_wide}. 

\begin{equation}
    \label{eq::distrib_vd_wide}
    v_d\sim \mathcal{N}(\mu=4,\,\sigma=2, low=0.1, up=10)
\end{equation}
This represent a 100 fold ratio between the minimum and maximum value as represented in Figure~\ref{fig:v_d_wide}.

\begin{figure}[h!]%
    \centering
    \includegraphics[width=1\linewidth]{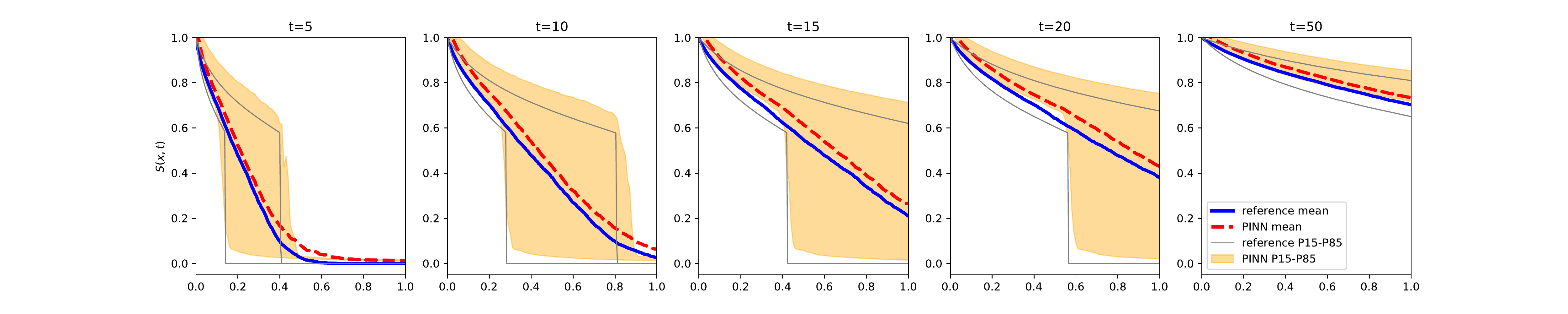}%
    \caption{Comparison of saturation distributions profiles at five different time steps for a Total velocity $v_d\sim \mathcal{N}(\mu=4,\,\sigma=2, low=0.1, up=10)$. The reference mean saturation (computed through Monte Carlo simulation of MOC) is in solid blue while the one computed with P-PINNs is in dashed red. The P15-P85 envelopes are represented for both reference and P-PINN}%
    \label{fig:saturation_vd_wide}%
\end{figure}

\begin{figure}[h!]%
    \centering
    \includegraphics[width=1\linewidth]{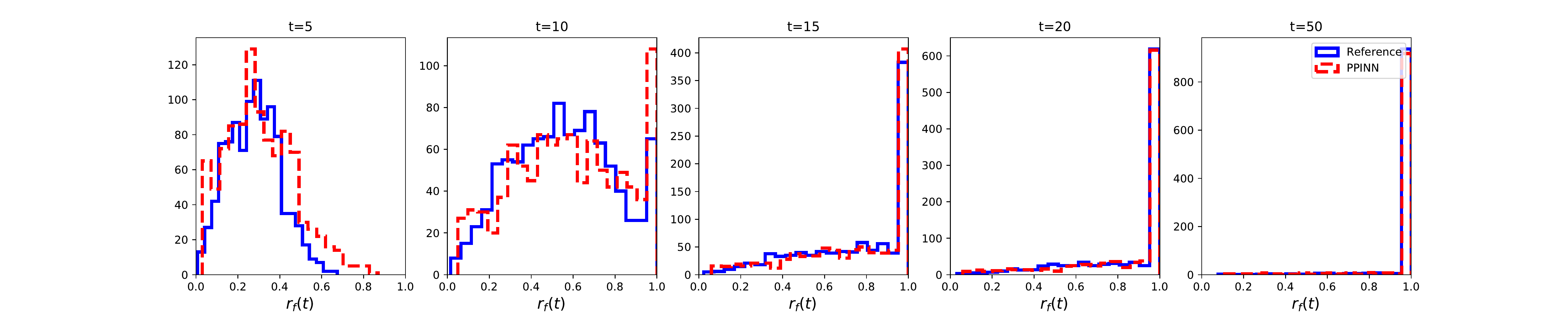}%
    \caption{Comparison of front radius distributions profiles at five different time steps for a Total velocity $v_d\sim \mathcal{N}(\mu=4,\,\sigma=2, low=0.1, up=10)$. The reference front radius (computed through Monte Carlo simulation with MOC) is in solid blue while the one computed with P-PINNs is in dashed red.}%
    \label{fig:front_radius_vd_wide}%
\end{figure}

We represent the breakthrough time and front radius distributions for the ensemble realizations and compare with the reference analytical solution. The distributions of shock radii are shown in Figure \ref{fig:front_radius_vd_wide}. 

\begin{figure}[h!]%
    \centering
    \includegraphics[width=1\linewidth]{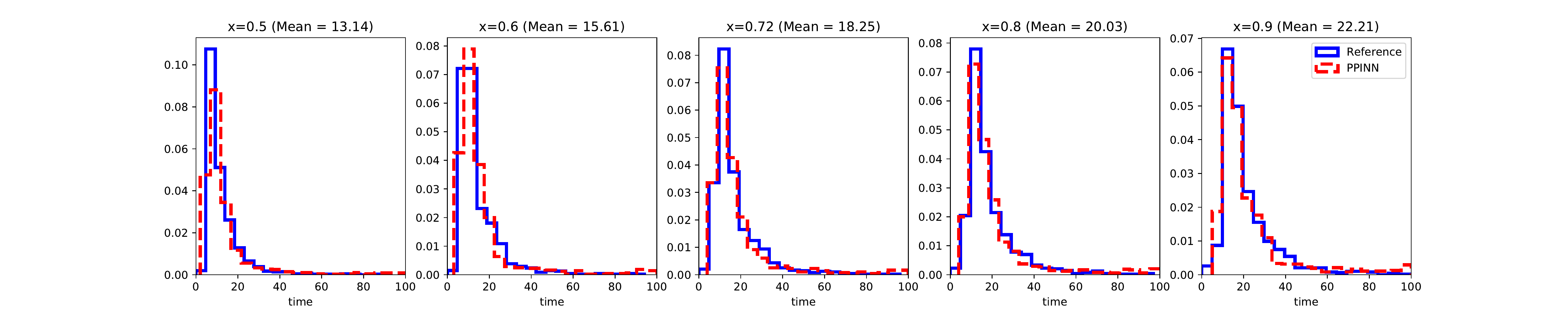}%
    \caption{Comparison of breakthrough time distributions profiles at five different locations in space for a Total velocity $v_d\sim \mathcal{N}(\mu=4,\,\sigma=2, low=0.1, up=10)$. The reference breakthrough time (computed through Monte Carlo simulation with MOC) is in solid blue while the one computed with P-PINNs is in dashed red.}%
    \label{fig:bt_vd_wide}%
\end{figure}

Once again, the distribution of the breakthrough time follows closely the reference as shown in Figure \ref{fig:bt_vd_wide} and the increase in overall variance of the input signal does not result in a larger output error.

Table~\ref{tab:vd_wide} shows the average Wasserstein distances between distributions computed using P-PINNS and the reference for  both front radius and breakthrough times.

\begin{table}[h!]
	\caption{Wasserstein distance average for distributions of QOIs for a Total velocity $v_d\sim \mathcal{N}(\mu=4,\,\sigma=2, low=0.1, up=10)$. U is a uniform distribution.}
	\centering
	\begin{tabular}{ccc}
		\toprule
		    & Front Radius     & Breakthrough time \\
	    \midrule
		$W_p(P_{MOC}, P_{PINN})$ & 0.02  & 2.01 \\
        $W_p(P_{MOC}, U)$     & 0.27 & 34.92 \\
        \midrule
        Relative difference & 7.9\% & 5.7\% \\
		\bottomrule
	\end{tabular}
	\label{tab:vd_wide}
\end{table}

\subsection{Residual and initial saturation effects}
So far, we only presented results that had trivial residual and injection saturation ($0$ and $1$) with injection saturation equal to residual water saturation.
In reality, fluid in place and displacing fluid interact with the rock and through trapping mechanisms can become immobile within the pore space. \cite{ERE221} offers a comprehensive analysis of these mechanisms. Residual saturation for the wetting and non wetting phase are accounted for in this example:

\begin{eqnarray}
    S(x,t=0) &= 0.15\\
    S(x=0,t) &= 1\\
    S_{wc} &= 0.1\\
    S_{or} &= 0.05\\
    M &= 2
\end{eqnarray}

\begin{figure}[h!]%
    \centering
    \includegraphics[width=1\linewidth]{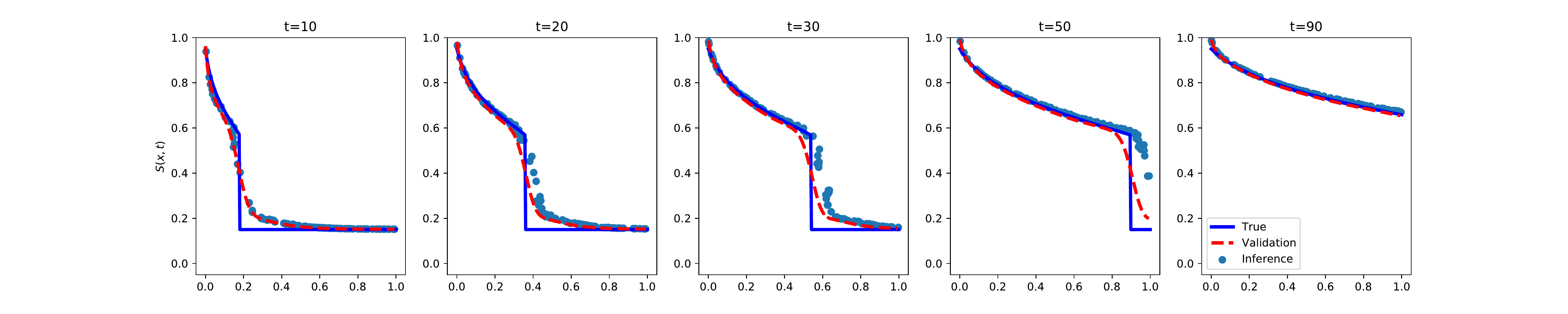}%
    \caption{Comparison of saturation profile at five different time steps for a Total velocity $v_d=1$ and non-trivial residual and injection saturation. The "True" saturation (computed through MOC) is in solid blue while the validation computed with P-PINN is in dashed red. The result of inference is dotted}%
    \label{fig:sat_realization_vd_residual}%
\end{figure}

\begin{figure}[h!]%
    \centering
    \includegraphics[width=1\linewidth]{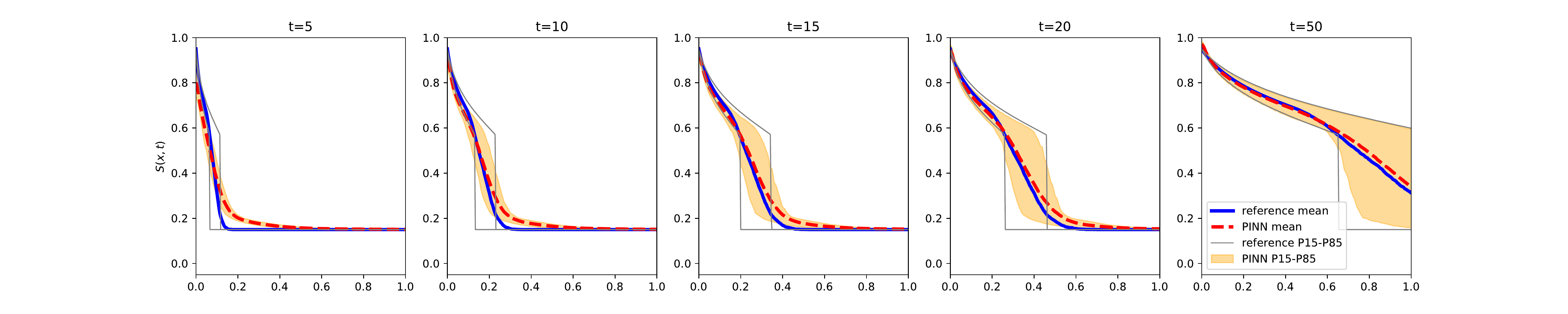}%
    \caption{Comparison of saturation distributions profiles at five different time steps for a Total velocity $v_d\sim \mathcal{N}(\mu=1,\,\sigma=0.3, low=0.5, up=2)$  and non-trivial residual and injection saturation. The reference mean saturation (computed through Monte Carlo simulation of MOC) is in solid blue while the one computed with P-PINNs is in dashed red. The P15-P85 envelopes are represented for both reference and P-PINN}%
    \label{fig:saturation_vd_residual}%
\end{figure}

We show that the handling of residual saturation is well handled by this formulation as seen in Figure \ref{fig:sat_realization_vd_residual} and \ref{fig:saturation_vd_residual}

\subsection{Bi-modal distribution}

\begin{figure}[h!]%
    \centering
    \includegraphics[width=0.5\linewidth]{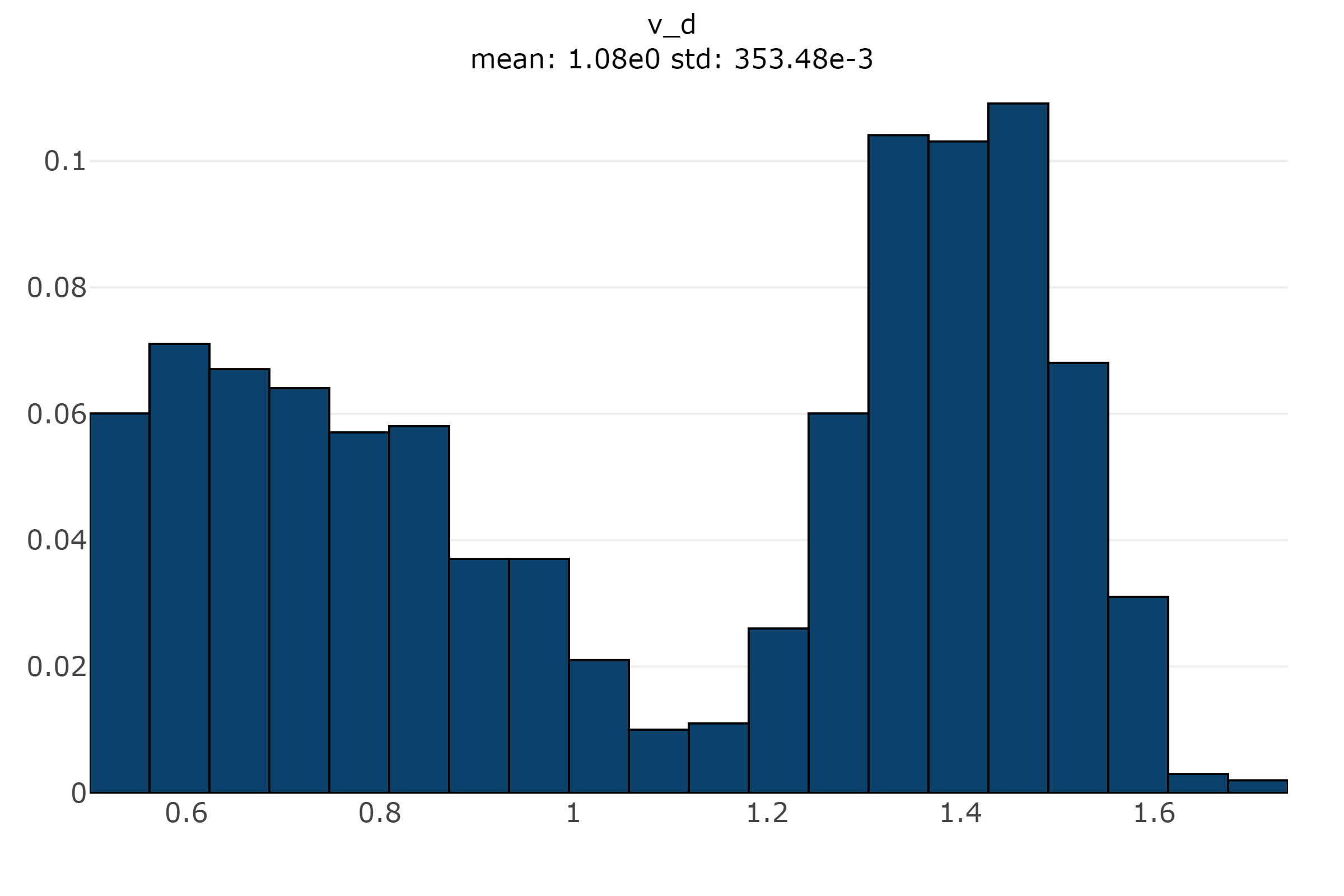}%
    \caption{Total (Darcy) velocity distribution for bi-modal transport modeling}%
    \label{fig:v_d_bimodal}%
\end{figure}

In this example, we are interested in computing the simulation for a non Gaussian velocity field. We model a bi-modal velocity distribution mimicking the behavior of a channelized reservoir. Since the properties are uniformly distributed, we assume two modes of transport: on-channel with a rapid breakthrough and off-channel for a delayed breakthrough. The distribution of velocities is presented in Figure \ref{fig:v_d_bimodal}

\begin{figure}[h!]%
    \centering
    \includegraphics[width=1\linewidth]{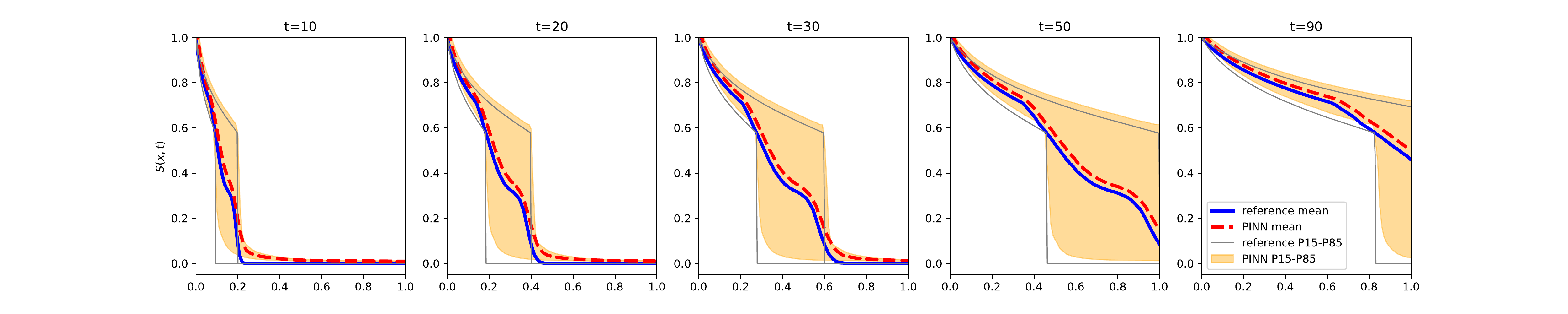}%
    \caption{Comparison of saturation distributions profiles at five different time steps for a bi-modal total velocity distribution. The reference mean saturation (computed through Monte Carlo simulation of MOC) is in solid blue while the one computed with P-PINNs is in dashed red. The P15-P85 envelopes are represented for both reference and P-PINN}%
    \label{fig:saturation_vd_bimodal}%
\end{figure}

We use the model trained on the normal distribution since the support of both distributions are similar. Results for the saturation envelope are presented in Figure \ref{fig:saturation_vd_bimodal}

\begin{figure}[h!]%
    \centering
    \includegraphics[width=1\linewidth]{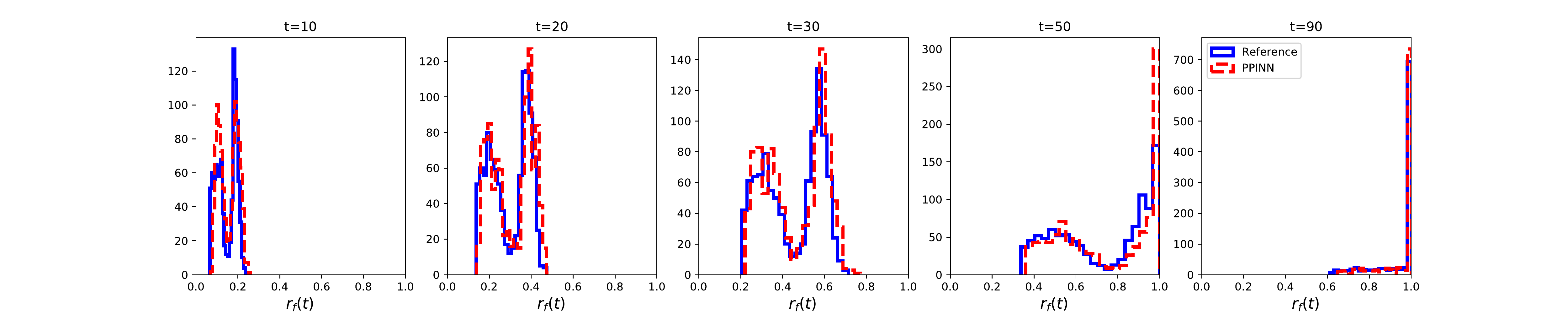}%
    \caption{Comparison of front radius distributions profiles at five different time steps for a bi-modal total velocity distribution. The reference front radius (computed through Monte Carlo simulation with MOC) is in solid blue while the one computed with P-PINNs is in dashed red.}%
    \label{fig:front_radius_vd_bimodal}%
\end{figure}

We represent the breakthrough time and front radius distributions and compare with the reference analytical solution. The distributions of shock radii are shown in Figure \ref{fig:front_radius_vd_bimodal}. They present close matches with the reference solution.

\begin{figure}[h!]%
    \centering
    \includegraphics[width=1\linewidth]{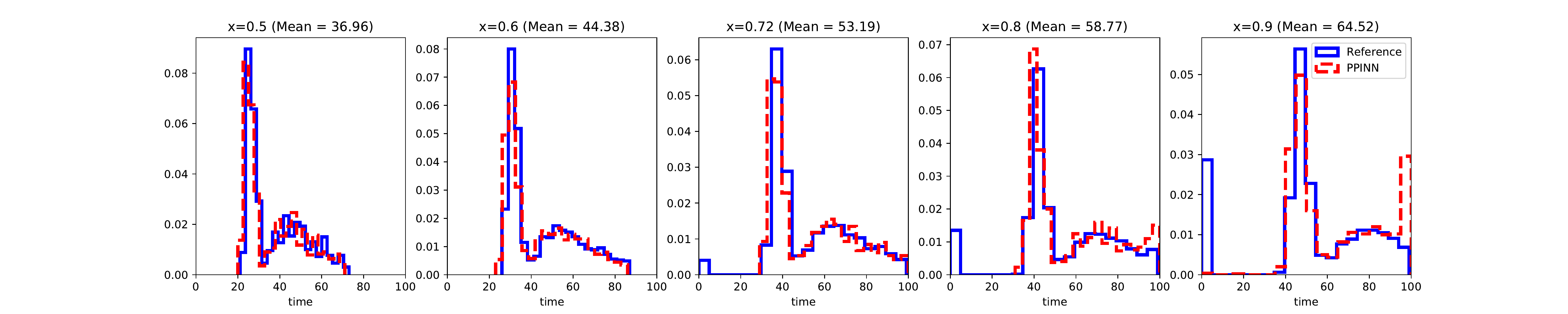}%
    \caption{Comparison of breakthrough time distributions profiles at five different locations in space for a bi-modal total velocity distribution. The reference breakthrough time (computed through Monte Carlo simulation with MOC) is in solid blue while the one computed with P-PINNs is in dashed red.}%
    \label{fig:bt_vd_bimodal}%
\end{figure}

The distribution of the breakthrough time matches closely too as shown in Figure \ref{fig:bt_vd_bimodal}. We note the the two modes of velocity used as input are found in the output quantity of interests (breakthrough time and radius of plume)

Table~\ref{tab:vd_bimodal} shows the average Wasserstein distances between distributions computed using P-PINNS and the reference for  both front radius and breakthrough times.

\begin{table}[h!]
	\caption{Wasserstein distance average for distributions of QOIs for a a bi-modal total velocity distribution. U is a uniform distribution.}
	\centering
	\begin{tabular}{ccc}
		\toprule
		    & Front Radius     & Breakthrough time \\
	    \midrule
		$W_p(P_{MOC}, P_{PINN})$ & 0.01  & 4.27 \\
        $W_p(P_{MOC}, U)$     & 0.28 & 47.98 \\
        \midrule
        Relative difference & 5.2\% & 8.9\% \\
		\bottomrule
	\end{tabular}
	\label{tab:vd_bimodal}
\end{table}

\subsection{Performance comparison}
\begin{figure}[h!]%
    \centering
    \includegraphics[width=1\linewidth]{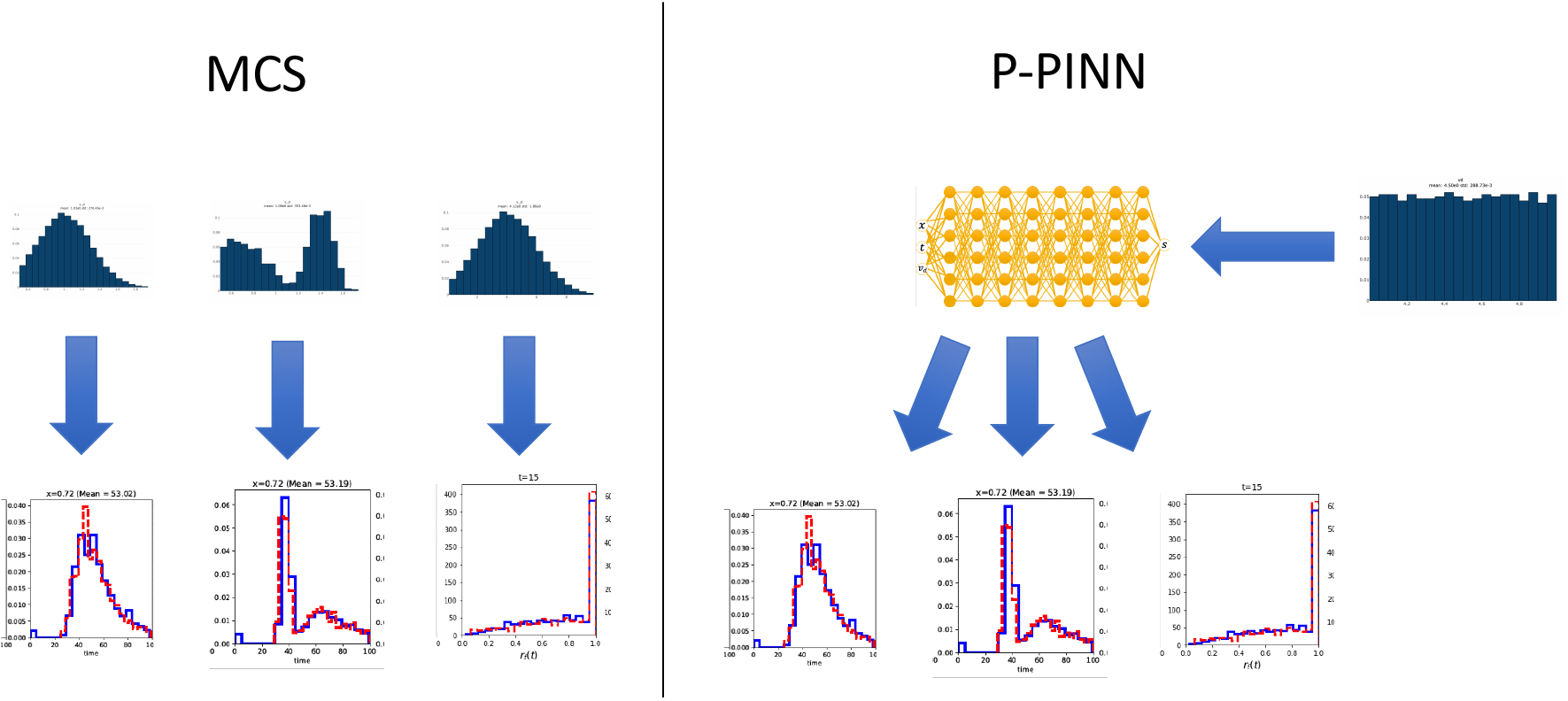}%
    \caption{Comparison of methodologies. Monte Carlo Simulation (MCS, on the left) relies on exhaustive simulation on input distributions using numerical methods. Parameterized PINN (P-PINN on the right) trains one neural network model that can be used to infer on a variety of distributions.}%
    \label{fig:MCSvsPPINN}%
\end{figure}

The parameterization of PINN allows to generalize the model to the ensemble realization. A single model can be used to make inference on series of realizations produced by a wide variety of distributions. Figure~\ref{fig:MCSvsPPINN} shows the difference between the two workflows. Note that in the MCS case, we use numerical methods to solve thousands of non-linear problems to eventually get an average result that averages out non-linearities. Each simulation results in convergence constraints that can significantly increase the time to get a result. The P-PINN approach on the other hand relies on a single training to infer these realizations. The error made in this interpolation is -as we demonstrated in the examples above, negligible for the uncertainty quantification we consider. This is one of the major motivations for the utilization of the parameterized approach. 

We compare the overall computation times for the two methods. We consider training and inference for the P-PINN approach. The problem is solved using 1000 samples for each run. The discretized model is solved on a 256 spatial grid blocks for 100 time steps. The problem is highly parallelizeable in both cases but the resolution of the discrete problem requires more CPU resources per run. The P-PINN on the other hand is trained on a GPU. Table~\ref{tab:MCSvsPPINN} presents the comparison of run times.

\begin{table}
	\caption{Performance comparison between MCS and P-PINN approach for Buckley-Leverett problem}
	\centering
	\begin{tabular}{lll}
		\toprule
		     & Monte Carlo Simulation     & Parameterized PINN \\
		\midrule
		Hardware & Intel-i7 3.2GHz (6 core)  & Tesla V-100 (16GB)     \\
		Training     & - & 15 min      \\
		Inference     & 3 min       & $\sim$10s  \\
		\midrule
		Total (per 1000 simulations) & 3 min & $\leq$16 min\\
		\bottomrule
	\end{tabular}
	\label{tab:MCSvsPPINN}
\end{table}
The benchmark is not very conclusive and shows that for the homogeneous problem, a numerical method will perform well on a 1000 batch sample. However, for each distribution of realizations, a new MCS has to be performed while the trained model can be re-used to perform inference on new distributions without any modifications. We will see that this advantage becomes clearer when the forward problem becomes more challenging (like in a heterogeneous velocity field).

\section{Heterogeneous 1D case}
In this section, the total velocity $v_d(x)$ is a random variable dependent upon the space dimension. This allows to consider random distributions of static properties (porosity, absolute permeability) along the x-axis.

\subsection{Formulation}
The conservation equation we solve is now defined in Eq. \ref{eq:conservation_vd_het}. We note that by comparison with Eq. \ref{eq:conservation_vd}, the total velocity term now depends on the space coordinate. We could also imagine a dependence on $t$ in case of geomechanical and chemical effects (compaction, diagenesis,...).

\begin{equation}
    \label{eq:conservation_vd_het}
    \frac{\partial S_w}{\partial t} + v_d(x) f_w^{\prime}(S) \frac{\partial S_w}{\partial x} = 0
\end{equation}

Eq. \ref{eq:conservation_vd_het} is typically solved numerically using a discretization scheme. The finite volume solution is computed using a Godunov scheme with an explicit treatment of the saturation and CFL constraint. 

Every realization's simulation takes 90 seconds on a 256 grid blocks mesh with $dt = dx/15$.

\subsection{Implementation}
In order to be able to model a Gaussian random field along the space dimension (see Figure \ref{fig:vd_het_distrib}), we sample uniformely a series of random variables $U$ and $V$ along with the $x$ and $t$ variables and convert them into a uniform distribution using a Box-Muller transform (\cite{Box_muller1958}). In this case the velocity field is written:
\begin{equation}
    v_d(x) = \sqrt{-2 \ln{U}}cos(2\pi V)
\end{equation}

A first idea has been to approach the heterogeneous case using approaches that had shown success previously:
\begin{itemize}
    \item Addition of artificial diffusivity
    \item Space-time weighting of loss function
    \item Sample based parameterization of fractional flow convex hull
\end{itemize}
The first is a well-known "trick" when using PINNS to solve hyperbolic problems. It consists in turning Eq.\ref{eq:conservation_vd_het} in a parabolic PDE by adding a viscous term as shown in Eq.~\ref{eq:conservation_vd_het_diff}.
\begin{equation}
    \label{eq:conservation_vd_het_diff}
    \frac{\partial S_w}{\partial t} + v_d(x) f_w^{\prime}(S) \frac{\partial S_w}{\partial x} = \epsilon\frac{\partial^2S_w}{\partial x^2}
\end{equation}
\cite{Fuks2020} covers this approach in details and explains how the added diffusive term leads to better convergence in training for the case with homogeneous boundary conditions. \cite{gasmi2021physics} extends the approach with a vanishing diffusivity to cases with a linear initial condition but shows this approach carries some limitations (manual tunning of the $\epsilon$ term, non-applicability in cases with non-monotonic boundary condition) and proposes a space dependent weighting of the loss function as an alternative. Both methods are applied to the heterogeneous case and results commented hereafter.

The third approach is an attempt at enforcing entropy at every location in the simulation space. Just like in a Godunov scheme, the exact (or approximate) Riemann problem is solved at each point in space. We achieve this approximation by parameterizing the convex hull of the fractional flow with the seepage velocity and interpolating for each location

\subsection{Normal distribution}
\begin{figure}[h!]%
    \centering
    \includegraphics[width=1\linewidth]{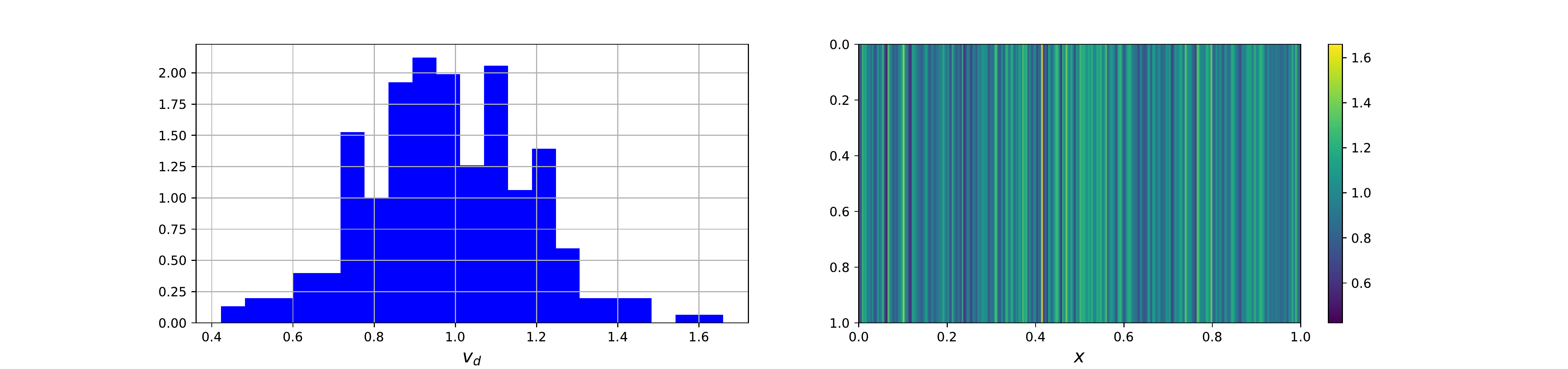}%
    \caption{Distribution of total velocities probability density (left) and along x-axis (right)}%
    \label{fig:vd_het_distrib}%
\end{figure}

In this example, we use a distribution with similar parameters as in the homogeneous case. The difference is that the velocity now varies along the x-axis. Figure \ref{fig:vd_het_distrib} shows the distribution along the x-axis.

\begin{figure}[h!]%
    \centering
    \includegraphics[width=1\linewidth]{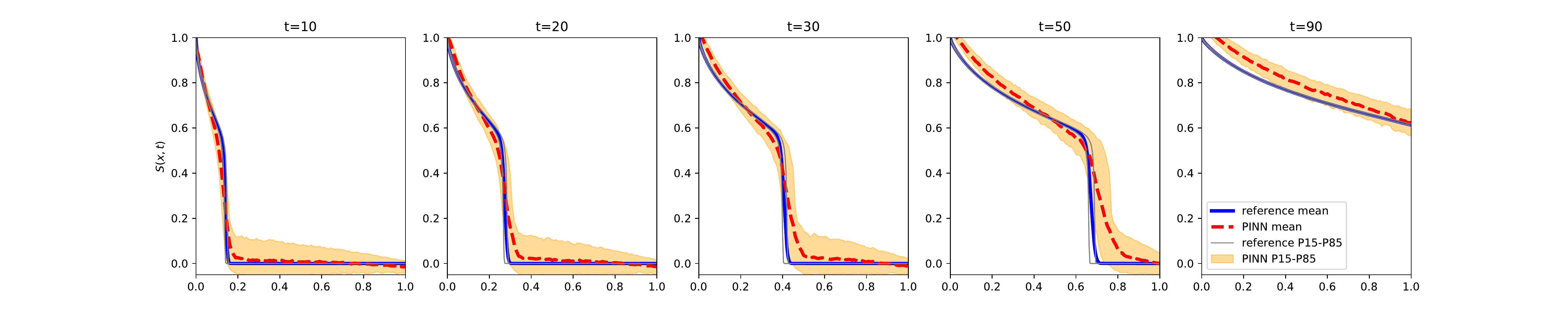}%
    \caption{Comparison of saturation distributions profiles at five different time steps for a heterogeneous total velocity distribution $v_d(x)\sim \mathcal{N}(\mu=1,\,\sigma=0.2, low=0.5, up=2)$. The reference mean saturation (computed through Monte Carlo simulation of a finite volume numerical model) is in solid blue while the one computed with P-PINNs is in dashed red. The P15-P85 envelopes are represented for both reference and P-PINN}%
    \label{fig:saturation_vd_het}%
\end{figure}

The results of the simulation are presented in Figure \ref{fig:saturation_vd_het}. We note that the distribution on the reference case is much narrower than for the homogeneous case. Heterogeneities tend to create barriers that act as buffers for the flow. The PINN case approximates the reference solution but not with the same level of accuracy we observed for the homogeneous case.

\begin{figure}[h!]%
    \centering
    \includegraphics[width=1\linewidth]{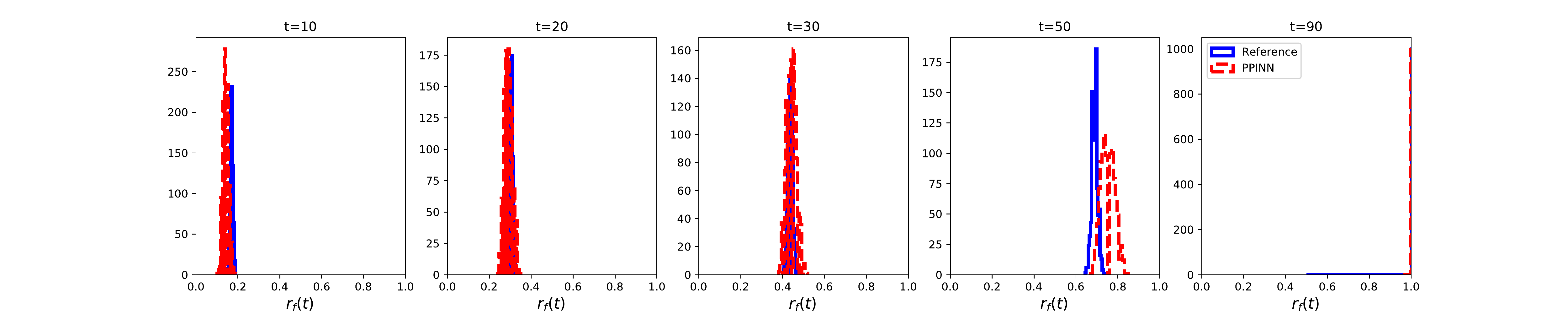}%
    \caption{Comparison of front radius distributions profiles at five different time steps for a heterogeneous total velocity distribution $v_d(x)\sim \mathcal{N}(\mu=1,\,\sigma=0.2, low=0.5, up=2)$. The reference front radius (computed through Monte Carlo simulation of a finite volume numerical model) is in solid blue while the one computed with P-PINNs is in dashed red.}%
    \label{fig:front_radius_vd_het}%
\end{figure}

We represent the breakthrough time and front radius distributions and compare with the reference analytical solution. The distributions of shock radii are shown in Figure \ref{fig:front_radius_vd_het}. They present a difference with the reference solution with a consistent delay in front progression and breakthrough for the finite volume simulation.

\begin{figure}[h!]%
    \centering
    \includegraphics[width=1\linewidth]{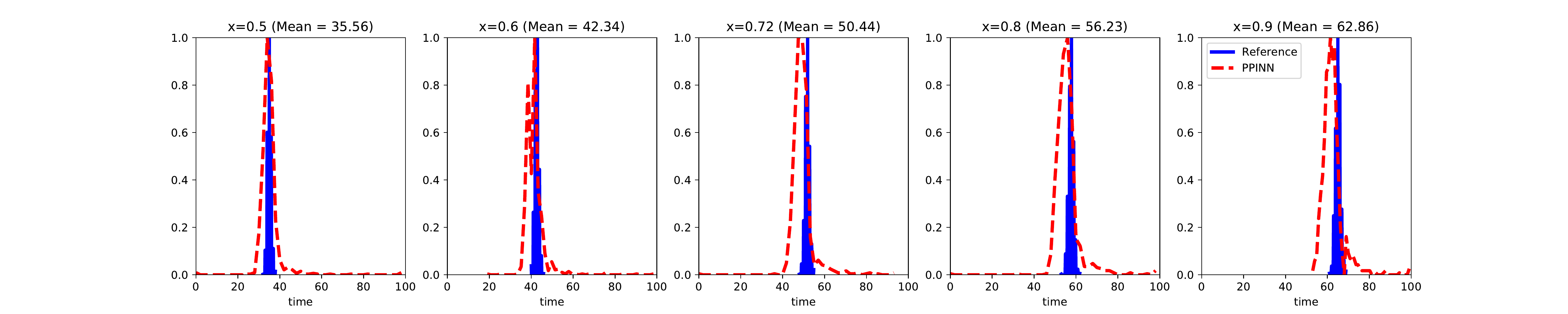}%
    \caption{Comparison of breakthrough time distributions profiles at five different locations in space for a heterogeneous total velocity distribution $v_d(x)\sim \mathcal{N}(\mu=1,\,\sigma=0.2, low=0.5, up=2)$. The reference breakthrough time (computed through Monte Carlo simulation of a finite volume numerical model) is in solid blue while the one computed with P-PINNs is in dashed red.}%
    \label{fig:bt_vd_het}%
\end{figure}

\begin{figure}[h!]%
    \centering
    \includegraphics[width=1\linewidth]{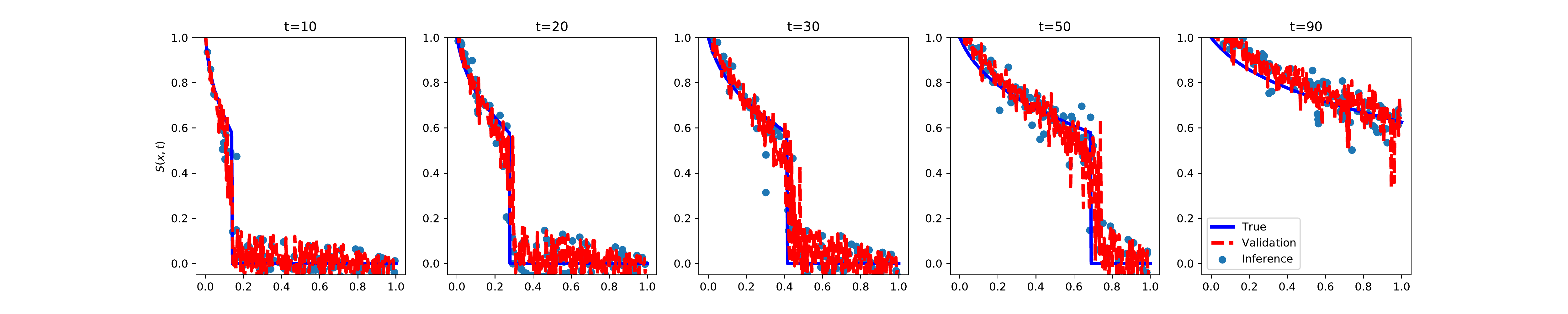}%
    \caption{Comparison of saturation profile at five different time steps for a heterogeneous total velocity distribution $v_d(x)\sim \mathcal{N}(\mu=1,\,\sigma=0.2, low=0.5, up=2)$. The "True" saturation (computed with finite volume numerical simulation) is in solid blue while the validation computed with P-PINN is in dashed red. The result of inference is dotted}%
    \label{fig:saturation_realization_het}%
\end{figure}

Both the distribution of the breakthrough time and the front radius (Figure \ref{fig:bt_vd_het}) display consistent and noticeable differences. The front distribution is wider with P-PINNS and advances slightly faster. Two reasons could be explaining these differences. The first one would be the numerical diffusion which tends to slow the shock progression. The second is that each realization of the P-PINN approach is solved approximately. We recall that in order to resolve the homogeneous case, we enforce an entropy condition constraint on the fractional flow curve (Figure \ref{fig:frac_flow_welge}). With a modified PDE including the total velocity term (see Eq.\ref{eq:conservation_vd_het}), the fractional flow term is now scaled by a term $v_d$ that depends on the location. This means we can no longer apply the convex hull constraint the same way we did in the homogeneous case. Our attempts at approximating. Figure \ref{fig:saturation_realization_het} shows such a comparison.

\subsection{Normal distribution with wide support}
\begin{figure}[h!]%
    \centering
    \includegraphics[width=1\linewidth]{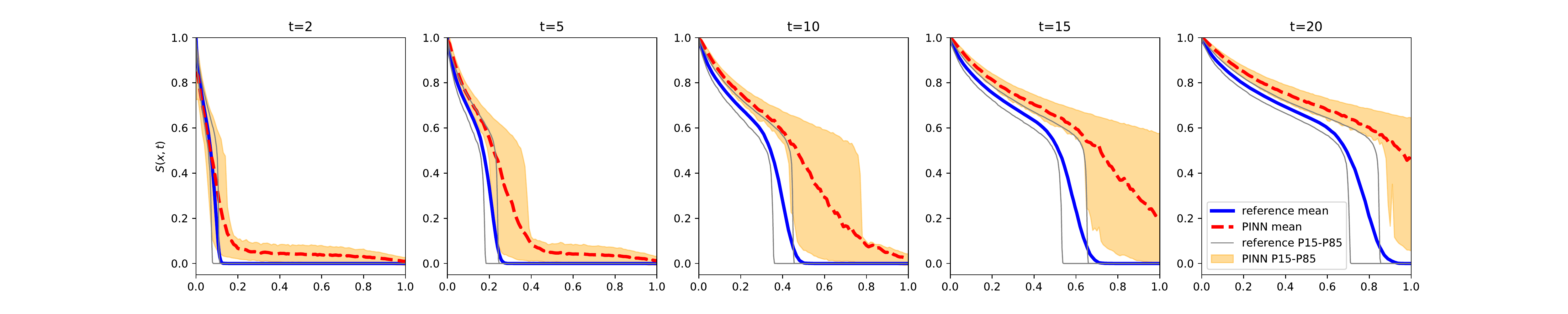}%
    \caption{Comparison of saturation distributions profiles at five different time steps for a heterogeneous total velocity distribution $v_d(x)\sim \mathcal{N}(\mu=4,\sigma=1, low=0.1, up=10)$. The reference mean saturation (computed through Monte Carlo simulation of a finite volume numerical model) is in solid blue while the one computed with P-PINNs is in dashed red. The P15-P85 envelopes are represented for both reference and P-PINN}%
    \label{fig:saturation_vd_het_wide}%
\end{figure}

Both the diffusion and parameterization of convex hull methods lead to errors in the results that scale with the uncertainty space. In cases where the velocity field is drawn from a wider distribution ($v_d(x)\sim \mathcal{N}(\mu=4,\sigma=1, low=0.1, up=10)$), the profiles produced using P-PINNs are very different from the reference using FVM. 
The results of the simulation are presented in Figure \ref{fig:saturation_vd_het_wide}

\begin{figure}[h!]%
    \centering
    \includegraphics[width=1\linewidth]{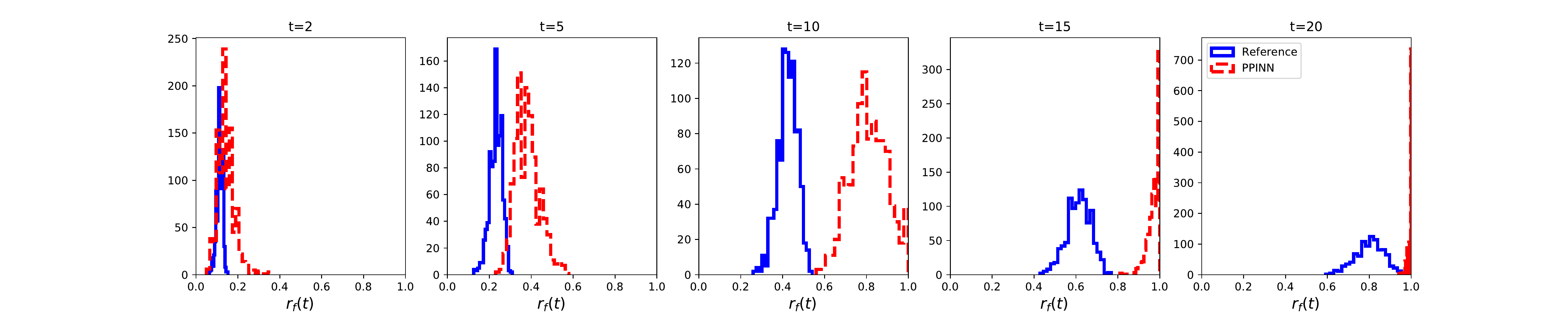}%
    \caption{Comparison of front radius distributions profiles at five different time steps for a heterogeneous total velocity distribution $v_d(x)\sim \mathcal{N}(\mu=4,\sigma=1, low=0.1, up=10)$. The reference front radius (computed through Monte Carlo simulation of a finite volume numerical model) is in solid blue while the one computed with P-PINNs is in dashed red.}%
    \label{fig:front_radius_vd_het_wide}%
\end{figure}

We represent the breakthrough time and front radius distributions and compare with the reference analytical solution. The distributions of shock radii are shown in Figure \ref{fig:front_radius_vd_het_wide}. They present a difference with the reference solution with a consistent delay in front progression and breakthrough for the finite volume simulation.

\begin{figure}[h!]%
    \centering
    \includegraphics[width=1\linewidth]{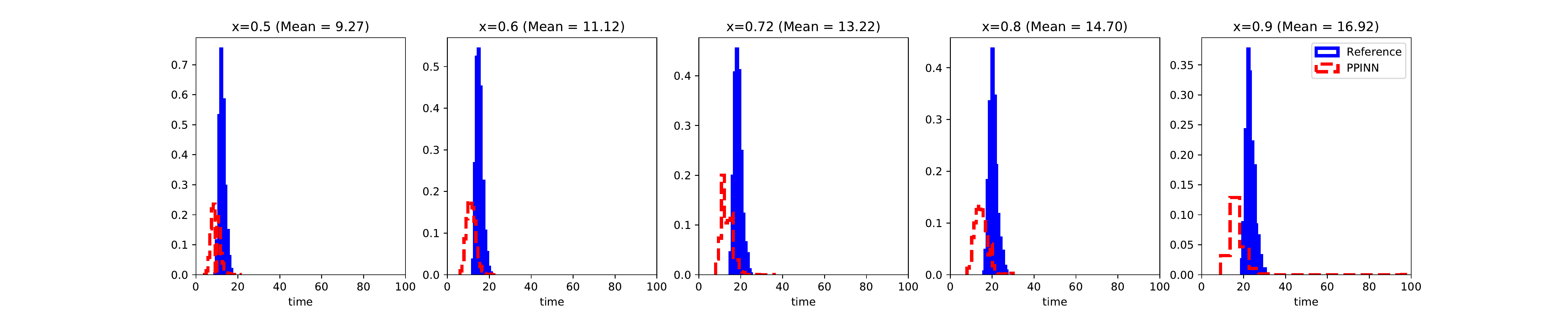}%
    \caption{Comparison of breakthrough time distributions profiles at five different locations in space for a heterogeneous total velocity distribution $v_d(x)\sim \mathcal{N}(\mu=4,\sigma=1, low=0.1, up=10)$. The reference breakthrough time (computed through Monte Carlo simulation of a finite volume numerical model) is in solid blue while the one computed with P-PINNs is in dashed red.}%
    \label{fig:bt_vd_het_wide}%
\end{figure}

Table~\ref{tab:bt_vd_het_wide} shows the average Wasserstein distances between distributions computed using P-PINNS and the reference for both front radius and breakthrough times.

\begin{table}[h!]
	\caption{Wasserstein distance average for distributions of QOIs for a for a heterogeneous total velocity distribution $v_d(x)\sim \mathcal{N}(\mu=4,\sigma=1, low=0.1, up=10)$. U is a uniform distribution.}
	\centering
	\begin{tabular}{ccc}
		\toprule
		    & Front Radius     & Breakthrough time \\
	    \midrule
		$W_p(P_{MOC}, P_{PINN})$ & 0.22  & 4.93 \\
        $W_p(P_{MOC}, U)$     & 0.29 & 34.00 \\
        \midrule
        Relative difference & 77.4\% & 14.5\% \\
		\bottomrule
	\end{tabular}
	\label{tab:bt_vd_het_wide}
\end{table}

The results obtained on a wider distribution emphasize the error incurred by using a single parameter to characterize the uncertainty in the problem with a heterogeneous field. The original idea of having a single neural network model that can interpolate saturation solutions within the uncertainty space regardless of the distribution on which it had been trained was appealing and this is why we tried (without success) to adapt it to heterogeneous fields. 

We document the various attempts made at solving this problem that were not successful. This hopefully will reveal useful to researchers in this field. For each of the experiments made, we performed a basic hyper-parameter tuning where various, learning rates, network depth and problem specific quantities variations were tested. This approach, although not comprehensive, is aligned with our focus on the physics of the problem rather than the parameters of the network. This allows us to present results that are more robust and explainable. 

\subsection{Diffusion term}
\begin{figure}[h!]%
    \centering
    \includegraphics[width=1\linewidth]{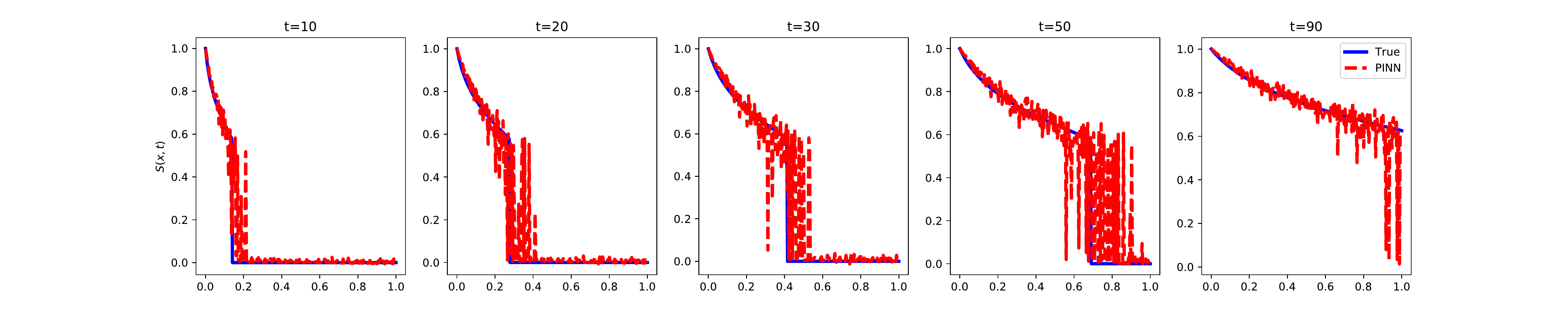}%
    \caption{Comparison of saturation profile (for a single realization) at five different time steps for a random heterogeneous total velocity distribution $v_d(x)\sim \mathcal{N}(\mu=1,\,\sigma=0.2, low=0.5, up=2)$. The solution computed with P-PINN (dashed red) where a diffusive term was added to the residual loss. The "True" saturation (computed with finite volume numerical simulation) is in solid blue.}%
    \label{fig:saturation_realization_diff}%
\end{figure}

This approach is inspired from previous work by Fuks and Tchelepi (\cite{Fuks2020}). The addition of a diffusion term to equation~\ref{eq::Conservation_w} is proven to have improved the resolution of the forward Riemann problem. We can re-write the conservation equation in 1D:
\begin{equation}
    \label{eq:conservation_vd_diff}
    \frac{\partial S_w}{\partial t} + v_d f_w^{\prime}(S) \frac{\partial S_w}{\partial x} = \nu \frac{\partial^2S_w}{\partial x^2}
\end{equation}
The term $\nu$ is an artificial diffusivity that enforces entropy condition. Figure~\ref{fig:saturation_realization_diff} shows one amongst several attempts at trying to produce a solution that would match the reference.

We see that the solution obtained on a realization is quite noisy especially around the shock. We observe this artefact on most of the other attempts that were made at formulating the problem a different way or at tuning the hyperparameters of the model (although it is the former rather than the later we are trying to promote).  

\subsection{Entropy constrained fractional flow}
We propose here to expand upon the formulation that was successfully implemented for the homogeneous case. We propose to use a convex hull to replace the original fractional flow as presented in \cite{gasmi2021physics}. For heterogeneous problems, we establish a linear correlation between the velocity field and the Buckley-Leverett shock saturation $S_{BL}$. We know that eq.~\ref{eq:conservation_vd_het} features a shock that travels with velocity $f(S_{BL})/S_{BL}$. We assume that when formed, a shock moves faster in sections of the domain where the total velocity is greater and slower when the total velocity is lower. This series of "acceleration-deceleration" can be approximated using a simple interpolation. We run a large number of finite volume simulations on a random velocity field and find a correlation relationship between the shock velocity rate of change and the Darcy velocity. at a given location. This leads to a new convex hull where:  
\begin{eqnarray}
    S_{BL} &\sim \alpha_s v_d + \beta_s\\
   f(S_{BL}) &\sim \alpha_f v_d + \beta_f 
\end{eqnarray}

This approach, although physically plausible, suffers several drawbacks amongst which the need to pre-compute these correlation and assume that the solution will necessarily have a shock (which may not always be the case). In practice, it led to noisy results similar to the ones presented in figure~\ref{fig:saturation_realization_diff}

\subsection{Weighted loss}
This approach is once again inspired from the deterministic case presented in \cite{gasmi2021physics} where the initial saturation is non-uniform. This case is solved using a loss weighting function that vanishes around sharp gradients (in $x$ or $t$). It helped solving the Buckley-Leverett problem with linear initial condition ($S(x,0) = x$) with the original fractional flow condition. We use the same loss weighting for the heterogeneous velocity field. 
\begin{equation}
    \omega(x,t) = \frac{1} {(\partial S/\partial x)^2 + (\partial S/\partial t)^2 + 1}
\end{equation}

\begin{figure}[h!]%
    \centering
    \includegraphics[width=1\linewidth]{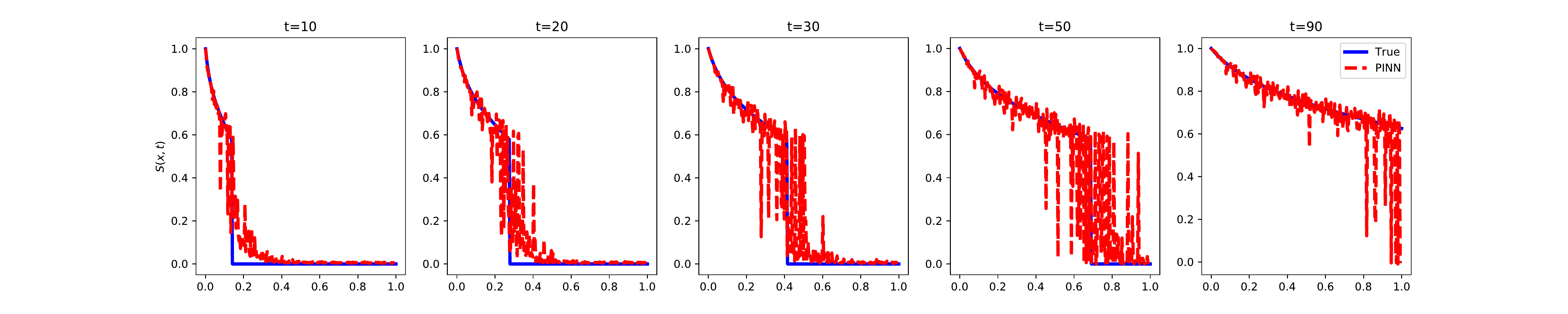}%
    \caption{Comparison of saturation profile (for a single realization) at five different time steps for a random heterogeneous total velocity distribution $v_d(x)\sim \mathcal{N}(\mu=1,\,\sigma=0.2, low=0.5, up=2)$. The solution computed with P-PINN (dashed red) where a weighting term was added to the residual loss to reduce the importance of samples around the shock. The "True" saturation (computed with finite volume numerical simulation) is in solid blue.}%
    \label{fig:saturation_realization_weight}%
\end{figure}

Results for one realization are presented in figure~\ref{fig:saturation_realization_weight}

We notice a similar noisy solution to the case with diffusion with the difference that it consistently smears out around the shock. This was observed for the homogeneous case as well. It looks like each approach that we attempt to translate to the heterogeneous case carries over the issues it had for the homogeneous case without solving the issue of noise.

\subsubsection{Neural network trained to mimic heterogeneous velocity field}
 One of the issues we encounter is the intractable form the velocity field can take. For each realization, the total velocity can be quite noisy as shown in Figure \ref{fig:vd_het_distrib}. One solution proposed is to train a secondary neural network jointly to the solution one to mimic the heterogeneous field and use it as a proxy for the velocity $v_d(x)$ in equation \ref{eq:conservation_vd_het}. In this approach we focus solely on deterministic solutions for a heterogeneous field. The idea being that if we can train a network to fit a velocity field realization, so can we for an ensemble. Empirical evidences show that provided enough data, neural networks can interpolate fairly efficiently. 
 We present a solution on two different velocity fields using the dual PINN approach.
 
 The first is a "stair" function emulated by a series of hyperbolic tangent functions
\begin{equation}
    v_d(x) = 2 - \tanh(80 * (x - 0.25)) / 4 + \tanh(80 * (x - 0.5)) / 4 + \tanh(80 * (x - 0.75)) / 4 + 0.75)
\end{equation}

\begin{figure}[h!]%
    \centering
    \includegraphics[width=0.6\linewidth]{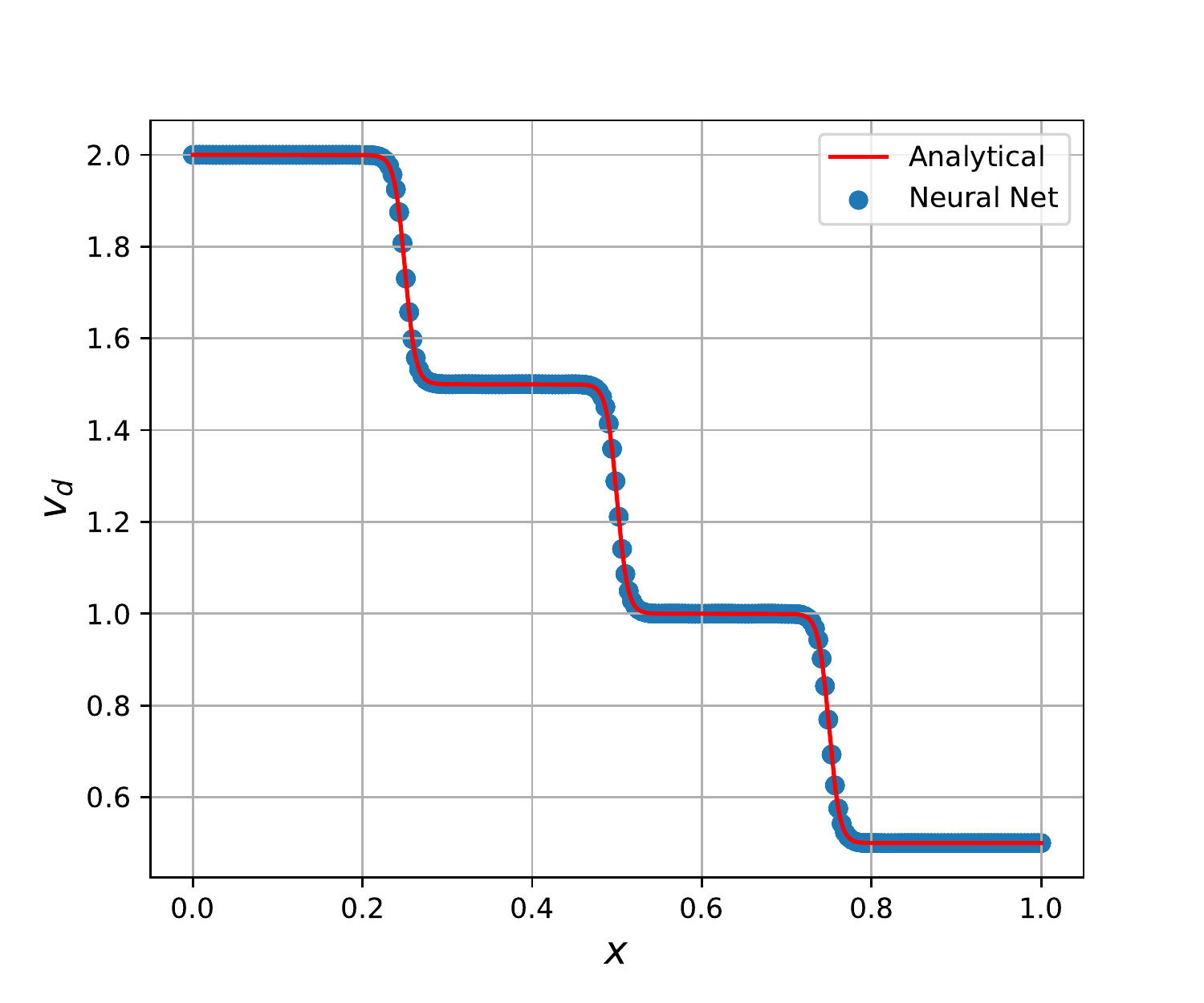}%
    \caption{Velocity field as a function of $x$ for an analytical stairs function (red) and fitted neural network model (dotted blue)}%
    \label{fig:velocity_field_tanh}%
\end{figure}

The velocity field defined by this formula along with the associated neural network are presented in figure~\ref{fig:velocity_field_tanh}

\begin{figure}[h!]%
    \centering
    \includegraphics[width=1\linewidth]{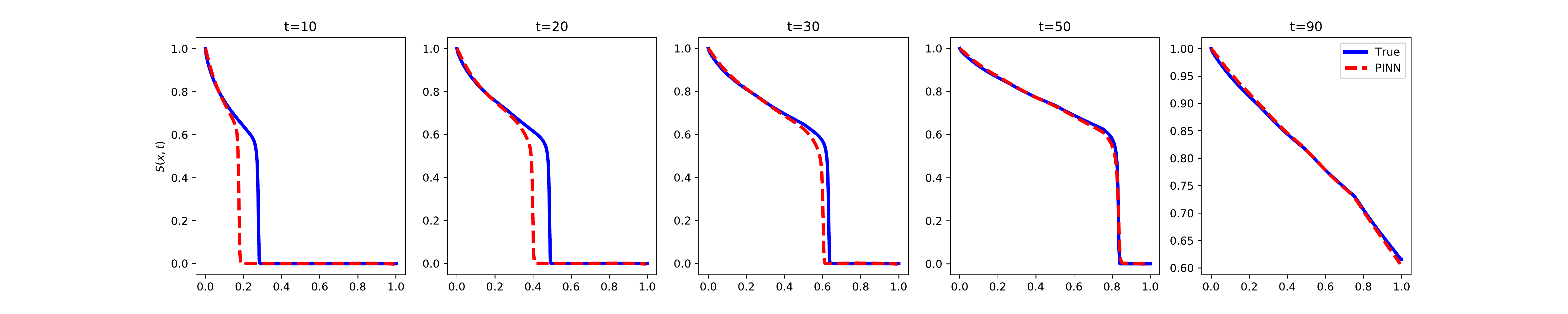}%
    \caption{Comparison of saturation profile at five different time steps for a deterministic heterogeneous (stairs shaped) total velocity field. The solution computed with dual network PINN (dashed red) is compared to the "True" saturation (computed with finite volume numerical simulation) in solid blue.}%
    \label{fig:saturation_neural_net_approx_stairs}%
\end{figure}

The results on the saturation simulation are presented in figure~\ref{fig:saturation_neural_net_approx_stairs}

The saturation profile produced using the two neural networks approach produces a solution that seems delayed compared with the theoretical one. We have a shock that catches up at later time and a rarefaction that captures the inflection due to the change in velocity at later times. This solution was produced as a result of a more advanced parameters tuning where a combination of all the methods presented earlier (weighting and diffusion) were implemented. Although the solution does not match the theoretical one, the noisiness has disappeared. 

We now try the same approach on a velocity field that displays a more complex behavior. We use a increasing cosine function with a high frequency coefficient defined as:
\begin{equation}
    v_d(x) = 1.5 + \cos(100x)
\end{equation}

\begin{figure}[h!]%
    \centering
    \includegraphics[width=0.6\linewidth]{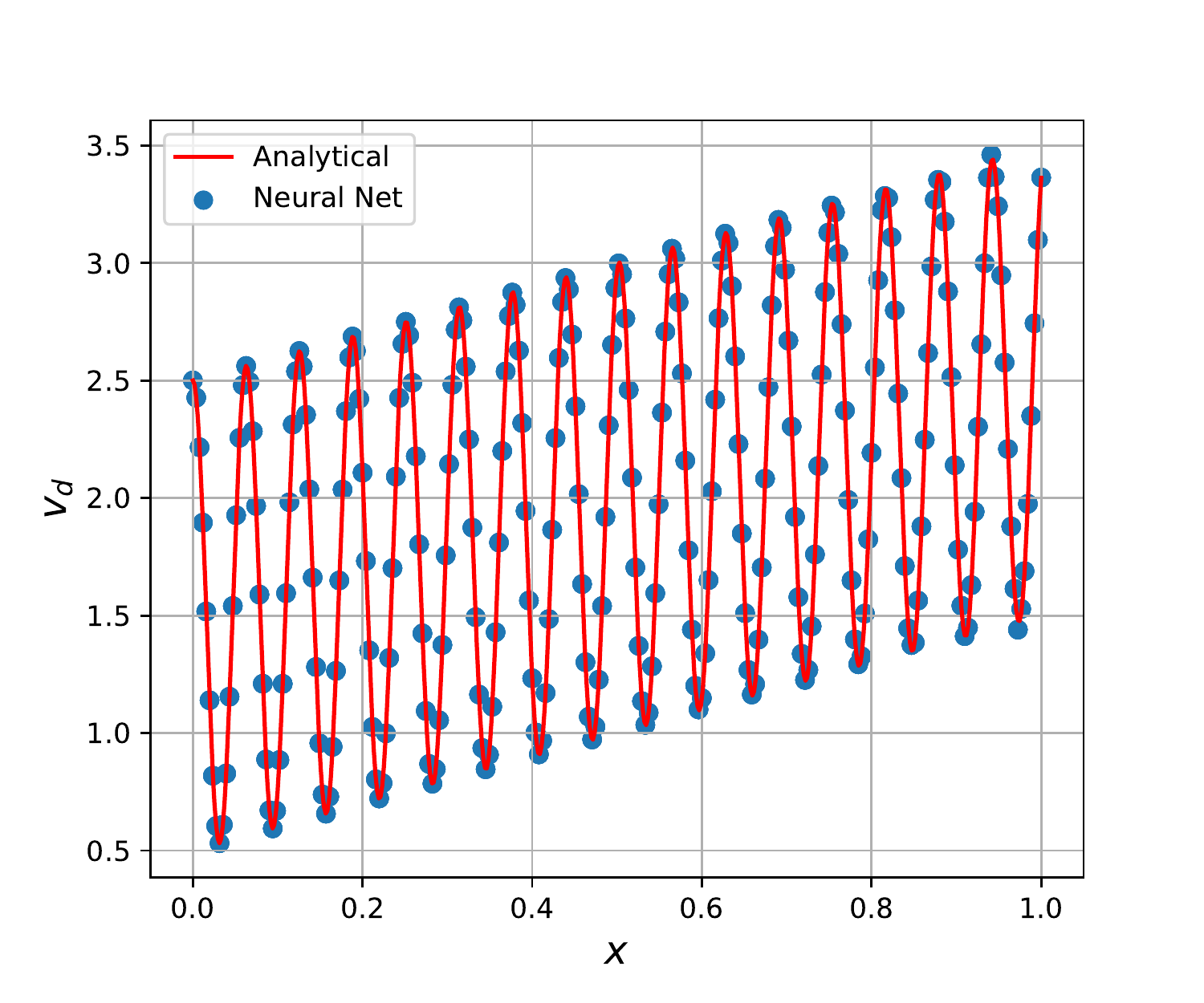}%
    \caption{Velocity field as a function of $x$ for an analytical high frequency cosine function (red) and fitted neural network model (dotted blue)}%
    \label{fig:velocity_field_sin}%
\end{figure}

The velocity field defined by this formula along with the associated neural network are presented in figure~\ref{fig:velocity_field_sin}

\begin{figure}[h!]%
    \centering
    \includegraphics[width=1\linewidth]{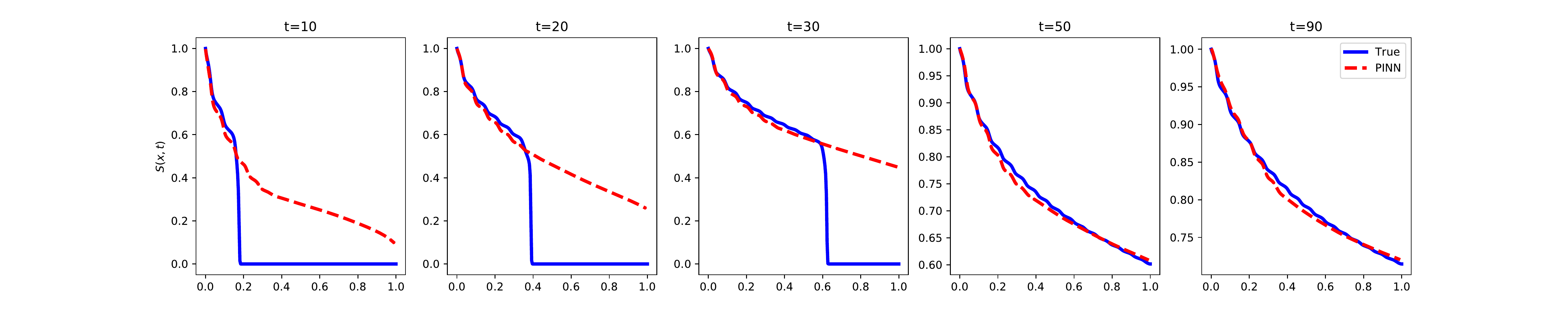}%
    \caption{Comparison of saturation profile at five different time steps for a deterministic heterogeneous (high frequency cosine shaped) total velocity field. The solution computed with dual network PINN (dashed red) is compared to the "True" saturation (computed with finite volume numerical simulation) in solid blue.}%
    \label{fig:saturation_neural_net_approx_sin}%
\end{figure}

The results on the saturation simulation are presented in figure~\ref{fig:saturation_neural_net_approx_sin}

The saturation profile produced is not able to capture the shock. This indicates a worsening of the performances of the PINN when faced with signals of higher frequency. This results point to the spectral bias (\cite{Rahaman2019}) than can be characterized as the tendency for neural networks to converge to lower frequency solutions.
 
The dual PINN approach presented, although unsuccessful informs us on the influence of the frequency of the input velocity. Velocity fields with a lower spacial range of randomness (higher frequency, more nugget effect) lead to poorer results in general.

\subsection{Integral continuity}
\begin{figure}[h!]%
    \centering
    \includegraphics[width=1\linewidth]{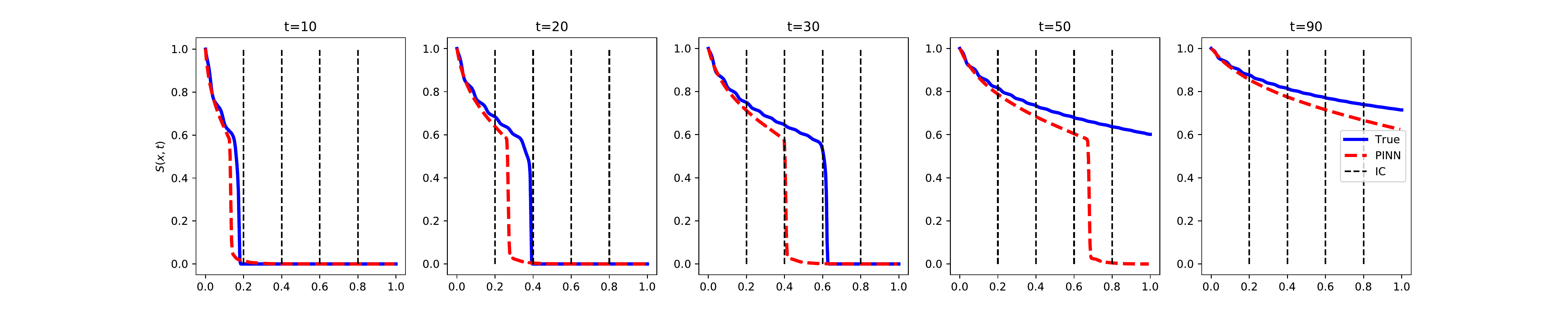}%
    \caption{Comparison of saturation profile at five different time steps for a deterministic heterogeneous (high frequency cosine shaped) total velocity field. The solution computed with integral continuity planes (in dotted black) is in dashed red. It is compared to the "True" saturation (computed with finite volume numerical simulation) in solid blue.}%
    \label{fig:saturation_solution_integral_cont}%
\end{figure}

For conservation problems described by hyperbolic PDEs, ensuring conservation of mass across the domain is of importance. Mesh based methods ensure that here is a conservation of mass at each grid block. When we solve reservoir problems using finite difference, it can occur that the result of a non-linear iteration leads to saturations that are negative or exceeding 1 which is non-physical. This is usually resolved in a post-processing step by correcting the saturation back to its "acceptable" solution. Sampling methods do not offer this safeguard. This is why adding continuity planes where we ensure that physical constraints are respected can help converging to correct solutions.
Another argument in favor of integral continuity planes for hyperbolic problems is that they allow to inform of a direction in the problem. Finite volume schemes like "Upwind" or Godunov use the direction of the flow as information. Sampling methods are lacking this notion.
We implement a series of continuity planes where we ensure that the total volumetric flowrate remains constant across each plane. This is formalized by writing the flow in and flow out of each plane. The $v_{sh}$ be the velocity of the shock. The construction of the displcement solution from the MOC tells us that this velocity is equal to
\begin{equation}
    v_{sh} = v_d\frac{\partial f_w}{\partial S_w}
\end{equation}
The volumetric flowrate across the plane is written:
\begin{equation}
    \label{eq:conservation_plane}
    q_Tf_w - v_{sh}\phi AS_w = cst
\end{equation}
If we differentiate eq~\ref{eq:conservation_plane} wrt $x$, we obtain:
\begin{equation}
    \frac{\partial f_w}{\partial x} - \frac{v_{sh}\phi A}{q_T}\frac{\partial S_w}{\partial x} = 0
\end{equation}
Or by using the Darcy velocity
\begin{equation}
\label{eq:residual_IC}
    \frac{\partial f_w}{\partial x} - \frac{v_{sh}}{v_d}\frac{\partial S_w}{\partial x} = 0
\end{equation}
We use the least square sum of eq~\ref{eq:residual_IC} as a loss added on each of the continuity planes we place in the domain. 
We introduce 4 such planes across the domain and enforce this condition. We simulate the saturation profile for the high frequency velocity field presented in figure~\ref{fig:velocity_field_sin}. The simulated saturation is presented in figure~\ref{fig:saturation_solution_integral_cont}.

The addition of integral continuity planes helps improve the solution as a sock is now captured for the high frequency velocity field (as opposed to the dual network model simulated in figure~\ref{fig:saturation_neural_net_approx_sin}) but the shock still appears delayed compared with the true solution. The addition of more continuity planes beyond 4 does not seem to improve upon the baseline we are presenting. 

The inconsistencies observed point to the limitations of a single parameter approach for heterogeneous velocity fields. We attempt to circumvent to some of these issues using a series of different approaches. We proceed to isolate the changes that were made from the homogeneous case to the heterogeneous case with random velocity field. We treat the heterogeneity separately from the randomness.

\section{Parameterizations of the uncertainty space}
\begin{figure}[h!]%
    \centering
    \includegraphics[width=0.6\linewidth]{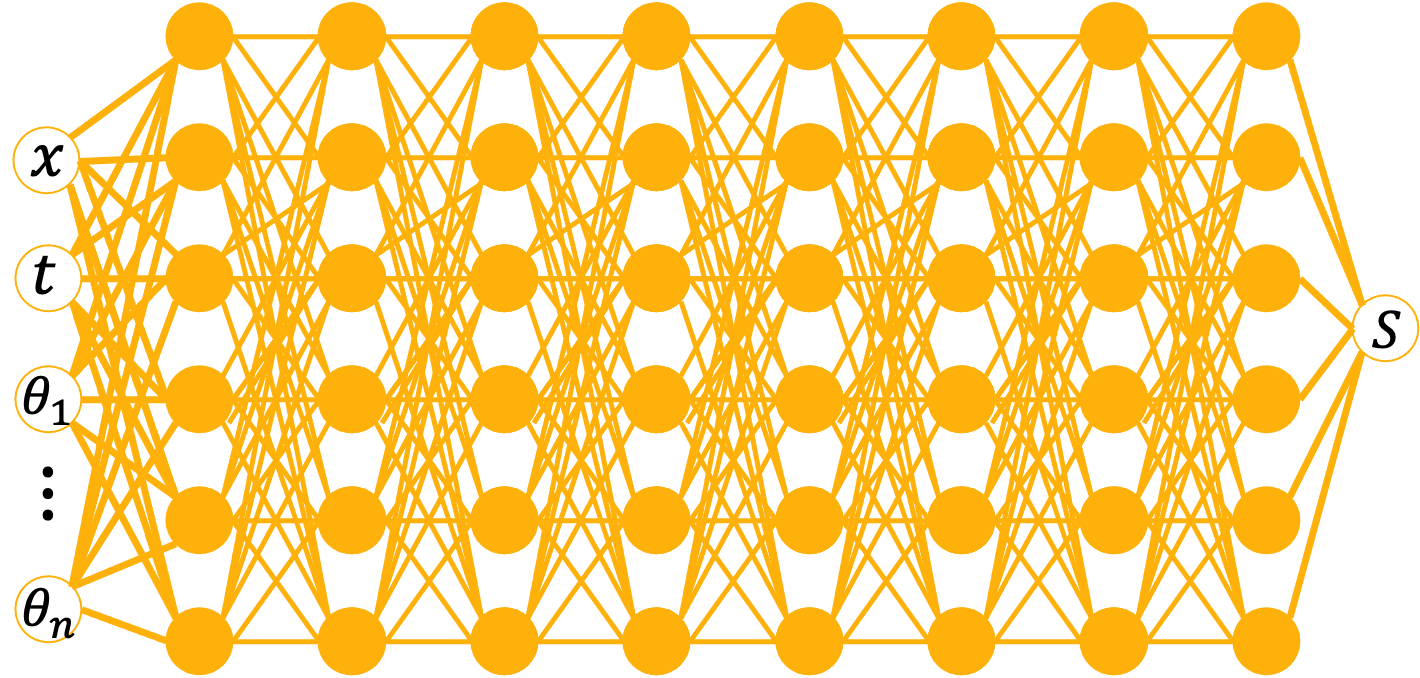}%
    \caption{Presentation of the model (input/output) for a heterogeneous velocity field parameterized by $\theta_k$}%
    \label{fig:mlp_proba_more_param}%
\end{figure}

We propose to address the problems presented in the previous section by treating separately the problem of heterogeneity from the one of randomness in the velocity field. Building from the homogeneous case, we test various parameterization techniques that effectively amount to a projection of the input variables into a finite dimension space with sufficient preservation of the original variability. Figure~\ref{fig:mlp_proba_more_param} shows how the problem can be formalized. We use $\theta_k$ parameters to encode the velocity field as opposed to a single velociy value.

\subsection{Stochastic Affine velocity function}
\begin{figure}[h!]%
    \centering
    \includegraphics[width=1\linewidth]{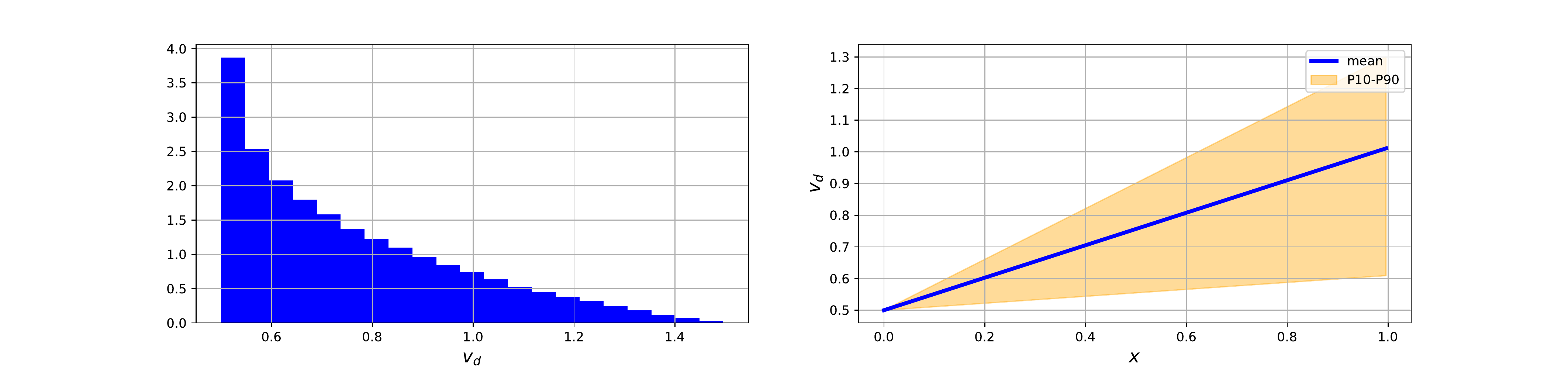}%
    \caption{Distribution of total velocities probability density (left) and along x-axis (right) where mean and P10-P90 envelope are represented}%
    \label{fig:v_d_het_affine_distrib}%
\end{figure}

In this scenario, a random Darcy velocity field that has a linear shape. The slope of the field is a random parameter $\theta$ with a Gaussian distribution.
\begin{equation}
    v_d(x) = \theta x + b
\end{equation}
With
\begin{equation}
    \theta \sim \mathcal{N}(\mu=1,\,\sigma^{2}=0.3)
\end{equation}
We plot the histogram of $v_d$ along with the ensemble realizations in figure~\ref{fig:v_d_het_affine_distrib}

\begin{figure}[h!]%
    \centering
    \includegraphics[width=1\linewidth]{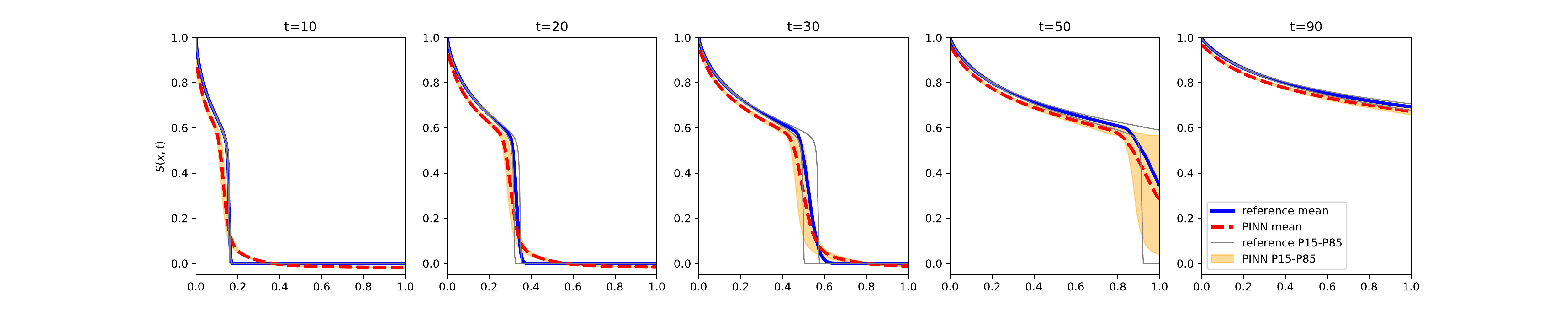}%
    \caption{Comparison of saturation distributions profiles at five different time steps for a heterogeneous affine total velocity distribution $v_d(x)\sim \mathcal{N}(\mu=1,\sigma=0.3)\times x + 0.5$. The reference mean saturation (computed through Monte Carlo simulation of a finite volume numerical model) is in solid blue while the one computed with P-PINNs is in dashed red. The P15-P85 envelopes are represented for both reference and P-PINN}%
    \label{fig:sat_distrib_vd_affine}%
\end{figure}

This represents the simplest non homogeneous field one can think of. It is part of a series of experiments meant to test the limits of the approach where we use a single parameter to encode $v_d$. The results of the simulated saturation are plotted in figure~\ref{fig:sat_distrib_vd_affine}

\begin{figure}[h!]%
    \centering
    \includegraphics[width=1\linewidth]{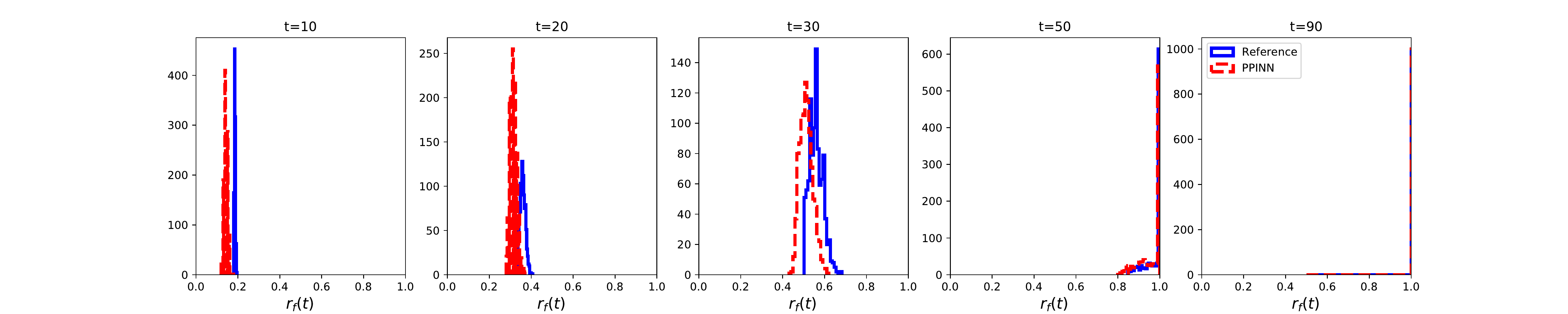}%
    \caption{Comparison of front radius distributions profiles at five different time steps for a heterogeneous total velocity distribution $v_d(x)\sim \mathcal{N}(\mu=1,\sigma=0.3)\times x + 0.5$. The reference front radius (computed through Monte Carlo simulation of a finite volume numerical model) is in solid blue while the one computed with P-PINNs is in dashed red.}%
    \label{fig:front_radius_vd_affine}%
\end{figure}

We represent the breakthrough time and front radius distributions and compare with the reference analytical solution. The distributions of shock radii are shown in Figure~\ref{fig:front_radius_vd_affine}. They present a slight difference with the reference solution but the uncertainty range is captured.

\begin{figure}[h!]%
    \centering
    \includegraphics[width=1\linewidth]{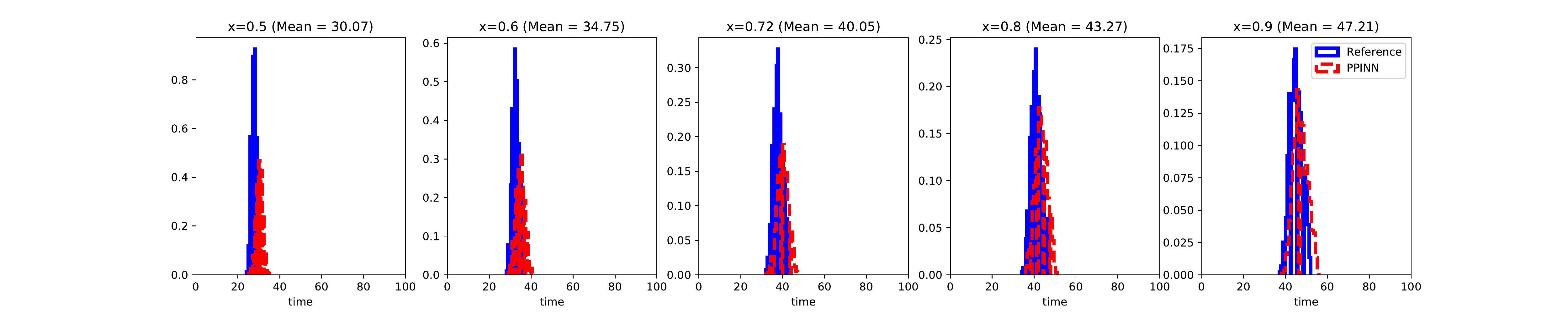}%
    \caption{Comparison of breakthrough time distributions profiles at five different locations in space for a heterogeneous total velocity distribution $v_d(x)\sim \mathcal{N}(\mu=1,\sigma=0.3)\times x + 0.5$. The reference breakthrough time (computed through Monte Carlo simulation of a finite volume numerical model) is in solid blue while the one computed with P-PINNs is in dashed red.}%
    \label{fig:bt_time_vd_affine}%
\end{figure}

The distribution of breakthrough times are presented in figure~\ref{fig:bt_time_vd_affine}

Table~\ref{tab:front_radius_vd_affine} shows the average Wasserstein distances between distributions computed using P-PINNS and the reference for both front radius and breakthrough times.

\begin{table}[h!]
	\caption{Wasserstein distance average for distributions of QOIs for a for a heterogeneous total velocity distribution $v_d(x)\sim \mathcal{N}(\mu=1,\sigma=0.3)\times x + 0.5$. U is a uniform distribution.}
	\centering
	\begin{tabular}{ccc}
		\toprule
		    & Front Radius     & Breakthrough time \\
	    \midrule
		$W_p(P_{MOC}, P_{PINN})$ & 0.03  & 2.30 \\
        $W_p(P_{MOC}, U)$     & 0.36 & 25.47 \\
        \midrule
        Relative difference & 8.2\% & 9.0\% \\
		\bottomrule
	\end{tabular}
	\label{tab:front_radius_vd_affine}
\end{table}

\subsection{Stochastic Periodic velocity function}
\begin{figure}[h!]%
    \centering
    \includegraphics[width=1\linewidth]{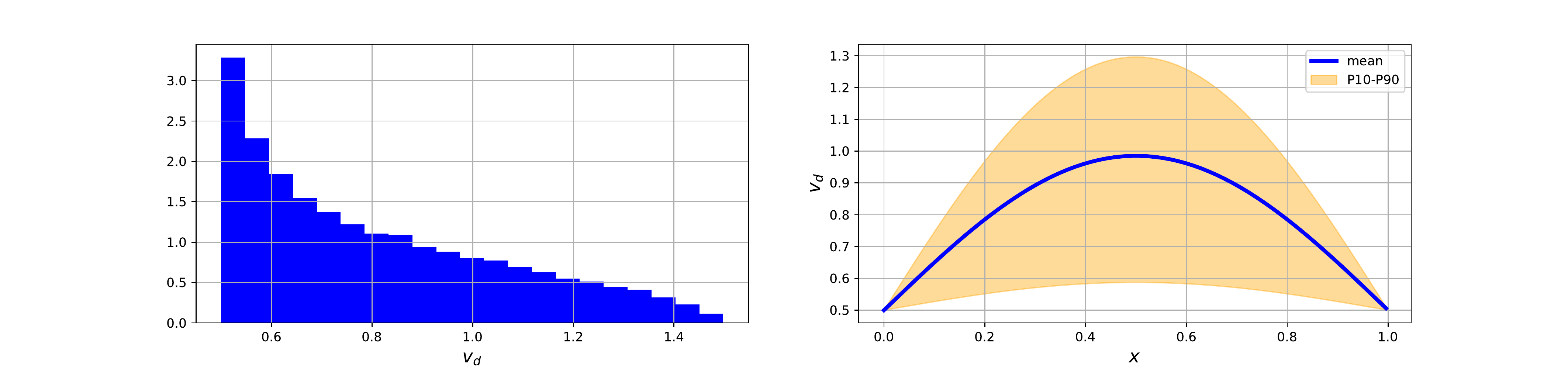}%
    \caption{Distribution of total velocities probability density (left) and along x-axis (right) where mean and P10-P90 envelope are represented}%
    \label{fig:v_d_het_sin_distrib}%
\end{figure}

In this scenario, we explore more complex random Darcy velocity fields with non monotonic and periodic behavior. The field can still be parameterized with a single random parameter $\theta$ with a Gaussian distribution.
\begin{equation}
    v_d(x) = \theta sin(x) + b
\end{equation}
With
\begin{equation}
    \theta \sim \mathcal{N}(\mu=1,\,\sigma^{2}=0.3)
\end{equation}
We plot the histogram of $v_d$ along with the ensemble realizations in figure~\ref{fig:v_d_het_sin_distrib}

\begin{figure}[h!]%
    \centering
    \includegraphics[width=1\linewidth]{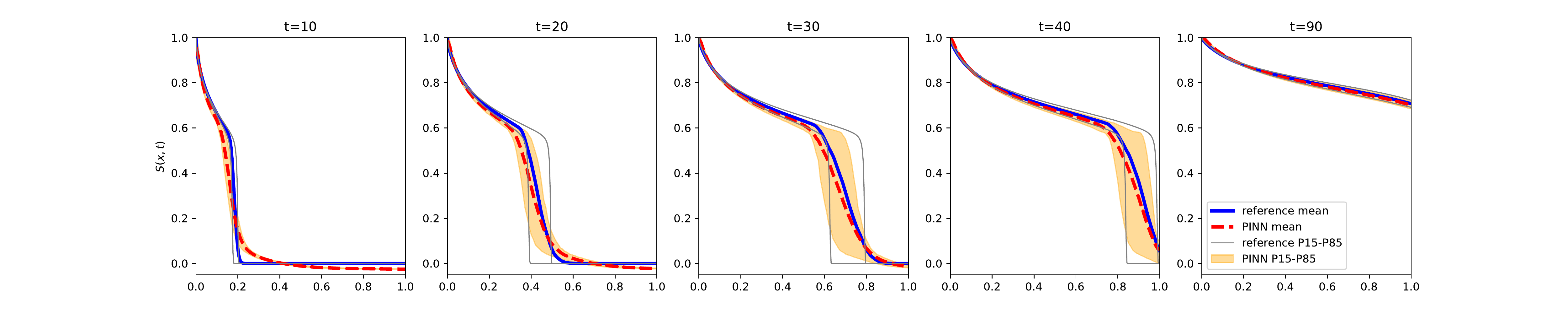}%
    \caption{Comparison of saturation distributions profiles at five different time steps for a heterogeneous affine total velocity distribution $v_d(x)\sim \mathcal{N}(\mu=1,\sigma=0.3)\times sin(x) + 0.5$. The reference mean saturation (computed through Monte Carlo simulation of a finite volume numerical model) is in solid blue while the one computed with P-PINNs is in dashed red. The P15-P85 envelopes are represented for both reference and P-PINN}%
    \label{fig:sat_distrib_vd_sin}%
\end{figure}

We represent the saturation solutions for the ensemble of realizations in figure~\ref{fig:sat_distrib_vd_sin}.

\begin{figure}[h!]%
    \centering
    \includegraphics[width=1\linewidth]{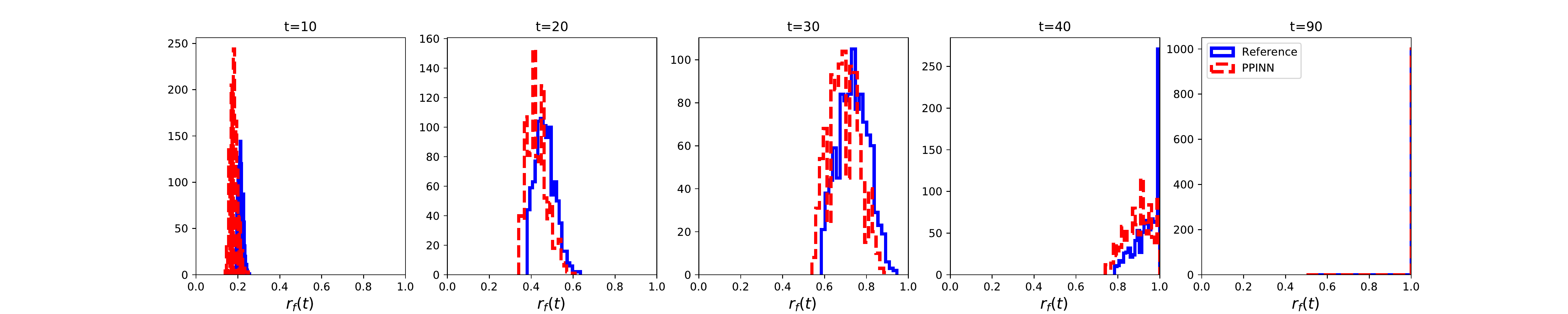}%
    \caption{Comparison of front radius distributions profiles at five different time steps for a heterogeneous total velocity distribution $v_d(x)\sim \mathcal{N}(\mu=1,\sigma=0.3)\times sin(x) + 0.5$. The reference front radius (computed through Monte Carlo simulation of a finite volume numerical model) is in solid blue while the one computed with P-PINNs is in dashed red.}%
    \label{fig:front_radius_vd_sin}%
\end{figure}

We represent the breakthrough time and front radius distributions and compare with the reference analytical solution. The distributions of shock radii are shown in Figure~\ref{fig:front_radius_vd_sin}. They present a slight difference with the reference solution but the uncertainty range is captured.

\begin{figure}[h!]%
    \centering
    \includegraphics[width=1\linewidth]{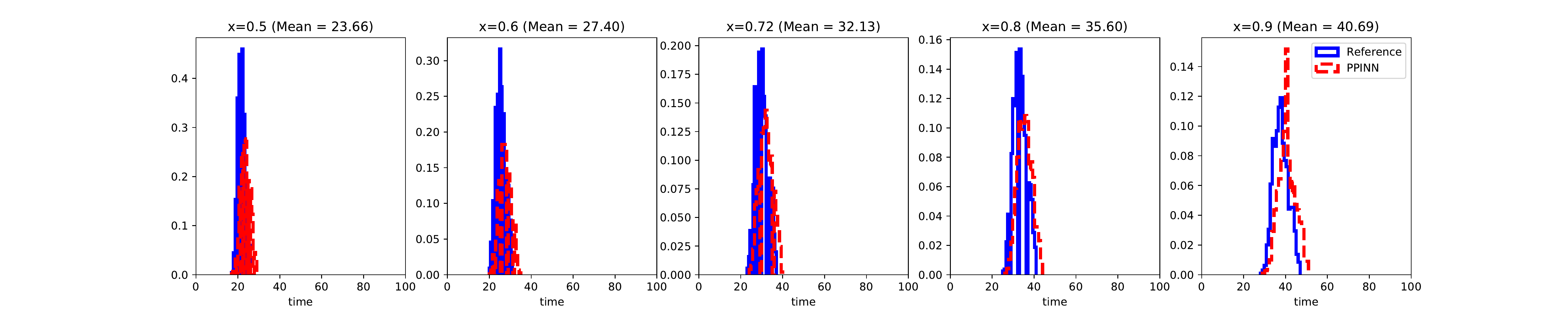}%
    \caption{Comparison of breakthrough time distributions profiles at five different locations in space for a heterogeneous total velocity distribution $v_d(x)\sim \mathcal{N}(\mu=1,\sigma=0.3)\times sin(x) + 0.5$. The reference breakthrough time (computed through Monte Carlo simulation of a finite volume numerical model) is in solid blue while the one computed with P-PINNs is in dashed red.}%
    \label{fig:bt_time_vd_sin}%
\end{figure}

The distribution of breakthrough times are presented in figure~\ref{fig:bt_time_vd_sin}

Table~\ref{tab:bt_time_vd_sin} shows the average Wasserstein distances between distributions computed using P-PINNS and the reference for both front radius and breakthrough times.

\begin{table}[h!]
	\caption{Wasserstein distance average for distributions of QOIs for a for a heterogeneous total velocity distribution $v_d(x)\sim \mathcal{N}(\mu=1,\sigma=0.3)\times sin(x) + 0.5$. U is a uniform distribution.}
	\centering
	\begin{tabular}{ccc}
		\toprule
		    & Front Radius     & Breakthrough time \\
	    \midrule
		$W_p(P_{MOC}, P_{PINN})$ & 0.03  & 2.17 \\
        $W_p(P_{MOC}, U)$     & 0.34 & 28.11 \\
        \midrule
        Relative difference & 8.7\% & 7.7\% \\
		\bottomrule
	\end{tabular}
	\label{tab:bt_time_vd_sin}
\end{table}

We increase the frequency of the velocity field function a 5 fold:
\begin{equation}
    v_d(x) = \theta sin(5x) + b
\end{equation}

\begin{figure}[h!]%
    \centering
    \includegraphics[width=1\linewidth]{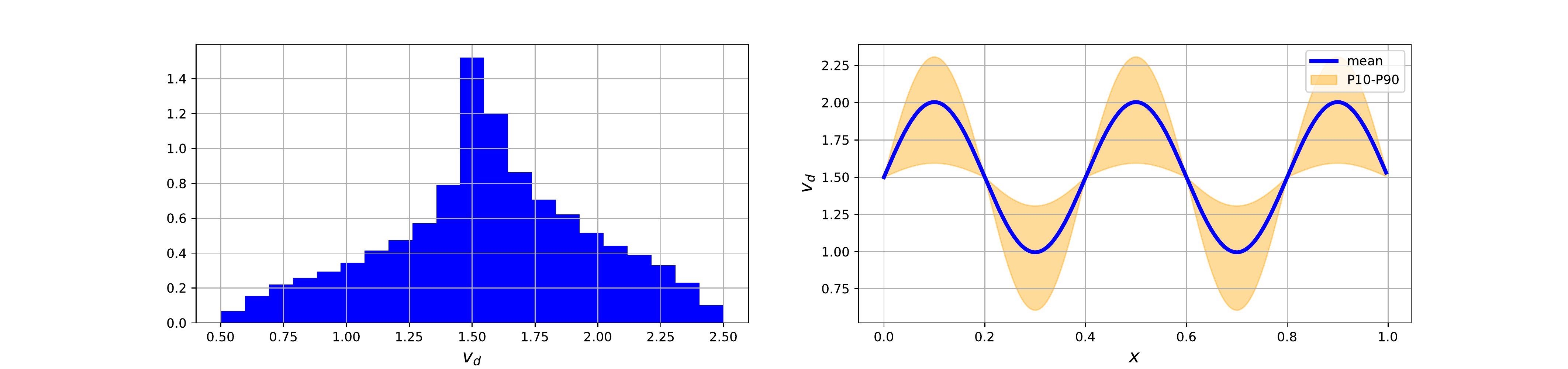}%
    \caption{Distribution of total velocities probability density (left) and along x-axis (right) where mean and P10-P90 envelope are represented}%
    \label{fig:v_d_het_sin_5_distrib}%
\end{figure}

The resulting field is represented in figure~\ref{fig:v_d_het_sin_5_distrib}

\begin{figure}[h!]%
    \centering
    \includegraphics[width=1\linewidth]{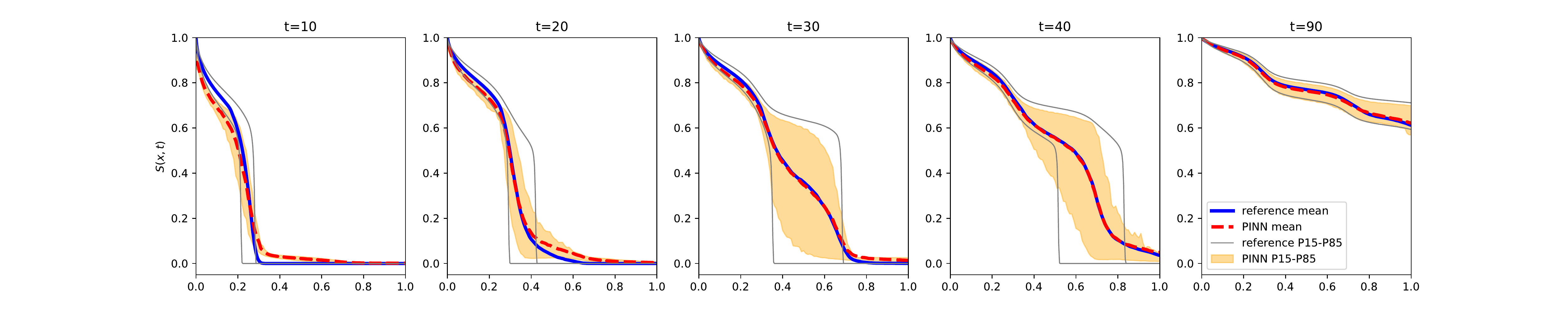}%
    \caption{Comparison of saturation distributions profiles at five different time steps for a heterogeneous affine total velocity distribution $v_d(x)\sim \mathcal{N}(\mu=1,\sigma=0.3)\times sin(5x) + 0.5$. The reference mean saturation (computed through Monte Carlo simulation of a finite volume numerical model) is in solid blue while the one computed with P-PINNs is in dashed red. The P15-P85 envelopes are represented for both reference and P-PINN}%
    \label{fig:sat_distrib_vd_sin5}%
\end{figure}

We represent the saturation solutions for the ensemble of realizations in figure~\ref{fig:sat_distrib_vd_sin5}.

\begin{figure}[h!]%
    \centering
    \includegraphics[width=1\linewidth]{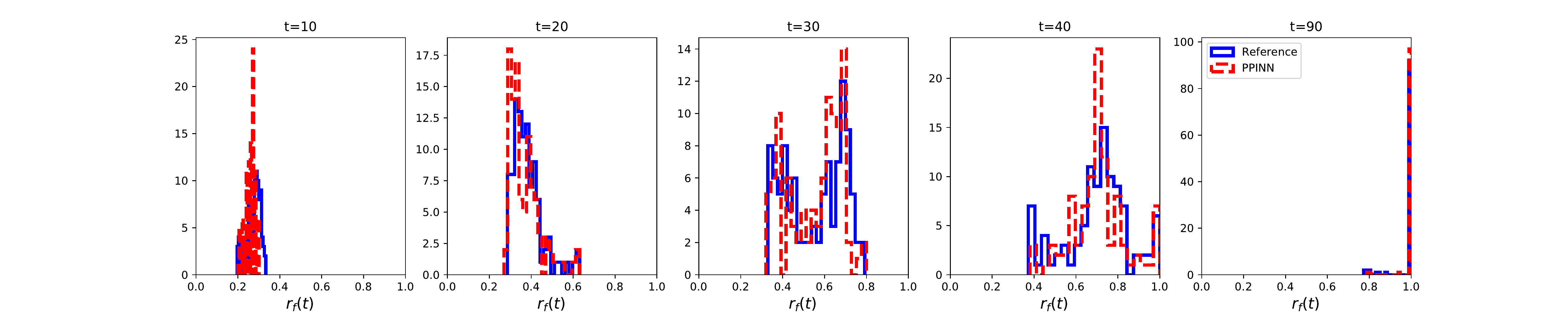}%
    \caption{Comparison of front radius distributions profiles at five different time steps for a heterogeneous total velocity distribution $v_d(x)\sim \mathcal{N}(\mu=1,\sigma=0.3)\times sin(5x) + 0.5$. The reference front radius (computed through Monte Carlo simulation of a finite volume numerical model) is in solid blue while the one computed with P-PINNs is in dashed red.}%
    \label{fig:front_radius_vd_sin5}%
\end{figure}

We represent the breakthrough time and front radius distributions and compare with the reference analytical solution. The distributions of shock radii are shown in Figure~\ref{fig:front_radius_vd_sin5}. They present a slight difference with the reference solution but the uncertainty range is captured.

\begin{figure}[h!]%
    \centering
    \includegraphics[width=1\linewidth]{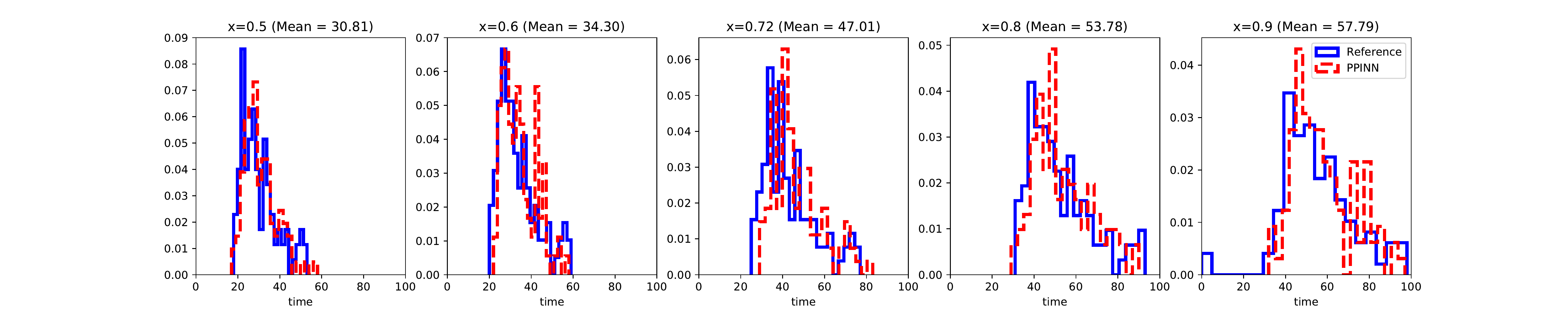}%
    \caption{Comparison of breakthrough time distributions profiles at five different locations in space for a heterogeneous total velocity distribution $v_d(x)\sim \mathcal{N}(\mu=1,\sigma=0.3)\times sin(5x) + 0.5$. The reference breakthrough time (computed through Monte Carlo simulation of a finite volume numerical model) is in solid blue while the one computed with P-PINNs is in dashed red.}%
    \label{fig:bt_time_vd_sin5}%
\end{figure}

The distribution of breakthrough times are presented in figure~\ref{fig:bt_time_vd_sin5}

Table~\ref{tab:bt_time_vd_sin5} shows the average Wasserstein distances between distributions computed using P-PINNS and the reference for both front radius and breakthrough times.

\begin{table}[h!]
	\caption{Wasserstein distance average for distributions of QOIs for a for a heterogeneous total velocity distribution $v_d(x)\sim \mathcal{N}(\mu=1,\sigma=0.3)\times sin(5x) + 0.5$. U is a uniform distribution.}
	\centering
	\begin{tabular}{ccc}
		\toprule
		    & Front Radius     & Breakthrough time \\
	    \midrule
		$W_p(P_{MOC}, P_{PINN})$ & 0.02  & 2.49 \\
        $W_p(P_{MOC}, U)$     & 0.26 & 18.15 \\
        \midrule
        Relative difference & 6.7\% & 13.7\% \\
		\bottomrule
	\end{tabular}
	\label{tab:bt_time_vd_sin5}
\end{table}

We increase the frequency of the velocity field function a 25 fold:
\begin{equation}
    v_d(x) = \theta sin(25x) + b
\end{equation}

\begin{figure}[h!]%
    \centering
    \includegraphics[width=1\linewidth]{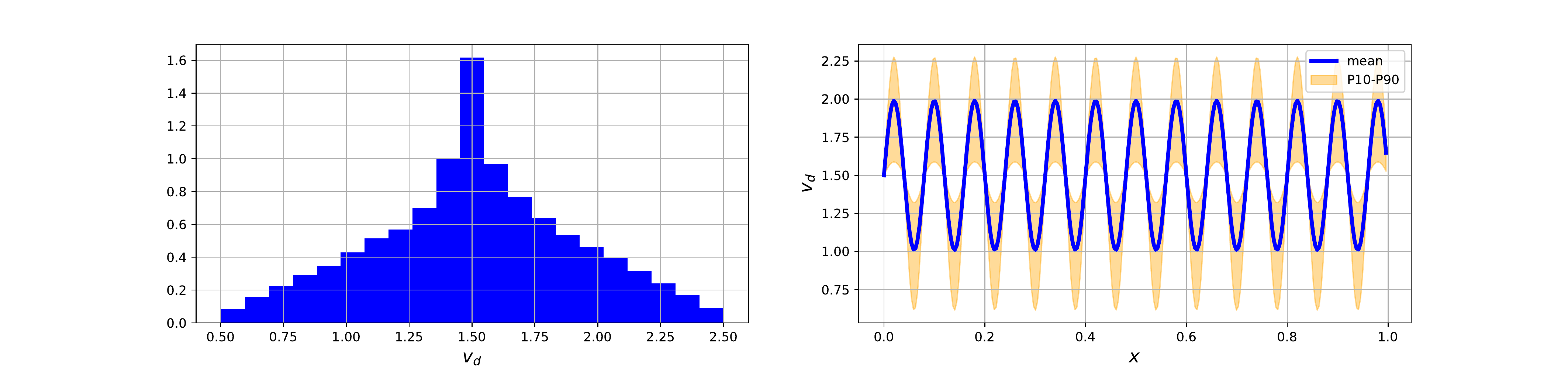}%
    \caption{Distribution of total velocities probability density (left) and along x-axis (right) where mean and P10-P90 envelope are represented}%
    \label{fig:v_d_het_sin_25_distrib}%
\end{figure}

The resulting field is represented in figure~\ref{fig:v_d_het_sin_25_distrib}

\begin{figure}[h!]%
    \centering
    \includegraphics[width=1\linewidth]{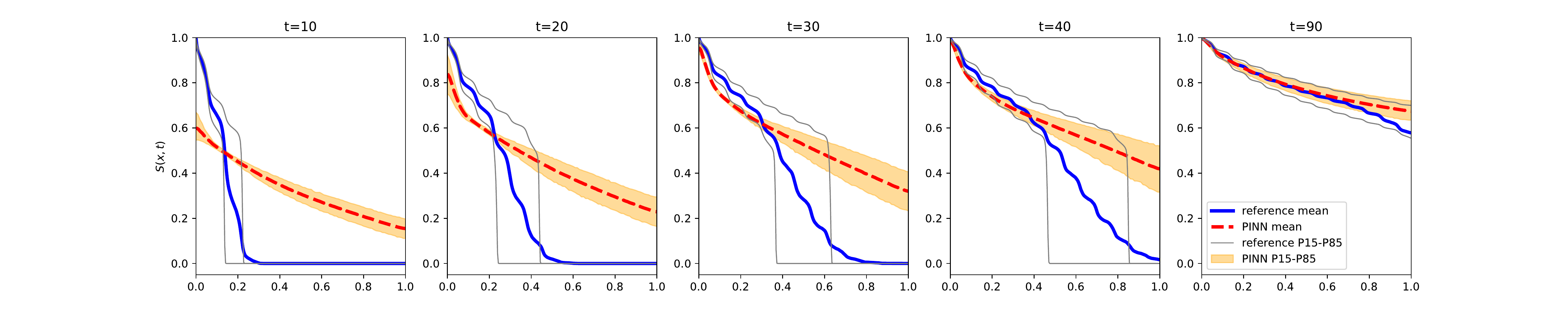}%
    \caption{Comparison of saturation distributions profiles at five different time steps for a heterogeneous affine total velocity distribution $v_d(x)\sim \mathcal{N}(\mu=1,\sigma=0.3)\times sin(25x) + 0.5$. The reference mean saturation (computed through Monte Carlo simulation of a finite volume numerical model) is in solid blue while the one computed with P-PINNs is in dashed red. The P15-P85 envelopes are represented for both reference and P-PINN}%
    \label{fig:sat_distrib_vd_sin_25_fail}%
\end{figure}

We represent the saturation solutions for the ensemble of realizations in figure~\ref{fig:sat_distrib_vd_sin_25_fail}.

\begin{figure}[h!]%
    \centering
    \includegraphics[width=1\linewidth]{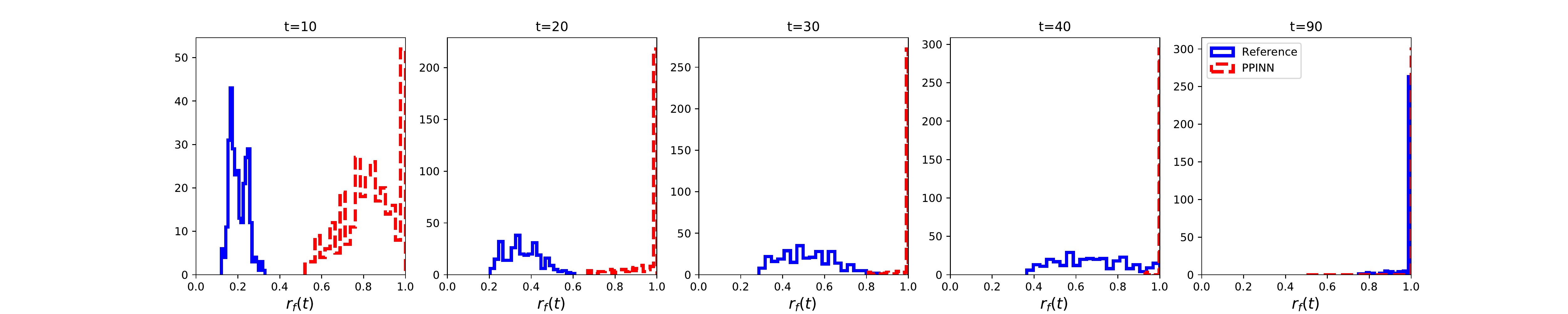}%
    \caption{Comparison of front radius distributions profiles at five different time steps for a heterogeneous total velocity distribution $v_d(x)\sim \mathcal{N}(\mu=1,\sigma=0.3)\times sin(25x) + 0.5$. The reference front radius (computed through Monte Carlo simulation of a finite volume numerical model) is in solid blue while the one computed with P-PINNs is in dashed red.}%
    \label{fig:front_radius_vd_sin_25_fail}%
\end{figure}

The figure shows that the field simulated using P-PINN fails to capture the front progression and the associated uncertainty. 
We represent the front radius distributions and compare with the reference analytical solution. The distributions of shock radii are shown in Figure~\ref{fig:front_radius_vd_sin_25_fail}. They present a slight difference with the reference solution but the uncertainty range is captured.

Table~\ref{tab:front_radius_vd_sin_25_fail} shows the average Wasserstein distances between distributions computed using P-PINNS and the reference for both front radius and breakthrough times.

\begin{table}[h!]
	\caption{Wasserstein distance average for distributions of QOIs for a for a heterogeneous total velocity distribution $v_d(x)\sim \mathcal{N}(\mu=1,\sigma=0.3)\times sin(25x) + 0.5$. U is a uniform distribution.}
	\centering
	\begin{tabular}{ccc}
		\toprule
		    & Front Radius     & Breakthrough time \\
	    \midrule
		$W_p(P_{MOC}, P_{PINN})$ & 0.41  & 18.59 \\
        $W_p(P_{MOC}, U)$     & 0.26 & 16.15 \\
        \midrule
        Relative difference & 153.9\% & 115.1\% \\
		\bottomrule
	\end{tabular}
	\label{tab:front_radius_vd_sin_25_fail}
\end{table}

The increase in frequency of the velocity field's function leads to the non-convergence of the P-PINN method. Although the distribution is still characterized by one parameter, it becomes more difficult for a neural network to model the solution when the input signal has a high frequency. This observation could be a consequence of spectral bias (\cite{Rahaman2019}. \cite{Tancik2020}, \cite{SiRen2018}). Proposals to remedy this bias involve the additional encoding of input via a higher dimensional space projection via Fourier transform (\cite{FourierNet2020}) of the input space. We achieve this by a adding an additional input layer to the original MLP model defined in eq.~\ref{eq:ffwd_form_base}. The resulting model can be written:

\begin{equation}
\label{eq:ffwd_form_fourier}
    \mathbf{\nu} \approx \mathbf{\nu}_{\theta} = \sigma\left[\mathbf{W}^{[n]}\times \sigma(\mathbf{W}^{[n-1]}(\dots\sigma(\mathbf{W}^{[0]}\phi_f([X,t,v_d]^T) + \mathbf{b}^{[0]}))\dots + \mathbf{b}^{[n-1]}) + \mathbf{b}^{[n]}\right]
\end{equation}

Where:

\begin{equation}
    \phi_f([x,t,v_d]^T) = [\cos(2\pi (\mathbf{W^{[f]}}[x,t,v_d]^T),\sin(2\pi (\mathbf{W^{[f]}}[x,t,v_d]^T)]^T 
\end{equation}
Where $\mathbf{W^{[f]}})\in \mathcal{R}^{N\times O(1)}$ is the weight matrix for the first layer and effectively amount to a Fourier expansion of the input arguments.
This effectively adds another layer to the neural network with an additional $3N$ training parameters. This transformation helps alleviate the issue of spectral bias as shown in the following results.

\begin{figure}[h!]%
    \centering
    \includegraphics[width=1\linewidth]{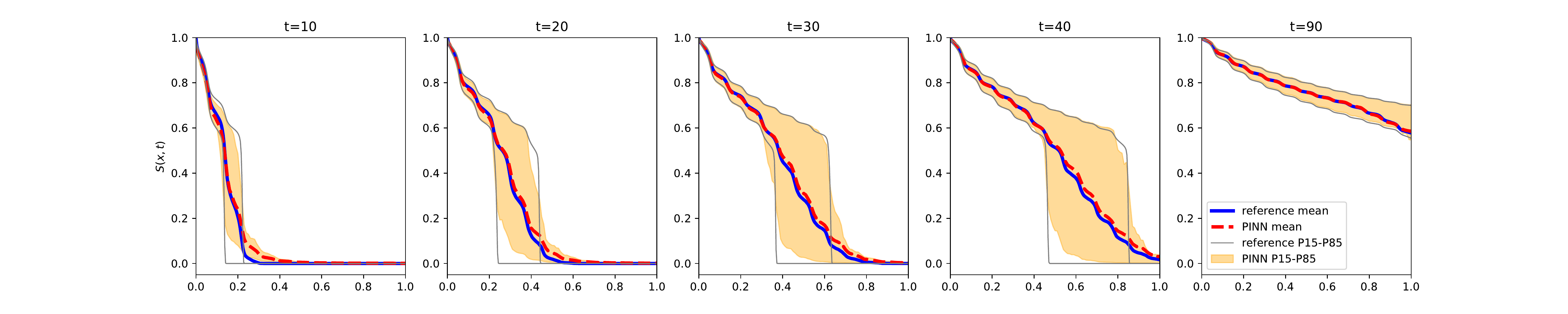}%
    \caption{Comparison of saturation distributions profiles at five different time steps for a heterogeneous affine total velocity distribution $v_d(x)\sim \mathcal{N}(\mu=1,\sigma=0.3)\times sin(25x) + 0.5$. The reference mean saturation (computed through Monte Carlo simulation of a finite volume numerical model) is in solid blue while the one computed with Fourier P-PINNs (FP-PINN) is in dashed red. The P15-P85 envelopes are represented for both reference and P-PINN}%
    \label{fig:sat_distrib_vd_sin25_fourier}%
\end{figure}

For the problem featuring the velocity field presented in figure~\ref{fig:v_d_het_sin_25_distrib}, the application of the Fourier layer transform leads to results that are largely improved as shown in figure~\ref{fig:sat_distrib_vd_sin25_fourier}

\begin{figure}[h!]%
    \centering
    \includegraphics[width=1\linewidth]{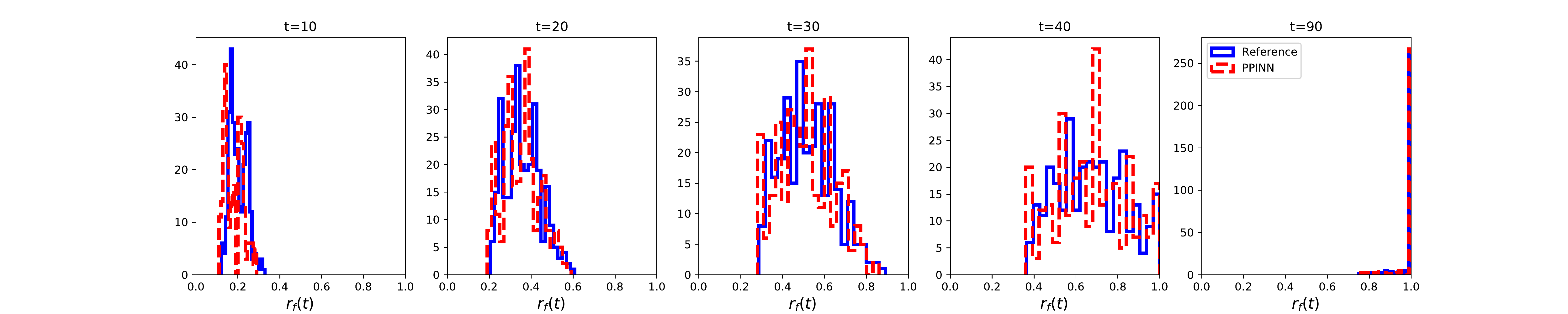}%
    \caption{Comparison of front radius distributions profiles at five different time steps for a heterogeneous total velocity distribution $v_d(x)\sim \mathcal{N}(\mu=1,\sigma=0.3)\times sin(25x) + 0.5$. The reference front radius (computed through Monte Carlo simulation of a finite volume numerical model) is in solid blue while the one computed with FP-PINNs is in dashed red.}%
    \label{fig:front_radius_vd_sin25_fourier}%
\end{figure}

We represent the breakthrough time and front radius distributions and compare with the reference analytical solution. The distributions of shock radii are shown in Figure~\ref{fig:front_radius_vd_sin25_fourier}. They present a slight difference with the reference solution but the uncertainty range is captured.

\begin{figure}[h!]%
    \centering
    \includegraphics[width=1\linewidth]{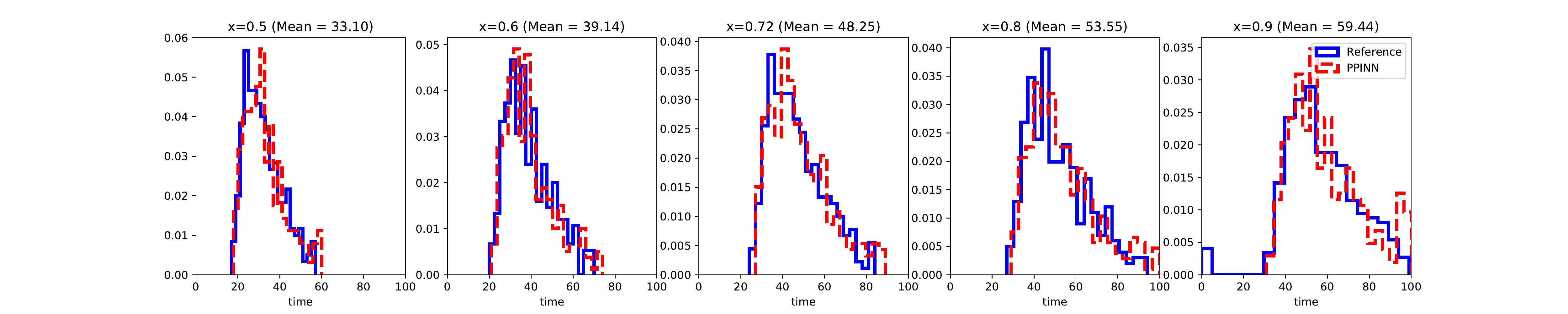}%
    \caption{Comparison of breakthrough time distributions profiles at five different locations in space for a heterogeneous total velocity distribution $v_d(x)\sim \mathcal{N}(\mu=1,\sigma=0.3)\times sin(25x) + 0.5$. The reference breakthrough time (computed through Monte Carlo simulation of a finite volume numerical model) is in solid blue while the one computed with FP-PINNs is in dashed red.}%
    \label{fig:bt_time_vd_sin25_fourier}%
\end{figure}

The distribution of breakthrough times are presented in figure~\ref{fig:bt_time_vd_sin25_fourier}

Table~\ref{tab:bt_time_vd_sin25_fourier} shows the average Wasserstein distances between distributions computed using P-PINNS and the reference for both front radius and breakthrough times.

\begin{table}[h!]
	\caption{Wasserstein distance average for distributions of QOIs for a for a heterogeneous total velocity distribution $v_d(x)\sim \mathcal{N}(\mu=1,\sigma=0.3)\times sin(25x) + 0.5$. U is a uniform distribution.}
	\centering
	\begin{tabular}{ccc}
		\toprule
		    & Front Radius     & Breakthrough time \\
	    \midrule
		$W_p(P_{MOC}, P_{PINN})$ & 0.01  & 1.68 \\
        $W_p(P_{MOC}, U)$     & 0.27 & 17.04 \\
        \midrule
        Relative difference & 4.8\% & 9.9\% \\
		\bottomrule
	\end{tabular}
	\label{tab:bt_time_vd_sin25_fourier}
\end{table}

\subsection{Fourier decomposition of the velocity function}
The results obtained previously allow us to model random fields of increasing complexity. Indeed, we can now approximate any random signal with its Fourier decomposition. Our parametrization has to account for the various modes since we have one random factor per mode as shown in eq.~\ref{eq:random_vd_fourier}.

\begin{equation}
\label{eq:random_vd_fourier}
    v_d(x) = \sum_{k=1}^N\theta_k\sin(2\pi kx) + b
\end{equation}
With $\theta_k$ modeled according to a uniform law:
\begin{equation}
    \theta_k \sim \mathcal{U}(low=0,\,high=1)
\end{equation}

\begin{figure}[h!]%
    \centering
    \includegraphics[width=1\linewidth]{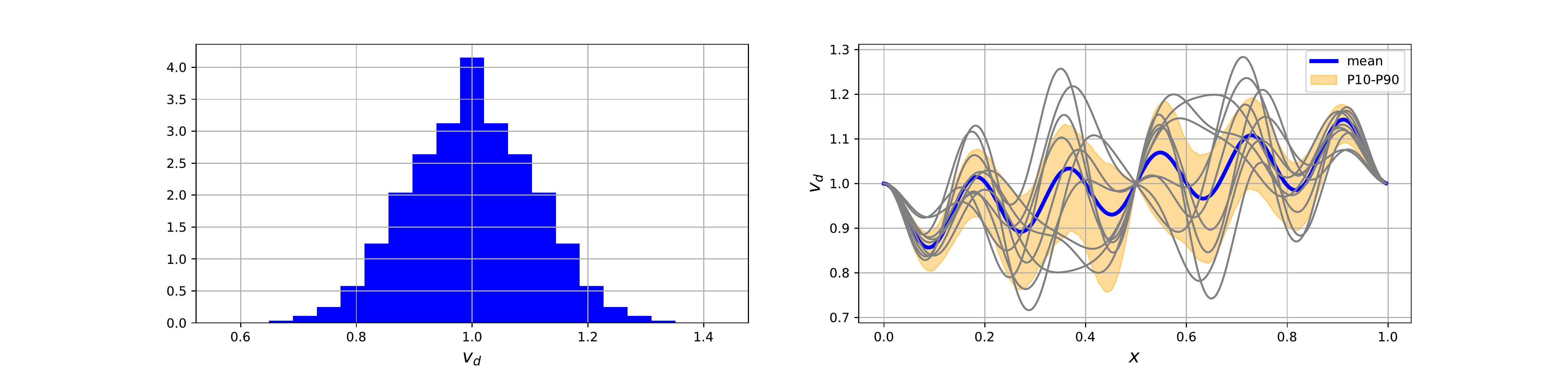}%
    \caption{Distribution of total velocities probability density (left) and along x-axis (right) for a random field modeled using Fourier series with 5 modes where mean and P10-P90 envelope are represented}%
    \label{fig:v_d_het_fourier5_distrib}%
\end{figure}

The fact that we can model Fourier components with higher modes than before helps considering short range variations in the velocity field. An example of such a field with 5 modes is presented in figure~\ref{fig:v_d_het_fourier5_distrib}. We represent a few realizations (in gray) in order to give a sense of the spatial variability one can expect from such a model.

\begin{figure}[h!]%
    \centering
    \includegraphics[width=1\linewidth]{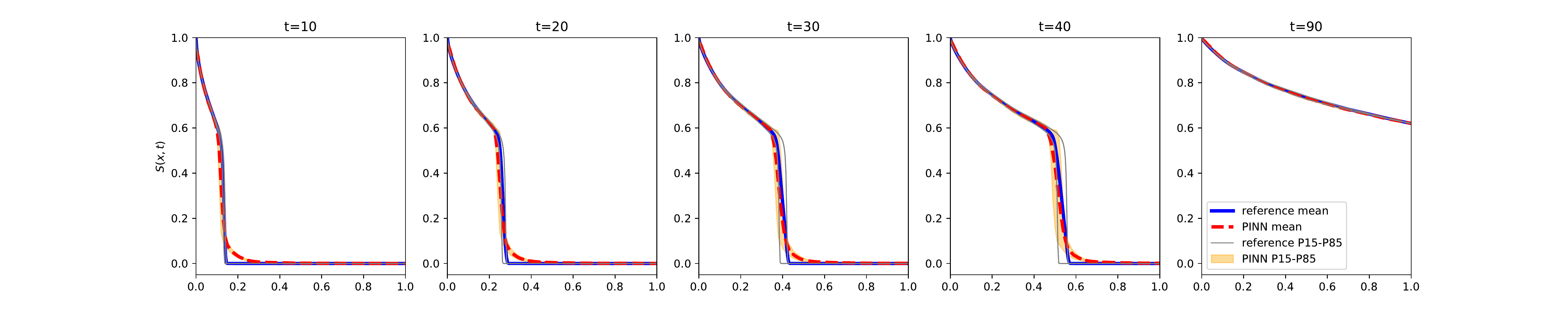}%
    \caption{Comparison of saturation distributions profiles at five different time steps for a heterogeneous affine total velocity distribution $v_d(x)\sim \sum_{k=1}^5\theta_k\sin(2\pi kx) + 1$. The reference mean saturation (computed through Monte Carlo simulation of a finite volume numerical model) is in solid blue while the one computed with Fourier P-PINNs (FP-PINN) is in dashed red. The P15-P85 envelopes are represented for both reference and P-PINN}%
    \label{fig:sat_distrib_vd_fourier5modes}%
\end{figure}

For the problem featuring the velocity field presented in figure~\ref{fig:v_d_het_fourier5_distrib}, using the FP-PINN approach with 5 input parameters $\theta_k$ for the velocity, we obtain saturation envelopes that match the finite volume solution as shown in figure~\ref{fig:sat_distrib_vd_fourier5modes}

\begin{figure}[h!]%
    \centering
    \includegraphics[width=1\linewidth]{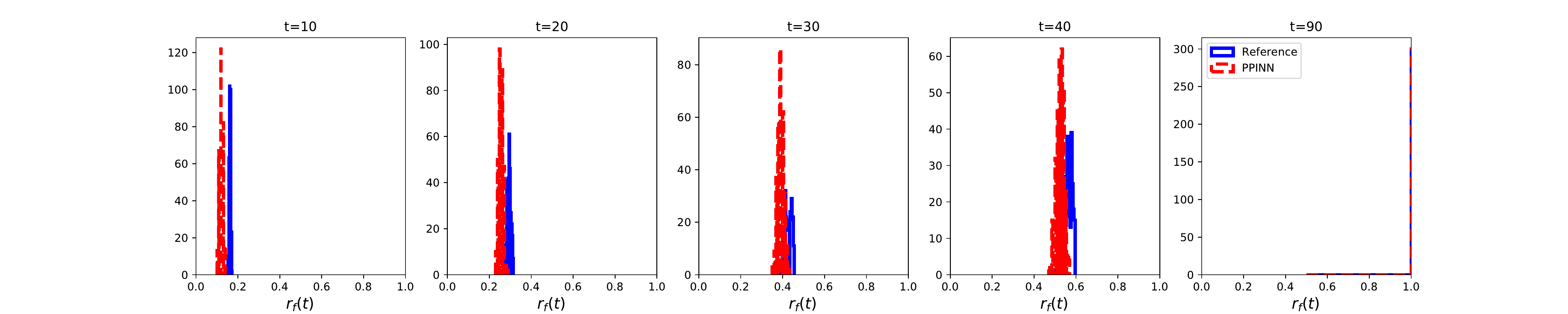}%
    \caption{Comparison of front radius distributions profiles at five different time steps for a heterogeneous total velocity distribution $v_d(x)\sim \sum_{k=1}^5\theta_k\sin(2\pi kx) + 1$. The reference front radius (computed through Monte Carlo simulation of a finite volume numerical model) is in solid blue while the one computed with FP-PINNs is in dashed red.}%
    \label{fig:front_radius_vd_fourier5modes}%
\end{figure}

We represent the breakthrough time and front radius distributions and compare with the reference analytical solution. The distributions of shock radii are shown in Figure~\ref{fig:front_radius_vd_fourier5modes}. They present a slight difference with the reference solution but the uncertainty range is captured.

\begin{figure}[h!]%
    \centering
    \includegraphics[width=1\linewidth]{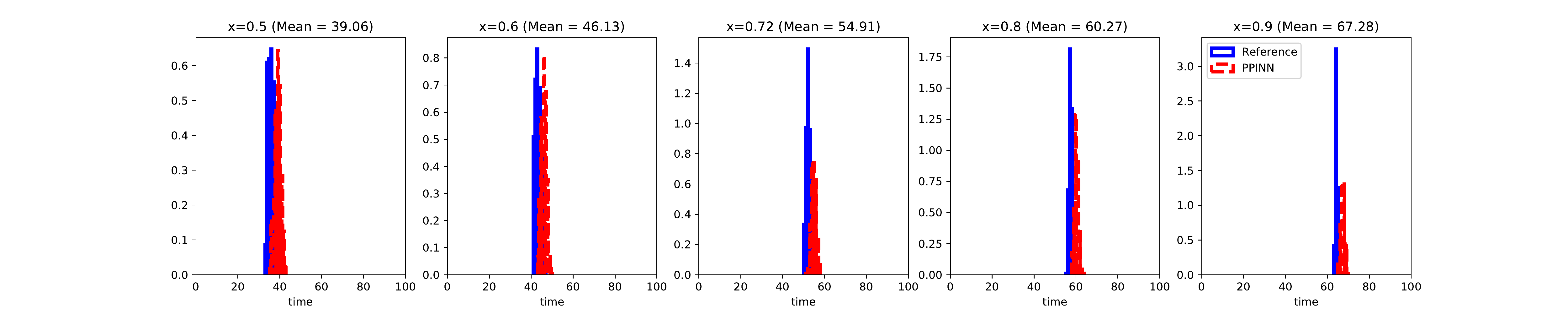}%
    \caption{Comparison of breakthrough time distributions profiles at five different locations in space for a heterogeneous total velocity distribution $v_d(x)\sim \sum_{k=1}^5\theta_k\sin(2\pi kx) + 1$. The reference breakthrough time (computed through Monte Carlo simulation of a finite volume numerical model) is in solid blue while the one computed with FP-PINNs is in dashed red.}%
    \label{fig:bt_time_vd_fourier5modes}%
\end{figure}

The distribution of breakthrough times are presented in figure~\ref{fig:bt_time_vd_fourier5modes}

Table~\ref{tab:bt_time_vd_fourier5modes} shows the average Wasserstein distances between distributions computed using P-PINNS and the reference for both front radius and breakthrough times.

\begin{table}[h!]
	\caption{Wasserstein distance average for distributions of QOIs for a for a heterogeneous total velocity distribution $v_d(x)\sim \sum_{k=1}^5\theta_k\sin(2\pi kx) + 1$. U is a uniform distribution.}
	\centering
	\begin{tabular}{ccc}
		\toprule
		    & Front Radius     & Breakthrough time \\
	    \midrule
		$W_p(P_{MOC}, P_{PINN})$ & 0.03  & 3.09 \\
        $W_p(P_{MOC}, U)$     & 0.33 & 25.38 \\
        \midrule
        Relative difference & 9.6\% & 12.2\% \\
		\bottomrule
	\end{tabular}
	\label{tab:bt_time_vd_fourier5modes}
\end{table}

The addition of parameters along with the Fourier networks allow to model transport uncertainty on a heterogeneous random field with substantial variability and offer a robust approach to both higher frequency and wider variability. To demonstrate, we present results for a wider distribution support on the random parameters $\theta_k$

\begin{equation}
    \theta_k \sim \mathcal{U}(low=0,\,high=3)
\end{equation}

\begin{figure}[h!]%
    \centering
    \includegraphics[width=1\linewidth]{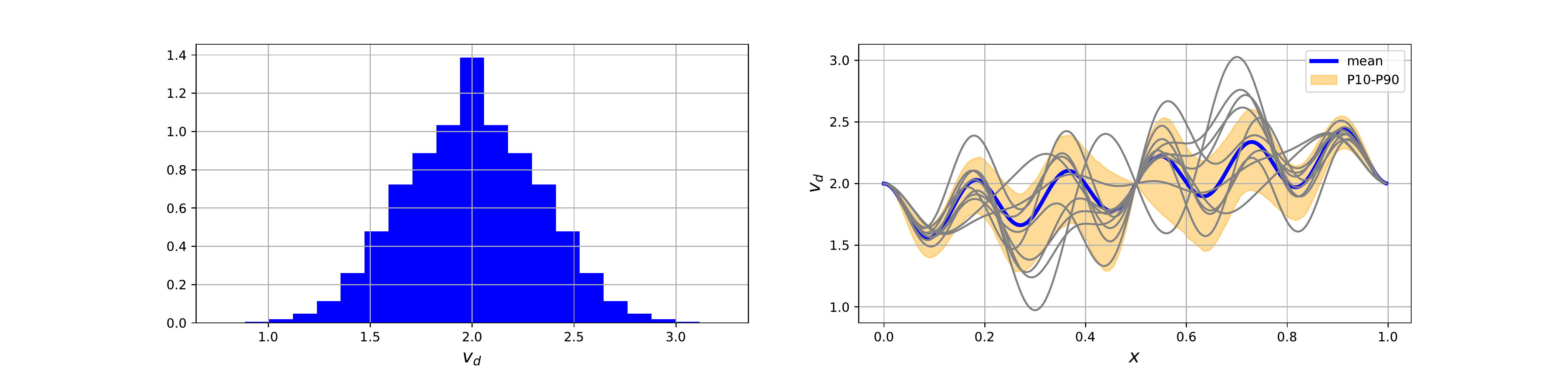}%
    \caption{Distribution of total velocities probability density (left) and along x-axis (right) for a random field modeled using Fourier series with 5 modes ($v_d(x)\sim 3\sum_{k=1}^5\theta_k\sin(2\pi kx) + 2$) where mean and P10-P90 envelope are represented}%
    \label{fig:v_d_het_fourier5_wide_distrib}%
\end{figure}

The random field is presented in figure~\ref{fig:v_d_het_fourier5_wide_distrib}. Note how the support of $v_d$ is 3 times as wide as in figure~\ref{fig:v_d_het_fourier5_distrib}.

\begin{figure}[h!]%
    \centering
    \includegraphics[width=1\linewidth]{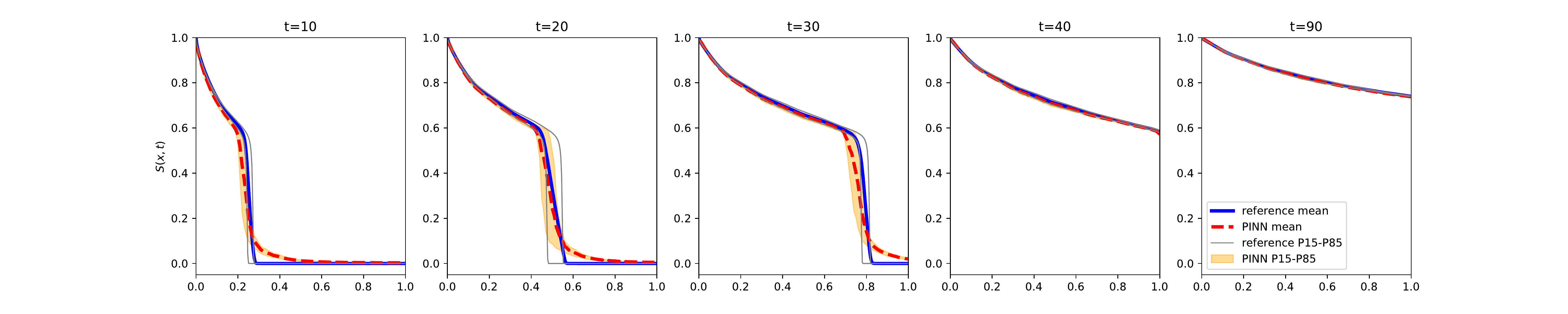}%
    \caption{Comparison of saturation distributions profiles at five different time steps for a heterogeneous affine total velocity distribution $v_d(x)\sim 3\sum_{k=1}^5\theta_k\sin(2\pi kx) + 2$. The reference mean saturation (computed through Monte Carlo simulation of a finite volume numerical model) is in solid blue while the one computed with Fourier P-PINNs (FP-PINN) is in dashed red. The P15-P85 envelopes are represented for both reference and P-PINN}%
    \label{fig:sat_distrib_vd_fourier5modes_wide}%
\end{figure}

Once again, we obtain saturation envelopes that match the finite volume solution as shown in figure~\ref{fig:sat_distrib_vd_fourier5modes_wide}

We represent the breakthrough time and front radius distributions and compare with the reference analytical solution. The distributions of shock radii are shown in Figure~\ref{fig:front_radius_vd_fourier5modes}. They present a slight difference with the reference solution but the uncertainty range is captured.
\begin{figure}[h!]%
    \centering
    \includegraphics[width=1\linewidth]{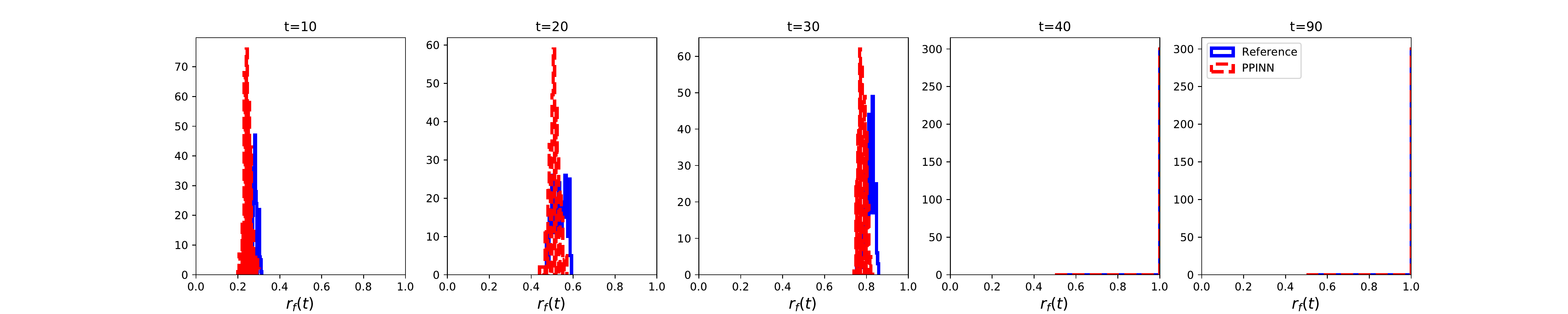}%
    \caption{Comparison of front radius distributions profiles at five different time steps for a heterogeneous total velocity distribution $v_d(x)\sim 3\sum_{k=1}^5\theta_k\sin(2\pi kx) + 2$. The reference front radius (computed through Monte Carlo simulation of a finite volume numerical model) is in solid blue while the one computed with FP-PINNs is in dashed red.}%
    \label{fig:front_radius_vd_fourier5modes_wide}%
\end{figure}

\begin{figure}[h!]%
    \centering
    \includegraphics[width=1\linewidth]{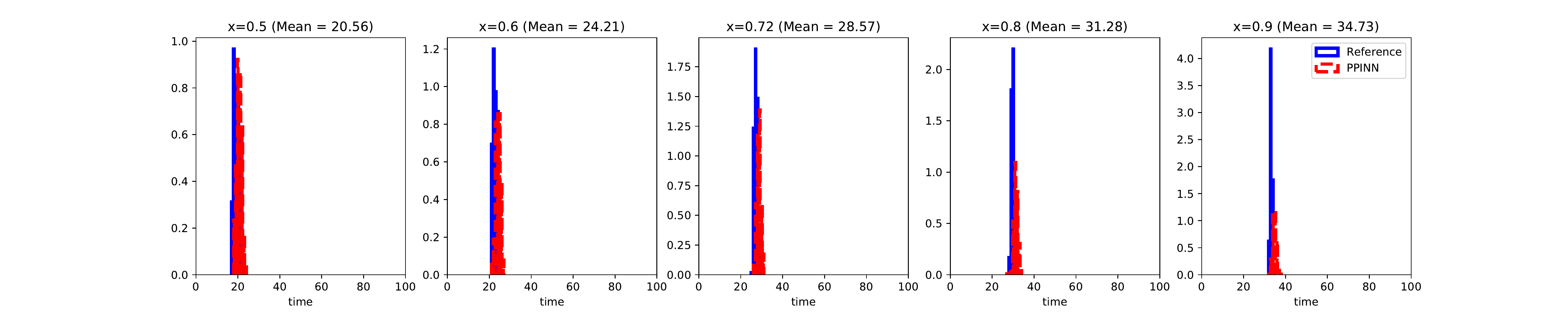}%
    \caption{Comparison of breakthrough time distributions profiles at five different locations in space for a heterogeneous total velocity distribution $v_d(x)\sim 3\sum_{k=1}^5\theta_k\sin(2\pi kx) + 2$. The reference breakthrough time (computed through Monte Carlo simulation of a finite volume numerical model) is in solid blue while the one computed with FP-PINNs is in dashed red.}%
    \label{fig:bt_time_vd_fourier5modes_wide}%
\end{figure}

The distribution of breakthrough times are presented in figure~\ref{fig:bt_time_vd_fourier5modes_wide}

Table~\ref{tab:bt_time_vd_fourier5modes_wide} shows the average Wasserstein distances between distributions computed using P-PINNS and the reference for both front radius and breakthrough times.

\begin{table}[h!]
	\caption{Wasserstein distance average for distributions of QOIs for a for a heterogeneous total velocity distribution $v_d(x)\sim 3\sum_{k=1}^5\theta_k\sin(2\pi kx) + 2$. U is a uniform distribution.}
	\centering
	\begin{tabular}{ccc}
		\toprule
		    & Front Radius     & Breakthrough time \\
	    \midrule
		$W_p(P_{MOC}, P_{PINN})$ & 0.02  & 1.49 \\
        $W_p(P_{MOC}, U)$     & 0.38 & 29.87 \\
        \midrule
        Relative difference & 5.3\% & 5.0\% \\
		\bottomrule
	\end{tabular}
	\label{tab:bt_time_vd_fourier5modes_wide}
\end{table}

All the results presented in this section show that a proper parameterization of the input space allow the modeling of the Buckley-Leverett problem with uncertain total velocity field for ensembles of realizations with increasing degrees of complexity. The problem remains one of interpolation in higher dimension where the network trained can be used to emulate each realization within the bounds of uncertainty on which it was trained. We propose an approach where the velocity field is first parameterized (we show an example with Fourier decomposition) and once a sufficient number of parameters has been established, the projected field is used to train the network. This requires an a-priori knowledge of the number of parameters required to encode the velocity field. The examples we have shown are synthetic and perfectly reproduce the uncertainty space. A more thorough error analysis would be required to establish a relationship between input and output error. Approaches based on principal elements decomposition (\cite{XuNumericalMethodsStochastic}) could be used to approximate the stochastic velocity field. We propose to explore these methods in section \ref{sec:moments_sim}.

\section{Simulation of moments}
\label{sec:moments_sim}
\begin{figure}[h!]%
    \centering
    \includegraphics[width=0.6\linewidth]{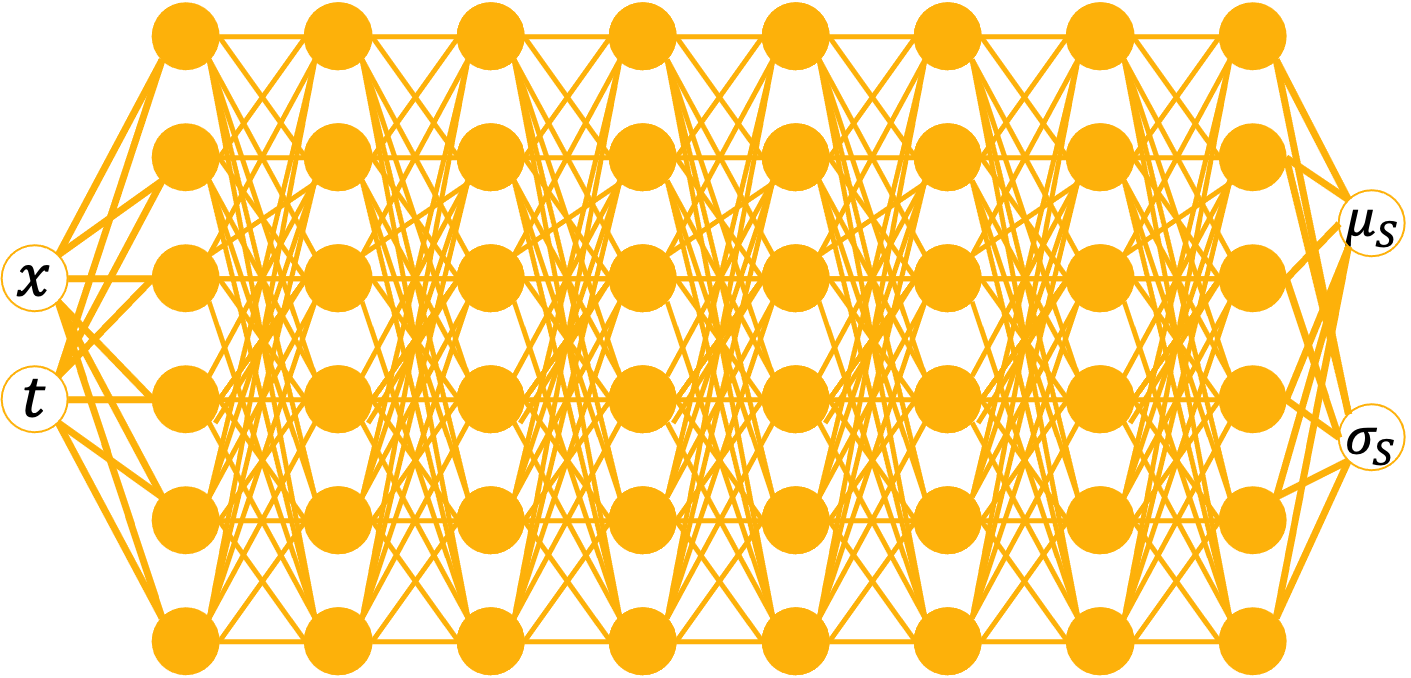}%
    \caption{Presentation of the model (input/output) used to model a Gaussian distribution of output parametrized by $\mu_s$ and $\sigma_s$ for a random velocity field}%
    \label{fig:mlp_proba_moments}%
\end{figure}

In this section, we present an application where PINN are used to simulate the moments of a distribution. Neural Networks are trained using least square techniques. Least Square methods can be formulated as a maximum likelihood (MLE) under the assumption of Gaussian distributions. Neural network training can be viewed as a generalization of MLE where a probability density describing data or a stochastic process can be approximated using a mixture of non-linear kernels (\cite{Bishop1994}). If generative and variational methods (\cite{Kingma_2014}, \cite{goodfellow2014generative}) have been employed in a data rich environment, no example -to our knowledge- has been presented in the context of forward PINN. We propose a PINN approach applied to solve the moments of the Buckley-Leverett approach. It is possible to derive partial differential equations from the moments of the hyperbolic equation \ref{eq:conservation_vd_het} and to solve for the moments of the random saturation solution (\cite{Jarman2000}, \cite{Jarman2006}, \cite{wang2013}, \cite{Langlo1994}). Analytical and numerical methods have been employed to solve this problem but never PINNs.  
An example of our new neural network model is presented in figure~\ref{fig:mlp_proba_moments}

\subsection{Formulation}
We use a perturbation method where the random terms are expanded using a two term Taylor series approximation. \cite{KAMINSKI2010272} presents a general framework for the application of perturbation techniques using such approximations. Each random field in the equation can be decomposed as the sum of an average and a random noise terms. Let the random saturation $\Tilde{S}_w(x,t)$ be:
\begin{equation}
    \Tilde{S}_w(x,t) = \langle\Tilde{S_w}(x,t)\rangle + \delta S_w(x,t)
\end{equation}
Where $\langle\Tilde{S_w}(x,t)\rangle =\Bar{S}_w$ is the expected value of the saturation and $\delta S_w$ a random fluctuation term with zero mean. Similarly, we write the random Darcy velocity field $v_d$ as:
\begin{equation}
    \Tilde{v}_d(x) = \Bar{v}_d + \delta v_d(x)
\end{equation}

We recall that the (now) Stochastic PDE we are solving is:

\begin{equation}
    \label{eq:conservation_stochastic}
    \frac{\partial \Tilde{S}_w}{\partial t} + \frac{\partial}{\partial x}(\Tilde{v}_d f_w(\Tilde{S}_w)) = 0
\end{equation}
By substituting the two terms in eq.~\ref{eq:conservation_stochastic}:

\begin{equation}
\label{eq:conservation_stochastic_two_terms}
    \frac{\partial}{\partial t}(\Bar{S}_w+\delta S_w) + \frac{\partial}{\partial x}((\Bar{v}_d + \delta v_d)f_w(\Bar{S}_w+\delta S_w)) = 0
\end{equation}

The fractional flow function is expanded into a Taylor series using a second order approximation:

\begin{equation}
    \label{eq:taylor_f_w}
    f_w(\Bar{S}_w+\delta S_w) \sim f_w(\Bar{S}_w) + f_w^{\prime}(\Bar{S}_w)(\delta S_w) + \frac{1}{2}f_w^{\prime\prime}(\Bar{S}_w)(\delta S_w)^2+O((\delta S_w)^3)
\end{equation}
Where $f_w^{\prime}(S_w) = df_w/dS_w$

Substituting in eq.~\ref{eq:conservation_stochastic_two_terms}:

\begin{equation}
\label{eq:conservation_stochastic_expanded}
\begin{split}
    \frac{\partial \Bar{S}_w}{\partial t}+\frac{\partial \delta S_w}{\partial t} + \frac{\partial}{\partial x}\left[\Bar{v}_d f_w(\Bar{S}_w)\right] + \frac{\partial}{\partial x}\left[ f_w^{\prime}(\Bar{S}_w)\Bar{v}_d\delta S_w + f_w(\Bar{S}_w)\delta v_d\right]  +  \\
    \frac{\partial}{\partial x}\left[f_w^{\prime}(\Bar{S}_w)\delta v_d\delta S_w + \frac{1}{2}f_w^{\prime\prime}(\Bar{S}_w)\Bar{v}_d(\delta S_w)^2\right] =& 0
\end{split}
\end{equation}

In order to compute the first moment (mean), we calculate the expected value of eq.~\ref{eq:conservation_stochastic_expanded}:

\begin{equation}
\label{eq:cons_stoch_expected}
    \frac{\partial \Bar{S}_w}{\partial t} + \frac{\partial}{\partial x}\left[\Bar{v}_d f_w(\Bar{S}_w)\right] + \frac{\partial}{\partial x}\left[f_w^{\prime}(\Bar{S}_w)\langle\delta v_d\delta S_w\rangle + \frac{1}{2}f_w^{\prime\prime}(\Bar{S}_w)\Bar{v}_d\langle(\delta S_w)^2\rangle\right]= 0
\end{equation}

The expected value of product terms $\langle\delta v_d\delta S_w\rangle$ between velocity and saturation are the co-variance. We neglect the second derivative term $f_w^{\prime\prime}(\Bar{S}_w)$. We can derive a similar PDE for the fluctuation term $\delta S_w$ by calculating the difference between  eq.~\ref{eq:conservation_stochastic_expanded} and eq.~\ref{eq:cons_stoch_expected}

\begin{equation}
\label{eq:cons_stoch_fluctuation}
    \frac{\partial \delta S_w}{\partial t} + \frac{\partial}{\partial x}\left[\delta v_d f_w(\Bar{S}_w) + f_w^{\prime}(\Bar{S}_w)\delta S_w\Bar{v}_d\right] + \frac{\partial}{\partial x}\left[(\delta v_d\delta S_w-\langle\delta v_d\delta S_w\rangle)f_w^{\prime}(\Bar{S}_w)\right]= 0
\end{equation}

Once again, if we neglect $\delta v_d\delta S_w-\langle\delta v_d\delta S_w\rangle$, eq.~\ref{eq:cons_stoch_fluctuation} becomes:

\begin{equation}
\label{eq:cons_stoch_fluctuation_simple}
    \frac{\partial \delta S_w}{\partial t} + f_w^{\prime}(\Bar{S}_w)\Bar{v}_d\frac{\partial\delta S_w}{\partial x} + f_w^{\prime}(\Bar{S}_w)\delta v_d\frac{\partial\Bar{ S}_w}{\partial x}= 0
\end{equation}

This equation can be solved using Green's function as shown in \cite{Langlo1994} with form:
\begin{equation}
    \label{eq:Green_S}
    \mathcal{G}(x,\xi,t,\tau) = \mathcal{H}(t-\tau)\delta(x-\xi-\int_{\tau}^{t}f_w^{\prime}(\Bar{S}_w)\Bar{v}_ddt)
\end{equation}

Where $\mathcal{H}$ is the Heavyside step function and $\delta$ the Dirac delta function. Both operators can be defined in terms of their relative derivatives:
\begin{equation}
    \delta(x) = \frac{d}{dx}H(x) = \frac{d}{dx}max(x,0)
\end{equation}

The integral in eq.~\ref{eq:Green_S} can be simplified if we assume that $f_w^{\prime}(\Bar{S}_w)$ is time independent. This leads to:
\begin{equation}
    \int_{\tau}^{t}f_w^{\prime}(\Bar{S}_w)\Bar{v}_ddt \sim f_w^{\prime}(\Bar{S}_w)\Bar{v}_d\Delta t
\end{equation}
with $\Delta t = t-\tau$

The solution of eq.~\ref{eq:cons_stoch_fluctuation_simple} can then be developed from Green's function \ref{eq:Green_S}:

\begin{equation}
\label{eq:Green_sol}
\begin{split}
    \delta S_w(x,t) &= \int\int\mathcal{G}(x,\xi,t,\tau)\left(-f_w^{\prime}(\Bar{S}_w)\delta v_d\frac{\partial\Bar{S}_w}{\partial \xi}\right)d\xi d\tau\\
    &= \int\int\mathcal{H}(t-\tau)\delta(x-\xi-f_w^{\prime}(\Bar{S}_w)\Bar{v}_d\Delta t)\left(-f_w^{\prime}(\Bar{S}_w)\delta v_d\frac{\partial\Bar{S}_w}{\partial \xi}\right)d\xi d\tau
    \end{split}
\end{equation}

Using the same argument of time independence for $f_w^{\prime}(\Bar{S}_w)$ and assuming $\frac{\partial\Bar{ S}_w}{\partial \xi}$ changes slowly on the characteristics (\cite{dagan1984}, \cite{kitanidis1988}), we can simplify eq.~\ref{eq:Green_sol}:
\begin{equation}
    \label{eq:deltaS}
    \delta S_w(x,t) = -f_w^{\prime}(\Bar{S}_w)\int_0^t\delta v_d\frac{\partial\Bar{S}_w}{\partial \Bar{x}}d\tau
\end{equation}
Where $\Bar{x} = x -f_w^{\prime}(\Bar{S}_w)\Bar{v}_d\Delta t$

Eq.~\ref{eq:deltaS} allows to evaluate the co-variance $\langle\delta v_d\delta S_w\rangle$:
\begin{equation}
\label{eq:S_vd_covariance}
    \begin{split}
        \langle\delta v_d\delta S_w\rangle &= -\langle f_w^{\prime}(\Bar{S}_w)\int_0^t\delta v_d(x)\frac{\partial\Bar{S}_w}{\partial x}\delta v_d(\Bar{x})d\tau\rangle\\
        &= - f_w^{\prime}(\Bar{S}_w)\frac{\partial\Bar{S}_w}{\partial x}\left(\int_0^t\langle \delta v_d(x)\delta v_d(\Bar{x})\rangle d\tau\right)\\
        &= -\frac{1}{\Bar{v}_d}\frac{\partial\Bar{S}_w}{\partial x}\int_0^x v_{xx}(\xi) d\xi
    \end{split}
\end{equation}

Assuming a stationary isotropic exponential log-permeability field defined by the log transmissivity $Y=ln(T)$:
\begin{equation}
    C_Y(|x|) = \sigma_Y^2\exp(-|x|/s)
\end{equation}
For convenience, we write $x=|x|/s$ where s is the scaling factor. The velocity co-variances can be derived (\cite{rubin1990}):

\begin{equation}
    v_{xx} = \frac{\Bar{v}_d^2}{2}\sigma_Y^2\left[e^{-x}\left(\frac{6}{x^2}+\frac{18}{x^3}+\frac{18}{x^4}\right)+\frac{3}{x^2}-\frac{18}{x^4}\right]
\end{equation}

This allows to explicitly compute the integral defined in eq.~\ref{eq:S_vd_covariance}:
\begin{equation}
    \int_0^x v_{xx}(\xi) d\xi = \frac{\Bar{v}_d^2}{2}\sigma_Y^2s\left[-e^{-x}\left(\frac{6}{x^2}+\frac{6}{x^3}\right)-\frac{3}{x}+\frac{6}{x^3}+2\right]
\end{equation}

Replacing the co-variance definition in eq.~\ref{eq:cons_stoch_expected} we obtain:

\begin{equation}
    \label{eq:stochastic_1d_parabolic}
    \frac{\partial \Bar{S}_w}{\partial t} + \frac{\partial}{\partial x}\left[\Bar{v}_d f_w(\Bar{S}_w)\right] - \frac{\partial}{\partial x}\left[f_w^{\prime,2}(\Bar{S}_w)\frac{\Bar{v}_d}{2}\sigma_Y^2s\left(-e^{-x}\left(\frac{6}{x^2}+\frac{6}{x^3}\right)-\frac{3}{x}+\frac{6}{x^3}+2\right)\frac{\partial \Bar{S}_w}{\partial x}\right]= 0
\end{equation}

Note that eq.~\ref{eq:stochastic_1d_parabolic} is parabolic and that the diffusion term is dependent on the square of the fractional flow. This equation can be solved using finite differences, We propose to use PINN to do so. We use the diffusion term to our advantage in order to establish a solution for the first moment for the saturation solution.

\subsection{Resolution using PINN}
We solve jointly eq.~\ref{eq:stochastic_1d_parabolic} and \ref{eq:cons_stoch_fluctuation_simple} using a deterministic PINN. We use a network architecture similar to the one used for the resolution of the deterministic problem but with two output (mean and variance). The problem is now entirely defined by the parameters $(x,t)$ as shown in figure~\ref{fig:mlp_proba_moments}.

The loss function used is:

\begin{equation}
    \mathcal{L} = \mathcal{L}_{\mu} + \mathcal{L}_{\sigma}
\end{equation}

where 
\begin{equation}
    \mathcal{L}_{\mu} = \sum_{i=1}^{N}\left\Vert\frac{\partial \mu_s^{(i)}}{\partial t^{(i)}} + \frac{\partial}{\partial x}\left[\Bar{v}_d f_w(\mu_s^{(i)})\right] - \frac{\partial}{\partial x^{(i)}}\left[f_w^{\prime,2}(\mu_s^{(i)})\frac{\Bar{v}_d}{2}\sigma_Y^2s\left(-exp(-x^{(i)})\left(\frac{6}{x^{(i),2}}+\frac{6}{x^{(i),3}}\right)-\frac{3}{x^{(i)}}+\frac{6}{x^{(i),3}}+2\right)\frac{\partial \mu_s^{(i)}}{\partial x^{(i)}}\right]\right\Vert^2
\end{equation}

and

\begin{equation}
    \mathcal{L}_{\sigma} = \sum_{i=1}^{N}\left\Vert\frac{\partial \sigma_s^{(i)}}{\partial t^{(i)}} + \frac{\partial}{\partial x^{(i)}}\left[f_w^{\prime}(\mu_s^{(i)})\Bar{v}_d\sigma_s^{(i)}+f_w(\mu_s^{(i)})\delta v_d\mu_s^{(i)}\right]\right\Vert^2
\end{equation}

\begin{figure}[h!]%
    \centering
    \includegraphics[width=1\linewidth]{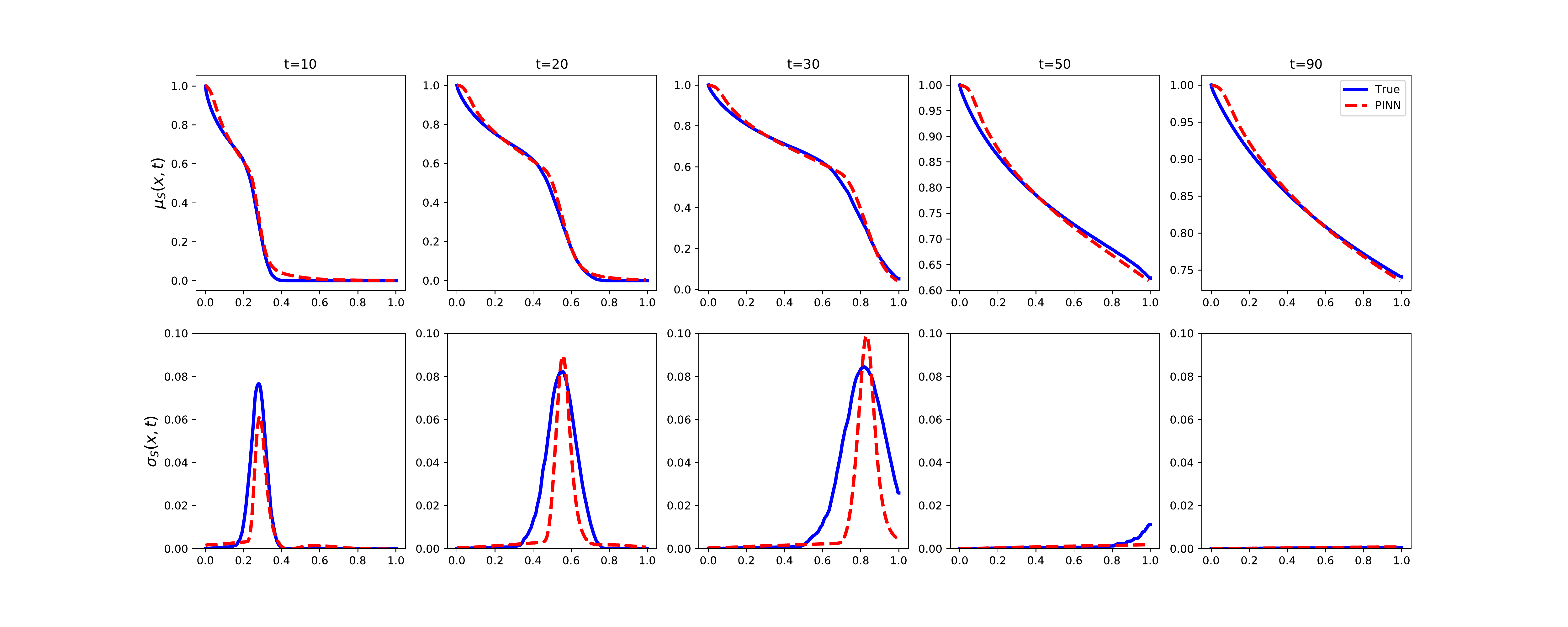}%
    \caption{Comparison of mean saturation profiles (top) and variance (bottom) at five different time steps for a heterogeneous random field simulated using a Gaussian process with exponential covariance $C_v(\|x\|) = \sigma_Y^2exp(-\|x\|/s)$ with $\sigma^2=0.1$. The reference saturation moments (computed through Monte Carlo simulation of a finite volume numerical model) is in solid blue while the ones computed with PINNs are in dashed red}%
    \label{fig:moments_BL_resolution}%
\end{figure}

We use the convex hull function of the fractional flow as defined in figure.~\ref{fig:frac_flow_welge}.
$(\mu_s(x,t),\sigma_s(x,t))$ are the output of a neural network (fig~\ref{fig:mlp_proba_moments}). We minimize $\mathcal{L}$ in a least square sense. Results are presented in figure~\ref{fig:moments_BL_resolution}

\begin{figure}[h!]%
    \centering
    \includegraphics[width=1\linewidth]{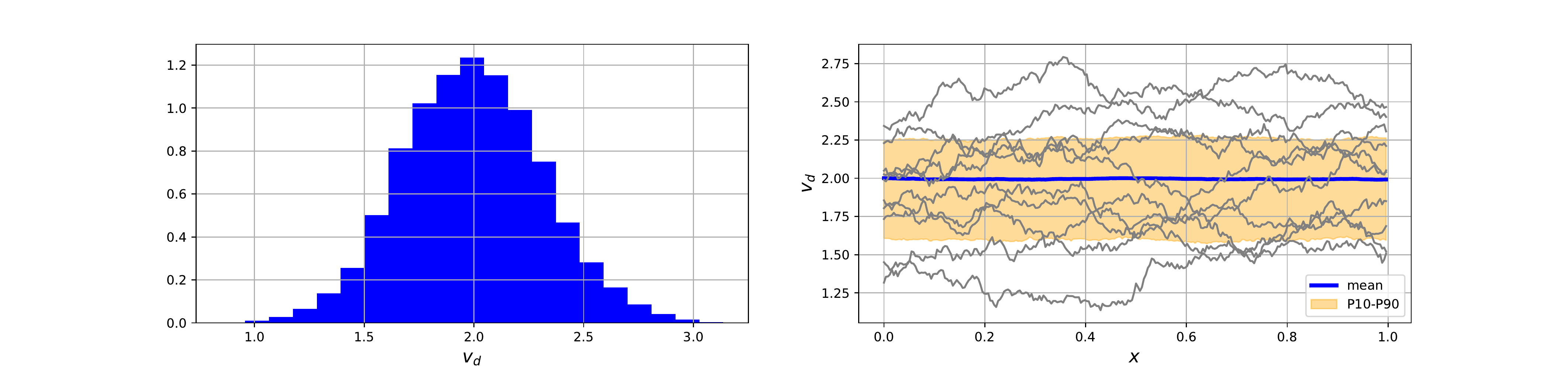}%
    \caption{Distribution of total velocities probability density (left) and along x-axis (right) for a random field modeled using a Gaussian process with exponential correlation structure, mean $s=2$ and a variance $\sigma^2=0.1$}%
    \label{fig:v_d_het_corr_exp}%
\end{figure}

The log transmissivity is simulated using a Gaussian field with exponential correlation:
\begin{equation}
    C_Y(\|x\|) = \sigma_Y^2exp(-\|x\|/s)    
\end{equation}
An example of such a field with $\sigma_Y^2=0.1$ and $s=2$ leads to a seepage velocity field presented in figure~\ref{fig:v_d_het_corr_exp}. We refer to \cite{rubin1990} to establish a correlation between log transmissivity $Y$, and seepage velocity. We represent a few realizations (in gray) in order to give a sense of the spatial variability one can expect from such a model.

We observe a close match between the reference mean solution and the one simulated with PINN while the variance displays slight differences (although peak variances are captured).

\begin{figure}[h!]%
    \centering
    \includegraphics[width=1\linewidth]{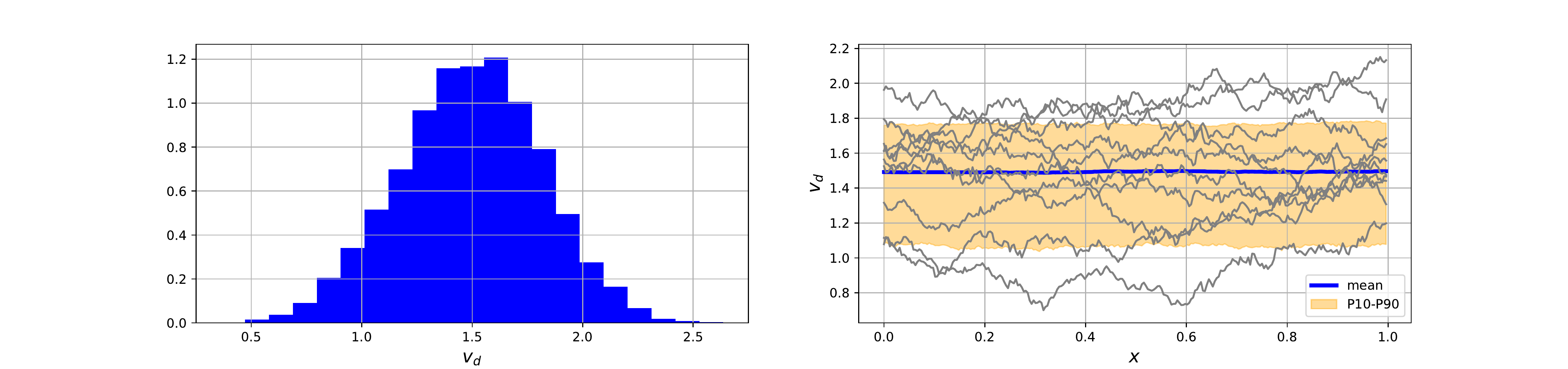}%
    \caption{Distribution of total velocities probability density (left) and along x-axis (right) for a random field modeled using a Gaussian process with exponential correlation structure and a mean $s=1.5$}%
    \label{fig:v_d_het_corr_exp_sigma_01}%
\end{figure}

We perform a sensitivity analysis around the mean of the velocity field. We decrease the mean velocity to ($s=1.5$). This corresponds to the velocity field represented in Figure~\ref{fig:v_d_het_corr_exp_sigma_01}

\begin{figure}[h!]%
    \centering
    \includegraphics[width=1\linewidth]{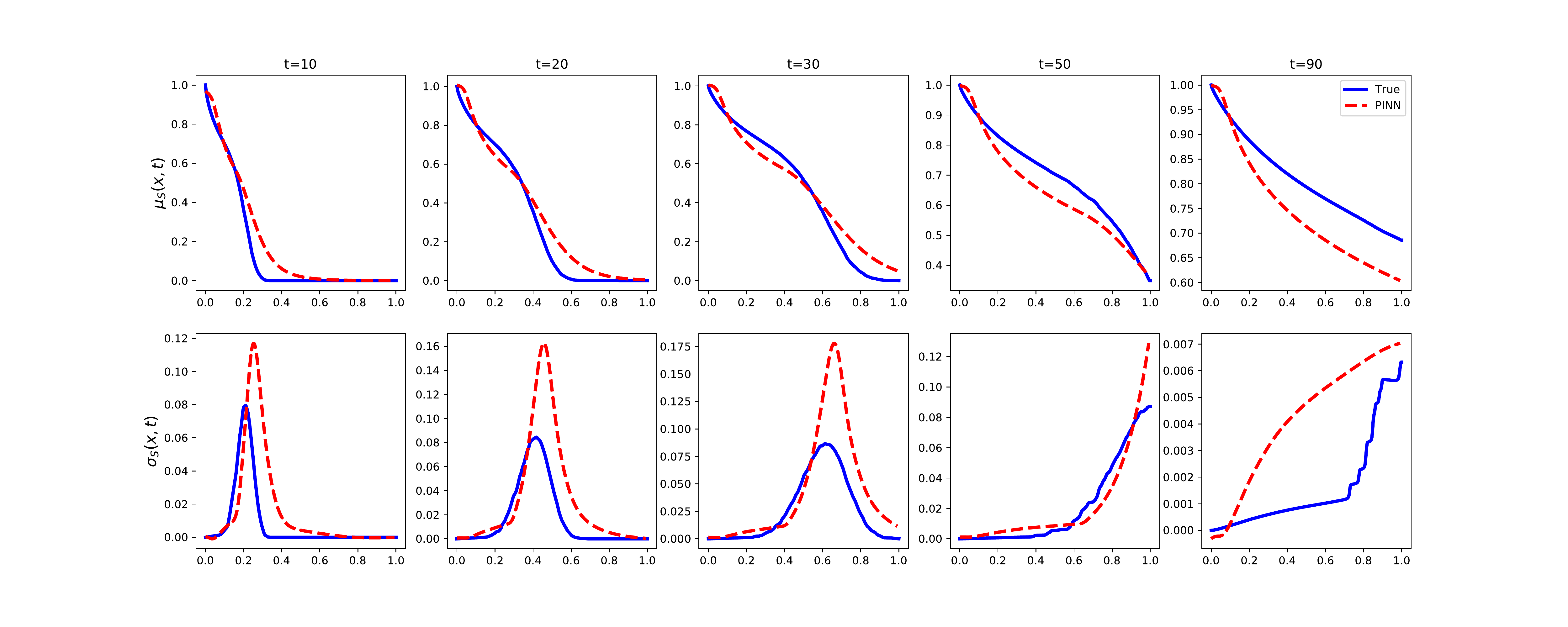}%
    \caption{Comparison of mean saturation profiles (top) and variance (bottom) at five different time steps for a heterogeneous random field simulated using a Gaussian process with exponential covariance $C_v(\|x\|) = \sigma_Y^2exp(-\|x\|/s)$ with $s=1.5$. The reference saturation moments (computed through Monte Carlo simulation of a finite volume numerical model) is in solid blue while the ones computed with PINNs are in dashed red}%
    \label{fig:moments_BL_resolution_01}%
\end{figure}

The corresponding stochastic moments are represented in Figure~\ref{fig:moments_BL_resolution_01}.

We notice that the performances degrade as we decrease the velocity of the velocity field. This is also true when we increase the variance of the velocity field. This could be due to the fact that the method of moments relies on a first order approximation for the random variables. This hypothesis is valid when the magnitude of the perturbation in the input and output are small. It is one of the main limitations of the method.

\subsection{Error Analysis}
\begin{figure}[h!]%
    \centering
    \includegraphics[width=1\linewidth]{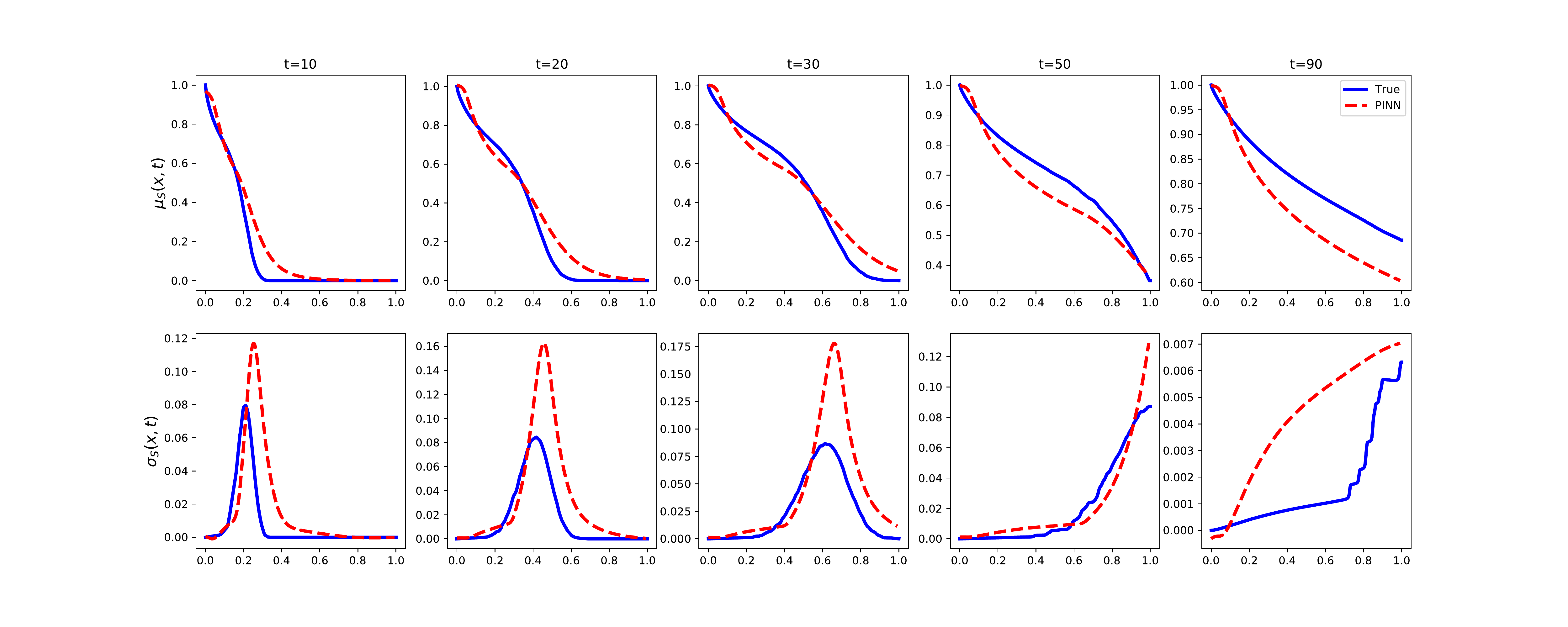}%
    \caption{Comparison of mean saturation profiles (top) and variance (bottom) at five different time steps for a heterogeneous random field simulated using a Gaussian process with exponential covariance $C_v(\|x\|) = \sigma_Y^2exp(-\|x\|/s)$ with $s=1.5$. The reference saturation moments (computed through Monte Carlo simulation of a finite volume numerical model) is in solid blue while the ones computed with PINNs are in dashed red}%
    \label{fig:moments_BLcomparison_error}%
\end{figure}

In this section, we present an error comparison analysis of the proposed approach as the moments to be simulated vary in magnitude. Our base case is the case presented in figure~\ref{fig:v_d_het_corr_exp}. We proceed to increase both the mean and standard deviation of the velocity field. Figure~\ref{fig:moments_BLcomparison_error} shows the variation in standard error ($e_{std}$) and the coefficient of determination $r^2$ defined as $e_{std} = \sqrt{(y-y^{ref})^2}$  and
\begin{equation}
    r = \frac{\sum_i(y_i-\Bar{y})(y^{ref}_i-\Bar{y^{ref}})}{\sqrt{\sum_i(y_i-\Bar{y})^2}\sqrt{\sum_i(y^{ref}_i-\Bar{y^{ref}})^2}}
\end{equation}
where $y$ is the QOI considered (saturation mean or variance in our case) for the reference case and the case computed with P-PINN and  $\Bar{y}$ is the mean of $y$.

\subsection{Performance comparison}

Table~\ref{tab:MCSvsPPINN_het} presents a performance comparison between the two methods. We do not include performances for the approach based on moments computations because the gain comes from the problem translation and we are not recovering the full realization ensemble in that case. The advantages of that approach have been thoroughly documented. The discrete problem is solved on a 256 spatial grid block mesh with 100 time steps using a finite volume method with Godunov scheme.

\begin{table}
	\caption{Performance comparison between MCS and P-PINN approach for Heterogeneous Buckley-Leverett problem}
	\centering
	\begin{tabular}{lll}
		\toprule
		     & Monte Carlo Simulation     & Parameterized PINN \\
		\midrule
		Hardware & Intel-i7 3.2GHz (6 core)  & Tesla V-100 (16GB)     \\
		Training     & - & 20 min      \\
		Inference     & 86 min       & $\sim$10s  \\
		\midrule
		Total (per 1000 simulations) & 86 min & $\leq$16 min\\
		\bottomrule
	\end{tabular}
	\label{tab:MCSvsPPINN_het}
\end{table}

All P-PINN examples run time are similar in time. We observe a sharp decline in loss terms in the first 5000 iterations (3 min) of each run and then let the model run for a larger number of iterations to ensure a better capture of the shock sharpness.

\section{Conclusion}
In this work, we explore the application of Physics Informed Neural Networks for the resolution of the stochastic Buckley-Leverett problem. The results we present indicate that provided an adequate parameterization of the input uncertainty space can be explicitly formulated, a single model can be trained to infer an ensemble of saturation realizations with very good accuracy. We use to our advantage the interpolation properties of neural networks to solve a problem where sharp gradients are traded for additional dimensions. In the case of a random velocity field, we show that additional information about the random process used to generate the realization can be used to solve the problem. This can be used to recover ensemble realizations in the case where we can explicit the parameters as input to the model or to recover the output statistical moments if we have information about the correlation structure of the data. This opens a series of possible extensions. One of them is to explore further the parametrization of the stochastic field using the principal components of the covariance matrix for the stochastic process at play. Another one if to extend the method to non parametric fields and to model the probability distribution function.








\newpage

\bibliographystyle{unsrt}
\bibliography{references}  






\end{document}